\documentclass{lmcs}

\usepackage[T1]{fontenc}
\usepackage{amssymb}
\usepackage{tikz}
\usetikzlibrary{calc}
\definecolor{scrK0}{HTML}{21BFAF}
\definecolor{scrK1}{HTML}{024873}
\definecolor{scrK2}{HTML}{001D55}
\definecolor{scrK3}{HTML}{03738C}
\definecolor{scrK4}{HTML}{12999E}
\definecolor{scrK5}{HTML}{013364}
\definecolor{scrK6}{HTML}{035E80}
\newcommand{\scrStar}[3]{%
  \pgfmathsetmacro{\scrR}{0.17*sqrt(#3)}\pgfmathsetmacro{\scrr}{0.24*\scrR}%
  \fill (#1,#2) +(0:\scrR) -- +(45:\scrr) -- +(90:\scrR) -- +(135:\scrr)
    -- +(180:\scrR) -- +(225:\scrr) -- +(270:\scrR) -- +(315:\scrr) -- cycle;}
\newcommand{\scrCapsule}[3]{%
  let \p1=($#2-#1$), \n1={atan2(\y1,\x1)} in
  ($#1+(\n1+90:#3)$) arc[start angle=\n1+90,end angle=\n1+270,radius=#3]
  -- ($#2+(\n1-90:#3)$) arc[start angle=\n1-90,end angle=\n1+90,radius=#3] -- cycle}
\newcommand{\scrDiamond}[1]{%
  \draw[line width=0.9pt] (0,-2.75) -- (2.75,0) -- (0,2.75) -- (-2.75,0) -- cycle;
  \draw[line width=0.3pt] (0,-2.9) -- (2.9,0) -- (0,2.9) -- (-2.9,0) -- cycle;
  \node[font=\Large] at (0,-3.45) {#1};}
\newcommand{\scrPts}{\foreach \u/\a in {-1/m,0/z,1/p} \foreach \v/\b in {-1/m,0/z,1/p}
  \coordinate (q\a\b) at ({(\u-\v)*0.78},{(\u+\v)*0.78});}
\newcommand{\scrStars}{\foreach \u in {-1,0,1} \foreach \v in {-1,0,1} {%
  \pgfmathtruncatemacro{\w}{ifthenelse(\u==0 && \v==0,2,1)}%
  \pgfmathsetmacro{\xx}{(\u-\v)*0.78}\pgfmathsetmacro{\yy}{(\u+\v)*0.78}%
  \scrStar{\xx}{\yy}{\w}}}
\tikzset{scrEdge/.style={line width=0.35pt,black!60},
         scrLink/.style={line width=0.45pt,black!65,dash pattern=on 2.2pt off 1.6pt},
         scrLab/.style={font=\large}}

\usetikzlibrary{arrows.meta,decorations.pathreplacing}
\newcommand{\scrStarP}[3]{%
  \pgfmathsetmacro{\scrR}{0.092*sqrt(#3)}\pgfmathsetmacro{\scrr}{0.24*\scrR}%
  \fill (#1,#2) +(0:\scrR) -- +(45:\scrr) -- +(90:\scrR) -- +(135:\scrr)
    -- +(180:\scrR) -- +(225:\scrr) -- +(270:\scrR) -- +(315:\scrr) -- cycle;}
\tikzset{scrVerb/.style={font=\tiny\itshape,inner sep=1pt}}
\usepackage{mathrsfs}

\makeatletter
\def\endfront@text{}
\def\enddoc@text{}
\def\@settitlecomment{}

\def\ps@firstpage{\let\@mkboth\@gobbletwo\def\@oddhead{\vbox to\headheight{\vss\hbox{\vrule width\hsize depth .4pt}}}%
  \def\@evenhead{}\def\@oddfoot{}\def\@evenfoot{}}
\makeatother

\newtheorem{theorem}{Theorem}[section]
\newtheorem{lemma}[theorem]{Lemma}
\newtheorem{proposition}[theorem]{Proposition}
\newtheorem{corollary}[theorem]{Corollary}
\theoremstyle{definition}
\newtheorem{definition}[theorem]{Definition}
\newtheorem{convention}[theorem]{Convention}
\theoremstyle{remark}

\newcommand{\Rr}{\mathbb{R}}
\newcommand{\Qq}{\mathbb{Q}}
\newcommand{\Cc}{\mathbb{C}}
\newcommand{\Nn}{\mathbb{N}}
\newcommand{\Ff}{\mathbb{F}_2}
\newcommand{\Ee}{\mathcal{E}}
\newcommand{\Mm}{\mathcal{M}}
\newcommand{\Aa}{\mathcal{A}}
\newcommand{\Se}{S_{\mathsf e}}
\def\sz(#1){N_{#1}}
\newcommand{\rf}{\varrho}
\newcommand{\Ob}{\mathrm{Ob}(T)}
\newcommand{\ms}[1]{\{\!\!\{\,#1\,\}\!\!\}}

\usepackage{etoolbox}
\makeatletter
\patchcmd{\@setauthors}{\author@andify\authors}{\def\and{\unskip,\space\ignorespaces}}{}{}
\DeclareRobustCommand\addrsym{\ifcase\value{addressctr}\or\dagger\or\ddagger\or\S\fi}
\patchcmd{\@setaddresses}{\alph{addressctr}}{\addrsym}{}{}
\makeatother

\begin{document}

\title[What an Observer Reads off a Holographic Screen]
      {What an Observer Reads off a Holographic Screen}
\author{Kenan Oggad}[\dagger]
\address{Universit\'{e} Paris-Saclay}
\email{kenan.oggad@universite-paris-saclay.fr}

\author{James F. Glazebrook}[\ddagger]
\address{Department of Mathematics and Computer Science, Eastern Illinois University}
\email{jfglazebrook@eiu.edu}

\author{Chris Fields}[\S]
\address{Allen Discovery Center, Tufts University}
\email{fieldsres@gmail.com}

\begin{abstract}
What does an observer read off a holographic screen that is an array of qubits rather than a
surface? In the generalized holographic principle the interaction between two separable systems acts
on such a screen as a weighted sum, up to an overall prefactor, of commuting observables with values plus and minus one. When the
weights are nonnegative and the prefactor is one and the joint eigenspaces are lines, everything a
probe reads off the spectrum depends only on the list of weights, which we call the profile. A
relabelling of the qubits leaves the spectrum unchanged at every profile while a regrouping can
change it. On the screen with equal weights summing to one the running dimension read by a diffusion
probe falls short of two at unit time by at most twice the reciprocal of the number of qubits, and the
peak of this dimension over time grows with that number; a geometric profile gives instead, through its level count, a plateau of any positive height. An observer correlates sets of qubits; its readout carries a profile at a given weighting when the levels it reads, counted with multiplicity, are those of a screen with one weight per set. We prove that a readout carries a profile at every weighting exactly when the distinct
nonzero traces of its qubits are linearly independent over the field with two elements, the trace of a
qubit being the list of sets that contain it; such a readout is a partition of the qubits it covers up to an invertible recoding, and a dependent one carries a profile only on finitely many proper linear subspaces of
weightings. Refining a grouping without end builds a tower whose boundary is a profinite set and hence
a condensed set, on which a strictly positive compatible profile separates the locally constant
functions by a distance. Applications certify probabilistic and strong contextuality and give the
leading shortfall of the distance between the poles of the fuzzy sphere; in a model of monogamous entanglement the pairs that weight-one threads representing operational ER=EPR can occupy at once are exactly the partial matchings.
\end{abstract}

\maketitle

\section{Introduction}\label{sec:intro}

\begin{figure}[!htb]
\centering
\resizebox{\linewidth}{!}{%
\begin{tikzpicture}
\begin{scope}
  \scrDiamond{(a)} \scrStars
\end{scope}
\begin{scope}[xshift=6.1cm]
  \scrDiamond{(b)} \scrPts
  \begin{scope}[fill opacity=0.5]
    \fill[scrK0] \scrCapsule{(qmm)}{(qmz)}{0.36};
    \fill[scrK1] (qmp) circle (0.36);
    \fill[scrK2] (qzm) circle (0.36);
    \fill[scrK3] (qzz) circle (0.36);
    \fill[scrK4] (qzp) circle (0.36);
    \fill[scrK5] (qpm) circle (0.36);
    \fill[scrK6] \scrCapsule{(qpz)}{(qpp)}{0.36};
  \end{scope}
  \draw[scrEdge] \scrCapsule{(qmm)}{(qmz)}{0.36};
  \foreach \c in {qmp,qzm,qzz,qzp,qpm} \draw[scrEdge] (\c) circle (0.36);
  \draw[scrEdge] \scrCapsule{(qpz)}{(qpp)}{0.36};
  \draw[scrLink] (qmm) -- (qmz);
  \draw[scrLink] (qpz) -- (qpp);
  \scrStars
  \node[scrLab,text=scrK0!85!black] at (-0.8,-1.55) {$A_0$};
  \node[scrLab,text=scrK1!85!black] at (-1.85,0.5) {$A_1$};
  \node[scrLab,text=scrK2!85!black] at (1.2,-1.2) {$A_2$};
  \node[scrLab,text=scrK3!85!black] at (0.62,0) {$A_3$};
  \node[scrLab,text=scrK4!85!black] at (-1.2,1.2) {$A_4$};
  \node[scrLab,text=scrK5!85!black] at (1.85,0.5) {$A_5$};
  \node[scrLab,text=scrK6!85!black] at (0.8,1.55) {$A_6$};
\end{scope}
\begin{scope}[xshift=12.2cm]
  \scrDiamond{(c)} \scrPts
  \begin{scope}[fill opacity=0.5]
    \fill[scrK0] \scrCapsule{(qmm)}{(qmp)}{0.36};
    \fill[scrK1] \scrCapsule{(qzm)}{(qzp)}{0.36};
    \fill[scrK2] \scrCapsule{(qpm)}{(qpp)}{0.36};
  \end{scope}
  \draw[scrEdge] \scrCapsule{(qmm)}{(qmp)}{0.36};
  \draw[scrEdge] \scrCapsule{(qzm)}{(qzp)}{0.36};
  \draw[scrEdge] \scrCapsule{(qpm)}{(qpp)}{0.36};
  \draw[scrLink] (qmm) -- (qmp);
  \draw[scrLink] (qzm) -- (qzp);
  \draw[scrLink] (qpm) -- (qpp);
  \scrStars
  \node[scrLab,text=scrK0!85!black] at (-2.0,0.44) {$A_0$};
  \node[scrLab,text=scrK1!85!black] at (-1.22,1.22) {$A_1$};
  \node[scrLab,text=scrK2!85!black] at (-0.44,2.0) {$A_2$};
\end{scope}
\begin{scope}[xshift=18.3cm]
  \scrDiamond{(d)} \scrPts
  \begin{scope}[fill opacity=0.4]
    \fill[scrK0] \scrCapsule{(qmp)}{(qpm)}{0.36};
    \fill[scrK1] \scrCapsule{(qmz)}{(qpz)}{0.36};
    \fill[scrK2] \scrCapsule{(qmm)}{(qpp)}{0.36};
    \fill[scrK3] \scrCapsule{(qzm)}{(qzp)}{0.36};
  \end{scope}
  \draw[scrEdge] \scrCapsule{(qmp)}{(qpm)}{0.36};
  \draw[scrEdge] \scrCapsule{(qmz)}{(qpz)}{0.36};
  \draw[scrEdge] \scrCapsule{(qmm)}{(qpp)}{0.36};
  \draw[scrEdge] \scrCapsule{(qzm)}{(qzp)}{0.36};
  \draw[scrLink] (qmp) -- (qpm);
  \draw[scrLink] (qmz) -- (qpz);
  \draw[scrLink] (qmm) -- (qpp);
  \draw[scrLink] (qzm) -- (qzp);
  \scrStars
  \node[scrLab,text=scrK0!85!black] at (-1.9,0.45) {$A_0$};
  \node[scrLab,text=scrK1!85!black] at (1.2,1.15) {$A_1$};
  \node[scrLab,text=scrK2!85!black] at (0.45,1.9) {$A_2$};
  \node[scrLab,text=scrK3!85!black] at (-1.2,1.15) {$A_3$};
\end{scope}
\end{tikzpicture}}
\caption{Each star is a qubit whose area is proportional to its weight, and each coloured region is
one set of a readout. (a) Nine qubits drawn in a three-by-three array that only names them, the central qubit twice as heavy as the others. (b) A grouping into seven sets, two pairs and five singletons, each qubit lying in exactly one of them (Proposition~\ref{prop:readout}). (c) The three rows of the array, which
are unions of sets of (b), namely $A_0\cup A_1$, then $A_2\cup A_3\cup A_4$, then $A_5\cup A_6$, and again form a grouping. (d) The four lines of the array through the
central qubit, each holding two of the eight outer qubits, so each outer qubit is on exactly one of them; an outer qubit on the $j$-th line has trace $\mathbf{e}_j$ and the central qubit has trace $\mathbf{e}_0+\mathbf{e}_1+\mathbf{e}_2+\mathbf{e}_3$ (Definition~\ref{def:trace}), so the five distinct traces sum to zero and
five is odd, so this readout carries a profile at no weighting with all weights positive
(Theorem~\ref{thm:oddgeneral}).}
\label{fig:sky}
\end{figure}
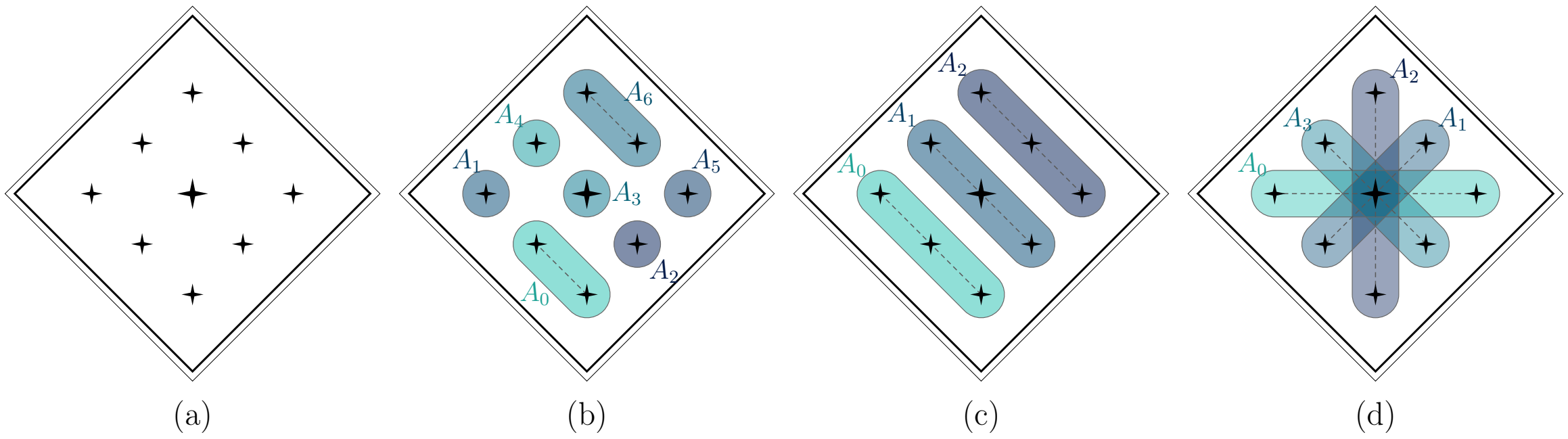

Consider two interacting systems that exchange information through a screen of qubits between them; each qubit
flashes plus or minus one, and the interaction may weigh some qubits more heavily than others. The screen can be metaphorically thought of as a sort of quantum-scale sky whose stars are the qubits, each as bright as its weight and none with a position (Figure~\ref{fig:sky}). An observer reads the sky through constellations fixed by its own measuring process; they may hold one star or many and may share stars. It perceives each constellation only as a whole, through the product of the values its stars flash. Each minus one flips the sign of that product, so it is negative exactly when an odd number of its stars flash minus one. A hundred stars with one at minus one therefore show the same value as a hundred with forty-nine, one and forty-nine being both odd. The single stars of that constellation tell the two apart, since the two differ at some star and a single star shows its own value. These constellations are arbitrary discretisations of the sky that may overlap, so two observers stargazing at the same sky can carry different constellations and read different levels off it.

Record for each star the list of constellations that contain it. Every choice of constellations registers a level equal to twice the total brightness of the stars that lie in an odd number of the chosen constellations. The observer reads faithfully when these levels with their multiplicities are those of a sky with one star per constellation, some of those stars possibly dark. This happens for every choice of brightnesses exactly when none of the different nonempty lists can be rebuilt by combining others, where
a constellation named an even number of times drops out; such an observer is then in effect cutting
the stars it sees into non-overlapping pieces. When some nonempty list can be rebuilt from others,
faithful readings need finely tuned brightnesses. When an odd number of different nonempty lists
cancel out together, no positive brightnesses give a faithful reading at all. As the observer's measuring process resolves each star into finer stars without end, the sky becomes a dust of points, and what the observer can hold on to there are values that are settled after finitely many resolutions. Two of these
settled values differ somewhere in the dust exactly when they are told apart by every weighing that gives every star at every stage a positive weight, with each star weighing as much as its finer stars together. We take reality for this observer to be separation and concordance, the distinctions its readings
draw and the agreement of those readings as its view is refined.

The screen belongs to a line of work begun by 't~Hooft, for whom the observable degrees of
freedom of quantum gravity at the Planck scale are best described as if they were Boolean variables on
a two-dimensional lattice \cite{tHooft1993}, and continued by Susskind, for whom the
three-dimensional world is an image of data stored on a two-dimensional projection
\cite{Susskind1995}, and by Maldacena and Susskind, who read two black holes joined through their
interior by an Einstein--Rosen bridge as an entangled EPR pair \cite{MaldacenaSusskind2013} (see also, e.g., \cite{Bousso1999}).
Fields, Glazebrook and Marcian\`o place the classical information that two separable systems exchange on a holographic screen $\mathscr{B}$ separating them, realised as an array of noninteracting qubits \cite{FieldsGlazebrookMarciano2022} (Figure~\ref{fig:screen}); the interaction acts on the screen as a weighted sum, up to an overall prefactor, of commuting observables with values plus and minus one, one for each qubit. We call the list of weights the profile of the screen and any assignment of real
weights to the qubits a weighting. The same authors state that the interaction is symmetric under every
permutation of the qubits \cite{FieldsGlazebrookMarciano2022}, with terms that can be listed in any
order \cite{FieldsGlazebrookMarciano2021}, and that a quantum reference frame breaks this symmetry
\cite{FieldsGlazebrookMarciano2022}. We ask what an observer reads off such a screen and how much of
what it reads belongs to the screen.

When the weights are nonnegative and the prefactor is one and each joint outcome of the observables
picks out a single state, the spectrum measured from its lowest level depends on the profile alone
whatever basis, reference angle per qubit or realisation of the observables is chosen
(Proposition~\ref{prop:cube}, Theorem~\ref{thm:signfamily}, Proposition~\ref{prop:phi}); a probe that
reads the screen through its spectrum therefore reads only the profile (Corollary~\ref{cor:split}). A
diffusion probe reads a running dimension from the rate at which the heat trace of the spectrum
decays. At equal weights summing to one this dimension stays below twice the elapsed time, which is
the line that Oggad \cite{Oggad2026} reads off a growing structure with a single relaxation time equal
to one; at unit time it falls short of two by at most twice the reciprocal of the number of qubits,
and its peak over time grows in proportion to that number (Corollary~\ref{cor:uniform-line},
Proposition~\ref{prop:uniform}). For a geometric profile count the levels up to a given value with their multiplicities; along a sequence of values this count grows like a fixed power of the value up to constant factors (Theorem~\ref{thm:plateau}). Convention~\ref{conv:dial} reads twice that exponent as the height of a plateau; this height equals two when each weight is half the one before and can be any positive number for a suitable ratio. At halving weights the running dimension itself stays below two at
every time, and on a large enough screen it comes as close to two as one likes over a stretch of time
that grows with the screen (Theorem~\ref{thm:dyadic}).

A relabelling of the qubits leaves the spectrum unchanged at every profile and so moves no reading
(Proposition~\ref{prop:relabel}), while once the observables are taken as spins along one axis the interaction itself is unchanged by swapping two qubits exactly when their weights are equal (Proposition~\ref{prop:swap}). A grouping sends each qubit to one of finitely many blocks and gives each block the total weight of its qubits; two groupings of one screen can give different spectra (Proposition~\ref{prop:groupings-separate}), and grouping screens of
equal weights summing to one produces exactly the nonnegative rational profiles of total one as the
number of qubits ranges over all sizes (Lemma~\ref{lem:rational}).

Formally a constellation is a set of qubits, and we use the constellations to model how a quantum reference frame breaks the swap symmetry \cite{FieldsGlazebrookMarciano2022}. Each qubit is read in its own single-qubit frame. In each run the observer obtains from a set $A$ one value, the product $\prod_{i\in A}s_i=\pm1$ of the outcomes of its qubits. Each outcome $-1$ flips the sign of this product, which is therefore $-1$ exactly when an odd number of them give $-1$. This value is the single-shot parity of $A$, and we call a family of such sets a readout. The sets play the part of the sectors of $\mathscr{B}$, proper subsets of the bits encoded on $\mathscr{B}$ \cite{FieldsGlazebrookMarciano2022} on which an observer's quantum reference frames act \cite{FieldsGlazebrookMarcianoZappala2025}; here a set may also be all of $\mathscr{B}$. A grouping is then the readout whose sets are its blocks, and at every weighting it reads the spectrum of the grouped screen (Proposition~\ref{prop:readout}).

In these terms the list of constellations containing a star is the trace of a qubit; a qubit enters every reading only through its weight and its trace (Lemma~\ref{lem:levelchar}). Lists combine by addition over the field with two elements. Reading faithfully is carrying a profile at a
weighting (Definition~\ref{def:readout}), where the screen with one qubit per set may have real weights that need not
sum to one. An observer reads faithfully at every weighting exactly when the distinct nonzero traces are linearly independent
over that field (Theorem~\ref{thm:dichotomy}), and the non-overlapping pieces arise after an invertible
recoding of the choices of sets (Corollary~\ref{cor:partition}). Independence is not needed for a
single weighting (Proposition~\ref{cor:quantifier}), but without it the weightings that still work lie
on finitely many proper linear subspaces (Corollary~\ref{cor:thin}). A readout with an odd number of distinct nonzero traces summing to zero carries a profile at no positive weighting (Theorem~\ref{thm:oddgeneral}, Figure~\ref{fig:sky}(d)).

Since groupings compose, refining one without end builds a tower of stages in which each stage is a finite list of sites grouped onto those of the stage before. The dust is the boundary of this tower, the compatible choices of one site per stage; this boundary is a profinite set and hence a condensed set in
the sense of Clausen and Scholze \cite{Scholze2026} (Theorem~\ref{thm:screen-profinite}). Passing to
condensed sets neither loses nor adds continuous maps between such boundaries, whether set-theoretic
size is handled by a Grothendieck universe or bounded by any uncountable strong limit cardinal
(Theorem~\ref{thm:screen-ff}). Over the boundary every principal bundle is trivial and every path is constant (Proposition~\ref{prop:boundary-trivial}). A compatible profile, a compatible family of profiles in Definition~\ref{def:profilefamily}, is one weighting per stage, and under each weighting a site carries the total weight of the next-stage sites mapped onto it; a compatible profile integrates the locally constant functions on the boundary as a finitely additive
measure, signed in general (Theorem~\ref{thm:sheaf-functional}, Corollary~\ref{cor:clopenmeasure}).
These functions are exactly the condensed maps into the discrete real line
(Theorem~\ref{thm:bridge-formal}), and pushing weights along a refinement is adjoint to pulling
functions back (Theorem~\ref{thm:adjoint-formal}). The settled values are the locally constant functions on the boundary, which we call the objects of the boundary; they are exactly the functions pulled back from assignments of a real value to each site of a finite stage, and they survive refinement (Theorem~\ref{thm:objects-formal},
Theorem~\ref{thm:persist-refine}). Weighings that give every site at every stage a positive weight are the strictly positive compatible profiles, which exist on every tower and separate any two different objects by the integral of the absolute value of their difference (Theorem~\ref{thm:metric-separates}, Lemma~\ref{lem:positive-family}), whereas a profile concentrated at one point of the boundary
cannot tell apart two objects that agree at that point (Theorem~\ref{thm:metric-degenerate}). For outcomes taken from a finite set of at least two elements a
strict refinement embeds the assignments of a coarser stage into those of a finer one injectively but
not onto, and the embedding preserves and reflects their labelled relations
(Theorem~\ref{thm:lts-embedding}, Theorem~\ref{thm:verdict}).

Abramsky and Brandenburger prove that strong contextuality implies possibilistic contextuality and
that possibilistic implies probabilistic, and they place the Bell model they consider at the weakest
level \cite{AbramskyBrandenburger2011}. A nonnegative real solution of a system of linear equations
with nonnegative coefficients yields a boolean solution of the system obtained by replacing every entry with whether it is nonzero (Proposition~\ref{prop:vvb}, Corollary~\ref{cor:placement}), which gives the same chain of implications for
the models used here. That Bell model is probabilistically but not strongly contextual. On a cover by nine contexts of four measurements, each of the eighteen measurements lying in exactly two contexts, every model whose possible outcomes make exactly one measurement true in each context is strongly contextual (Corollary~\ref{cor:placement}).

A web is our model of monogamous entanglement among several parties; it is a symmetric table of
nonnegative entries indexed by the parties with zeros on the diagonal and rows summing to at most one,
a normalisation modelled on the monogamy inequality that Coffman, Kundu and Wootters conjectured
\cite{CoffmanKunduWootters2000} and Osborne and Verstraete proved \cite{OsborneVerstraete2006}. A
thread of a party is an entry of its row in the column of another party; in a web with at least two
parties every party has a thread no heavier than one over the number of other parties
(Theorem~\ref{thm:thin}), and no party has two threads heavier than one half
(Theorem~\ref{thm:bridge-matching}). Threads of weight one model the monogamous pair of the
operational $\mathrm{ER}=\mathrm{EPR}$ \cite{FieldsGlazebrookMarcianoZappala2025} and by this second
bound never meet at a party; since every partial matching is realised by some web (Proposition~\ref{prop:matching-bridges}), the sets of pairs that weight-one threads can occupy at once are exactly the partial matchings.

D'Andrea, Lizzi and V\'arilly give in closed form the Connes distance between the two poles of the
fuzzy sphere for its own Dirac operator \cite{DAndreaLizziVarilly2013}, and this distance tends to the
distance between the poles of the round unit sphere while falling short of it by an amount that decays
as the inverse square root of the matrix size with leading coefficient minus twice the Riemann zeta
function at one half (Theorem~\ref{thm:zeta}). For an AF algebra with a faithful tracial state
Aguilar and Latr\'emoli\`ere bound the quantum propinquity between a stage and the limit using metric
structures on both that are built from a chosen positive sequence tending to zero
\cite{AguilarLatremoliere2015}; their bound is the value of that sequence at the stage
\cite{AguilarLatremoliere2015}, so the bound decays exactly as the chosen sequence does, and no single
sequence of numbers is that bound for every admissible choice (Proposition~\ref{prop:af}, Corollary~\ref{cor:kind}).

\section{The screen is its spectrum}\label{sec:spectrum}

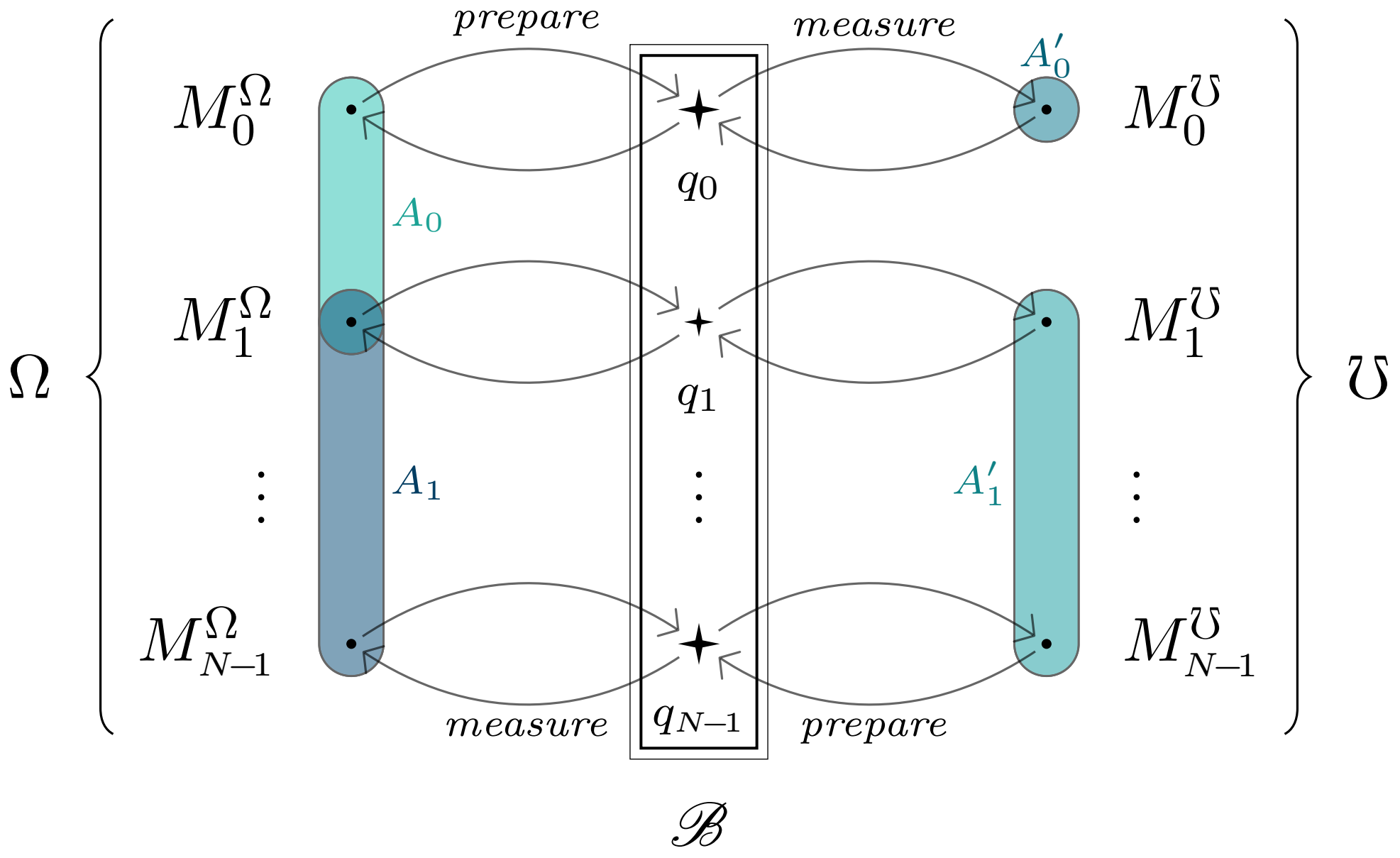
\begin{figure}[!htb]
\centering
\scalebox{1.5}{%
\begin{tikzpicture}[scale=0.8]
\def\scrRows{3.3/0/2, 1.65/1/1, -0.85/{\scriptscriptstyle N\!-\!1}/2}
\tikzset{scrCap/.style={line width=0.35pt,draw=black!60,fill opacity=0.5,draw opacity=1,rounded corners=0.2cm}}
\fill[scrCap,fill=scrK0] (-2.95,1.4) rectangle (-2.45,3.55);
\fill[scrCap,fill=scrK1] (-2.95,-1.1) rectangle (-2.45,1.9);
\draw[scrCap] (-2.95,1.4) rectangle (-2.45,3.55);
\draw[scrCap] (-2.95,-1.1) rectangle (-2.45,1.9);
\fill[scrCap,fill=scrK3] (2.45,3.05) rectangle (2.95,3.55);
\fill[scrCap,fill=scrK4] (2.45,-1.1) rectangle (2.95,1.9);
\draw[scrCap] (2.45,3.05) rectangle (2.95,3.55);
\draw[scrCap] (2.45,-1.1) rectangle (2.95,1.9);
\node[font=\tiny,text=scrK0!85!black] at (-2.18,2.48) {$A_0$};
\node[font=\tiny,text=scrK1!85!black] at (-2.18,0.4) {$A_1$};
\node[font=\tiny,text=scrK3!85!black] at (2.7,3.72) {$A'_0$};
\node[font=\tiny,text=scrK4!85!black] at (2.18,0.4) {$A'_1$};
\tikzset{scrMor/.style={line width=0.4pt,draw opacity=0.6,
  -{Straight Barb[length=0.85mm,width=2mm,line cap=round,line join=round]}}}
\newcommand{\scrArcUp}[3]{\draw[scrMor,#3] (#1) .. controls ($(#1)!0.333!(#2)+(0,0.6)$) and ($(#1)!0.667!(#2)+(0,0.6)$) .. (#2);}
\newcommand{\scrArcDn}[3]{\draw[scrMor,#3] (#1) .. controls ($(#1)!0.333!(#2)+(0,-0.6)$) and ($(#1)!0.667!(#2)+(0,-0.6)$) .. (#2);}
\foreach \y/\i/\w in \scrRows {
  \coordinate (dl) at (-2.7,\y); \coordinate (q) at (0,\y); \coordinate (dr) at (2.7,\y);
  \fill (dl) circle (1.1pt); \fill (dr) circle (1.1pt);
  \scrArcUp{dl}{q}{shorten <=2.4pt,shorten >=4.2pt}
  \scrArcDn{q}{dl}{shorten <=4.2pt,shorten >=2.4pt}
  \scrArcUp{q}{dr}{shorten <=4.2pt,shorten >=2.4pt}
  \scrArcDn{dr}{q}{shorten <=2.4pt,shorten >=4.2pt}
}
\draw[line width=0.5pt]  (-0.45,-1.66) rectangle (0.45,3.72);
\draw[line width=0.17pt] (-0.535,-1.745) rectangle (0.535,3.805);
\node at (0,-2.25) {$\mathscr{B}$};
\draw[line width=0.4pt,decorate,decoration={brace,amplitude=4.5pt,raise=0pt}]
  (-4.55,-1.55) -- (-4.55,4.0);
\draw[line width=0.4pt,decorate,decoration={brace,amplitude=4.5pt,raise=0pt}]
  (4.55,4.0) -- (4.55,-1.55);
\node at (-5.2,1.225) {$\Omega$};
\node at ( 5.2,1.225) {$\mho$};
\foreach \y/\i/\w in \scrRows {
  \scrStarP{0}{\y}{1.5625*\w}
  \node[font=\scriptsize,inner sep=0pt] at (0,\y-0.6) {$q_{\i}$};
}
\node[anchor=east] at (-3.12,3.3)  {$M^{\Omega}_{0}$};
\node[anchor=east] at (-3.12,1.65) {$M^{\Omega}_{1}$};
\node[anchor=east] at (-3.12,-0.85) {$M^{\Omega}_{\scriptscriptstyle N\!-\!1}$};
\node[anchor=west] at ( 3.12,3.3)  {$M^{\mho}_{0}$};
\node[anchor=west] at ( 3.12,1.65) {$M^{\mho}_{1}$};
\node[anchor=west] at ( 3.12,-0.85) {$M^{\mho}_{\scriptscriptstyle N\!-\!1}$};
\node[scrVerb,above] at (-1.35,3.82) {prepare};
\node[scrVerb,above] at ( 1.35,3.82) {measure};
\node[scrVerb,below] at (-1.35,-1.38) {measure};
\node[scrVerb,below] at ( 1.35,-1.38) {prepare};
\foreach \x in {-3.4,0,3.4} \node at (\x,0.42) {$\vdots$};
\end{tikzpicture}}
\caption{A holographic screen $\mathscr{B}$ separating systems $\Omega$ and $\mho$ can be realised by an
array of $N$ noninteracting qubits $q_0,\dots,q_{N-1}$, drawn as stars whose areas are proportional to the
weights $\alpha^k_i$ of \eqref{eq:ham} for one side $k$, as in Figure~\ref{fig:sky}. Each operator
$M^k_i$, drawn as a dot, is joined to the star of $q_i$ by two arrows: along the upper
arrows $\Omega$ prepares $q_i$ through $M^\Omega_i$ and $\mho$ measures it through
$M^\mho_i$, and along the lower arrows $\mho$ prepares and $\Omega$ measures. The
capsules around operators mark readouts as in Figure~\ref{fig:sky}, each side reading the parity of
the outcomes of its operators in every set: the sets $A_0=\{q_0,q_1\}$ and $A_1=\{q_1,\dots,q_{N-1}\}$ of
$\Omega$ share $q_1$, while the sets $A'_0=\{q_0\}$ and $A'_1=\{q_1,\dots,q_{N-1}\}$ of $\mho$ are
disjoint and form a grouping (Proposition~\ref{prop:readout}). $\Omega$ and $\mho$ need not share preparation and measurement bases, that is, quantum
reference frames. Adapted from Fields, Glazebrook and Marcian\`o \cite[Fig.~1]{FieldsGlazebrookMarciano2022}.}
\label{fig:screen}
\end{figure}

Writing $k$ for either of the two systems $\Omega$ and $\mho$ the screen $\mathscr{B}$ (Figure~\ref{fig:screen}) separates, the interaction Hamiltonian of
the generalized holographic principle is \cite{FieldsGlazebrookMarciano2022}
\begin{equation}\label{eq:ham}
H_{\Omega\mho} \;=\; \mathfrak{e}_k \sum_{i\in[N]} \alpha^{k}_{i} M^{k}_{i},
\end{equation}
where the prefactor $\mathfrak{e}_k>0$ is the product of an inverse measure of the thermodynamic
efficiency of $k$, Boltzmann's constant and the temperature of the environment,
$[N]:=\{0,1,\dots,N-1\}$ indexes the $N$ sites of $\mathscr{B}$, the $M^{k}_{i}$ are pairwise commuting with
eigenvalues in $\{-1,+1\}$, and the weights lie in $[0,1]$ with $\sum_i\alpha^k_i=1$; the same
weighted form is equation~(2) of \cite{FieldsGlazebrookMarciano2021}, and the operational
$\mathrm{ER}=\mathrm{EPR}$ theorem takes the screen as a finite array of qubits
\cite{FieldsGlazebrookMarcianoZappala2025}. The prefactor enters as an overall scale and a change of
basis conjugates the $M_i$, so when the joint eigenspaces of the $M_i$ are one-dimensional the profile
$\alpha=(\alpha_0,\dots,\alpha_{N-1})$ is what the form leaves free in the shape of the spectrum
(Proposition~\ref{prop:cube}); with every weight equal to $1/N$ the form is
$(\mathfrak{e}_k/N)\sum_iM^k_i$, the unweighted sum of the $M^k_i$ up to its prefactor. This
form is the equal-weight screen, of profile $u_N\equiv1/N$, and writing each $M^k_i$ as a $\phi$-spin operator leaves
its spectrum unchanged (Proposition~\ref{prop:phi}). The nonnegative profiles of total one with rational weights are exactly
those obtained from some $u_N$ by grouping its sites into blocks, each block carrying the total
weight of its sites (Definition~\ref{def:push}, Lemma~\ref{lem:rational}). The
form \eqref{eq:ham} gives each side $k$ its own profile $\alpha^k$, and we fix one side. For a proposition $P$, $[P]$ is $1$ when $P$ holds and $0$ otherwise.

\begin{lemma}\label{lem:freedoms}
Let $H$ be an operator on a finite-dimensional space. For $c\neq0$ the eigenspace of $cH$ at $c\mu$ is
the eigenspace of $H$ at $\mu$, and for every invertible $U$ the operator $UHU^{-1}$ has the
characteristic polynomial of $H$, hence its eigenvalues with their algebraic multiplicities.
\end{lemma}

\begin{proof}
$cHv=c\mu v$ holds exactly when $Hv=\mu v$, multiplying by $c^{-1}$; and
$\det(X-UHU^{-1})=\det\bigl(U(X-H)U^{-1}\bigr)=\det(X-H)$.
\end{proof}

So the prefactor $\mathfrak{e}_k>0$ rescales the spectrum without changing its shape and can be set
to one once and for all, and a change of basis, which replaces each $M_i$ by $UM_iU^{-1}$ for one
invertible $U$ and hence $H_{\Omega\mho}$ by $\sum_i\alpha_iUM_iU^{-1}=UH_{\Omega\mho}U^{-1}$, moves no eigenvalue
and no multiplicity. We
drop the index $k$ from now on.

\begin{proposition}\label{prop:cube}
Set the prefactor of \eqref{eq:ham} to one, let $\lambda_{\min}(H_{\Omega\mho})$ be the least eigenvalue of
$H_{\Omega\mho}$ and $L := H_{\Omega\mho} - \lambda_{\min}(H_{\Omega\mho})$, let every
$\alpha_i\ge0$, and take the $M_i$ to be commuting self-adjoint single-qubit observables on $N$
independent qubits, each with spectrum $\{\pm1\}$, whose $2^{N}$ joint eigenspaces are all
one-dimensional. Then
\[
\operatorname{spec}(L) \;=\; \Bigl\{\, 2\sum_{i \in S} \alpha_i \;:\; S \subseteq [N] \,\Bigr\},
\]
the multiplicity of a value being the number of subsets realising it; term for term this is the
Laplacian spectrum of the $N$-cube whose edges in direction $i$ carry weight $\alpha_i$.
\end{proposition}

\begin{proof}
The $M_i$ are self-adjoint and commute pairwise, so by the spectral theorem for a commuting family
of self-adjoint operators on a finite-dimensional space they are simultaneously diagonalisable: the
space is the orthogonal direct sum of the joint eigenspaces
$E_s:=\{v : M_iv=s_iv\ \text{for all } i\}$. Each $M_i$ has spectrum in $\{\pm1\}$, so the label
$s$ of a nonzero joint eigenspace is a sign vector $s\in\{\pm1\}^N$; by hypothesis each of the
$2^N$ spaces $E_s$ is a line, spanned by a unit vector $e_s$, and the $e_s$ form an orthonormal basis.

On $E_s$ the operator $\sum_i\alpha_iM_i$ acts by the scalar
$\langle s,\alpha\rangle:=\sum_i s_i\alpha_i$, so in the basis $(e_s)$ it is diagonal with these
entries, and its eigenvalues with algebraic multiplicity are the $2^N$ numbers
$\langle s,\alpha\rangle$, one per sign vector. Since $s_i\ge-1$ and $\alpha_i\ge0$, every term
satisfies $s_i\alpha_i\ge-\alpha_i$, hence $\langle s,\alpha\rangle\ge-\sum_i\alpha_i$, with
equality at $s=(-1,\dots,-1)$; so $\lambda_{\min}=-\sum_i\alpha_i$ and it is attained. Therefore
$L$ acts on $e_s$ by
\[
\langle s,\alpha\rangle-\lambda_{\min}\;=\;\sum_i(s_i+1)\alpha_i\;=\;2\sum_{i:s_i=+1}\alpha_i ,
\]
since $s_i+1$ is $2$ when $s_i=+1$ and $0$ when $s_i=-1$. The map $s\mapsto S=\{i:s_i=+1\}$ is a
bijection from $\{\pm1\}^N$ onto the subsets of $[N]$, its inverse setting $s_i=+1$
exactly for $i\in S$. Hence the multiplicity of a value $\mu$ in $\operatorname{spec}(L)$, the
number of basis vectors $e_s$ on which $L$ acts by $\mu$, equals the number of subsets $S$ with
$2\sum_{i\in S}\alpha_i=\mu$.

For the cube, let $Q_N$ have vertex set $\{\pm1\}^N$ and an edge of weight $\alpha_i$ between $s$ and
the vector $s^{(i)}$ obtained by flipping its $i$-th coordinate. Its Laplacian acts on functions by
$(\Delta f)(s)=\sum_i\alpha_i\bigl(f(s)-f(s^{(i)})\bigr)$. For $S\subseteq[N]$ the Walsh
character $\chi_S(s):=\prod_{i\in S}s_i$ satisfies $\chi_S(s^{(i)})=-\chi_S(s)$ if $i\in S$ and
$\chi_S(s^{(i)})=\chi_S(s)$ otherwise, so $\Delta\chi_S=\bigl(2\sum_{i\in S}\alpha_i\bigr)\chi_S$.
The $2^N$ characters are pairwise orthogonal: for $S\ne S'$ the product $\chi_S\chi_{S'}$ is
$\chi_{S\triangle S'}$ because $s_i^2=1$, and flipping one coordinate $i\in S\triangle S'$ is a
bijection of $\{\pm1\}^N$ that negates $\chi_{S\triangle S'}$, so its sum over $\{\pm1\}^N$ equals its
own negative and vanishes. Being $2^N$ orthogonal nonzero vectors in a space of dimension $2^N$, the
characters form an eigenbasis of $\Delta$, and the eigenvalue at $\chi_S$ is the eigenvalue of $L$
at $e_s$ for the $s$ corresponding to $S$.
\end{proof}

\begin{corollary}\label{cor:heat}
Under the hypothesis of Proposition~\ref{prop:cube}, the heat trace of $L$ is a product over the
weights,
\[
\Theta_{\mathrm{full}}(t) \;=\; \operatorname{Tr} e^{-tL} \;=\; \prod_{i\in[N]}\bigl(1 + e^{-2t\alpha_i}\bigr),
\qquad
\Theta_{\mathrm{nz}}(t) \;:=\; \Theta_{\mathrm{full}}(t) - 1 \;=\; \sum_{S\neq\varnothing}e^{-2t\sum_{i\in S}\alpha_i},
\]
the trace with the ground state $S=\varnothing$ removed; and the running spectral dimension
$D_S(t) := -2\,\mathrm{d}\log\Theta/\mathrm{d}\log t$ is, the second formula holding for $N\ge1$,
\begin{equation}\label{eq:ds}
D_S^{\mathrm{full}}(t) \;=\; \sum_{i\in[N]} \frac{4t\alpha_i}{1 + e^{2t\alpha_i}},
\qquad
D_S^{\mathrm{nz}}(t) \;=\; D_S^{\mathrm{full}}(t)\cdot\frac{\Theta_{\mathrm{full}}(t)}{\Theta_{\mathrm{full}}(t)-1}.
\end{equation}
\end{corollary}

\begin{proof}
In the orthonormal eigenbasis $(e_s)$ of the proof of Proposition~\ref{prop:cube}, $L$ is diagonal
with entries $\lambda_s=2\sum_{i\in S}\alpha_i$, so the exponential series gives $e^{-tL}$ diagonal
with entries $e^{-t\lambda_s}$, and the trace, which does not depend on the basis, is
\[
\operatorname{Tr}e^{-tL}=\sum_{S}e^{-2t\sum_{i\in S}\alpha_i}=\sum_S\prod_{i\in S}e^{-2t\alpha_i}
=\prod_{i\in[N]}\bigl(1+e^{-2t\alpha_i}\bigr),
\]
the last equality being the expansion of the product, in which each subset $S$ records the factors
from which the second term $e^{-2t\alpha_i}$ is taken; the term of $S=\varnothing$ is $1$, so the
sum over $S\neq\varnothing$ is $\Theta_{\mathrm{full}}-1$. Every factor exceeds one, so
$\Theta_{\mathrm{full}}>0$ and $\log\Theta_{\mathrm{full}}=\sum_i\log(1+e^{-2t\alpha_i})$. Since
$\mathrm{d}/\mathrm{d}\log t=t\,\mathrm{d}/\mathrm{d}t$,
\[
-2t\,\frac{\mathrm{d}}{\mathrm{d}t}\log\Theta_{\mathrm{full}}
=-2t\sum_i\frac{-2\alpha_ie^{-2t\alpha_i}}{1+e^{-2t\alpha_i}}
=\sum_i\frac{4t\alpha_i}{e^{2t\alpha_i}+1},
\]
multiplying numerator and denominator by $e^{2t\alpha_i}$ in the last step. For $N\ge1$ the product
exceeds one, so $\Theta_{\mathrm{nz}}=\Theta_{\mathrm{full}}-1>0$, and
$\tfrac{\mathrm{d}}{\mathrm{d}t}\log(\Theta-1)=\Theta'/(\Theta-1)
=\tfrac{\Theta}{\Theta-1}\cdot\tfrac{\mathrm{d}}{\mathrm{d}t}\log\Theta$; multiplying by $-2t$ gives
the second formula of \eqref{eq:ds}.
\end{proof}

\begin{lemma}\label{lem:rescale}
Under the hypothesis of Proposition~\ref{prop:cube}, for every $c>0$ and $t>0$ the profile
$c\alpha$ has $\Theta_{\mathrm{full}}(t;c\alpha)=\Theta_{\mathrm{full}}(ct;\alpha)$,
$D_S^{\mathrm{full}}(t;c\alpha)=D_S^{\mathrm{full}}(ct;\alpha)$ and, for $N\ge1$,
$D_S^{\mathrm{nz}}(t;c\alpha)=D_S^{\mathrm{nz}}(ct;\alpha)$.
\end{lemma}

\begin{proof}
The profile $c\alpha$ again has nonnegative weights, so Corollary~\ref{cor:heat} applies to it
and to $\alpha$, and gives
$\Theta_{\mathrm{full}}(t;c\alpha)=\prod_i(1+e^{-2tc\alpha_i})=\Theta_{\mathrm{full}}(ct;\alpha)$ and
$D_S^{\mathrm{full}}(t;c\alpha)=\sum_i4tc\alpha_i/(1+e^{2tc\alpha_i})=D_S^{\mathrm{full}}(ct;\alpha)$.
The second formula of \eqref{eq:ds} expresses $D_S^{\mathrm{nz}}$ at one time through
$D_S^{\mathrm{full}}$ and $\Theta_{\mathrm{full}}$ at the same time, so it transforms in the same way.
\end{proof}

The graph of either running dimension against $\log t$ is therefore translated by
$-\log c$ and otherwise unchanged: a positive multiple $c\alpha$ of a profile reads the same heat
trace and running dimensions as $\alpha$ on the clock rescaled by $c$; in particular the unweighted
screen, every weight equal to one, has the heat trace and running
dimensions of the equal-weight profile $\alpha_i=1/N$ on a clock rescaled by $N$. The hypothesis of Proposition~\ref{prop:cube} names one
realisation, and four of its properties define the class of realisations.

\begin{definition}\label{def:signfamily}
Let $E$ be a finite-dimensional inner product space over $\mathbb{K}\in\{\Rr,\Cc\}$. A family
$M : [N] \to \mathrm{End}(E)$ is a \emph{sign family} when each $M_i$ is self-adjoint, the family
commutes pairwise, each $M_i$ squares to the identity, and every one of the $2^{N}$ joint
eigenspaces is a line.
\end{definition}

Here the joint eigenspace at $s\in\{\pm1\}^N$ is $E_s:=\{v\in E: M_iv=s_iv \text{ for all } i\}$.

\begin{proposition}\label{prop:signfamily-inhabited}
On the space with orthonormal basis $(e_s)_{s\in\{\pm1\}^N}$ over $\mathbb{K}$, the operators
$Z_ie_s:=s_ie_s$ form a sign family. The space has dimension $2^N$, and $E_s$ is the line
through $e_s$.
\end{proposition}

\begin{proof}
The operators are diagonal with real entries, hence self-adjoint and pairwise commuting; they square
to the identity because $s_i^2=1$; and a vector $v=\sum_tc_te_t$ with $Z_iv=s_iv$ for every $i$ has
$c_t(t_i-s_i)=0$ for all $i$ and $t$, so $c_t=0$ unless $t=s$, and $E_s$ is the line through $e_s$.
The basis has $2^N$ elements.
\end{proof}

The $z$-spin reading of the source is this family \cite{FieldsGlazebrookMarciano2022}.

For $\alpha:[N]\to\Rr$ write $\Sigma_N(\alpha)$ for the multiset of the $2\sum_{i\in S}\alpha_i$,
one entry per subset $S\subseteq[N]$.

\begin{theorem}\label{thm:signfamily}
Let $M$ be a sign family on $E$ and $\alpha : [N] \to \Rr$. Then $\dim E = 2^{N}$, and with
$\widetilde L := \sum_i \alpha_i M_i + \sum_i \alpha_i$ the conclusions of Proposition~\ref{prop:cube} and
Corollary~\ref{cor:heat} hold for $\widetilde L$ in place of $L$: the spectrum of $\widetilde L$ with
multiplicity is $\Sigma_N(\alpha)$, the heat trace $\operatorname{Tr}e^{-t\widetilde L}$ is
$\prod_i (1 + e^{-2t\alpha_i})$, and both forms of \eqref{eq:ds} follow. If every
$\alpha_i \geq 0$ then $\lambda_{\min}(\sum_i\alpha_iM_i) = -\sum_i \alpha_i$, attained, so
$\widetilde L = \sum_i \alpha_i M_i - \lambda_{\min}$. For a profile of either sign
$\lambda_{\min} = -\sum_i |\alpha_i|$, and shifting by it instead moves every value of
$\Sigma_N(\alpha)$ up by $2\sum_{\alpha_i<0}|\alpha_i|$, so the two shifts agree exactly when every
$\alpha_i \geq 0$.
\end{theorem}

\begin{proof}
\emph{Eigenvalues of each $M_i$.} If $M_iv=\mu v$ with $v\ne0$ then $v=M_i^2v=\mu^2v$, so
$\mu^2=1$; self-adjointness makes $\mu$ real, so $\mu\in\{-1,+1\}$.

\emph{The joint eigenspaces exhaust $E$.} We show by induction on $n\le N$ that $E$ is the sum of the
spaces $E^{(n)}_{s}:=\{v:M_iv=s_iv,\ i<n\}$, $s\in\{\pm1\}^n$. For $n=0$ there is one space, $E$
itself. Suppose the claim holds for $n-1$, where $1\le n\le N$. Each $E^{(n-1)}_s$ is invariant under
$M_{n-1}$, because $M_{n-1}$ commutes with $M_0,\dots,M_{n-2}$: if $M_iv=s_iv$ then
$M_iM_{n-1}v=M_{n-1}M_iv=s_iM_{n-1}v$. The restriction of $M_{n-1}$ to $E^{(n-1)}_s$ is self-adjoint,
so by the spectral theorem that space is the sum of its eigenspaces for $M_{n-1}$, whose eigenvalues lie in $\{\pm1\}$ by the first step; these eigenspaces are
the spaces $E^{(n)}_{(s,\pm1)}$. This completes the induction, and at $n=N$ the spaces are the $E_s$.

\emph{Orthogonality.} Let $v\in E_s$, $w\in E_{s'}$ with $s\ne s'$, and choose $i$ with
$s_i\ne s'_i$. Then $s_i\langle v,w\rangle=\langle M_iv,w\rangle=\langle v,M_iw\rangle
=s'_i\langle v,w\rangle$, the eigenvalues being real, so $(s_i-s'_i)\langle v,w\rangle=0$ and
$\langle v,w\rangle=0$.

\emph{Basis and dimension.} Each $E_s$ is a line by hypothesis; choose a unit vector $e_s\in E_s$.
The $2^N$ vectors $e_s$ are orthonormal by the previous step and span $E$ because the $E_s$ do, so
they form an orthonormal basis and $\dim E=2^N$.

\emph{Spectrum and heat trace.} In this basis every $M_i$ is diagonal with entries $s_i$, so $\widetilde L$ is
diagonal with entries $\langle s,\alpha\rangle+\sum_i\alpha_i=\sum_i(s_i+1)\alpha_i
=2\sum_{i:s_i=+1}\alpha_i$, for any real $\alpha$. Its characteristic polynomial is
$\prod_s\bigl(X-2\sum_{i:s_i=+1}\alpha_i\bigr)$, and through the bijection $s\mapsto\{i:s_i=+1\}$
its roots with multiplicity are $\Sigma_N(\alpha)$. The trace of $e^{-t\widetilde L}$, computed in the same
basis, is $\sum_S e^{-2t\sum_{i\in S}\alpha_i}=\prod_i(1+e^{-2t\alpha_i})$ as in the proof of
Corollary~\ref{cor:heat}, and the two formulas \eqref{eq:ds} follow from it by the differentiation
carried out there, which used nothing but this product. Over $\Cc$ the roots are real, being the
values $2\sum_{i\in S}\alpha_i$, so the heat trace and both running dimensions are real.

\emph{The shift.} In the same basis $\sum_i\alpha_iM_i$ is diagonal with entries
$\langle s,\alpha\rangle$, so $\lambda_{\min}$ is the least of these numbers. Termwise
$s_i\alpha_i\ge-|\alpha_i|$, with equality when $s_i=-\operatorname{sgn}\alpha_i$ (either sign when
$\alpha_i=0$); choosing $s$ in this way attains $-\sum_i|\alpha_i|$, which is therefore
$\lambda_{\min}$. When $\alpha\ge0$ this is $-\sum_i\alpha_i$, attained at $s=(-1,\dots,-1)$, and
then $\widetilde L=\sum_i\alpha_iM_i+\sum_i\alpha_i=\sum_i\alpha_iM_i-\lambda_{\min}$. In general
$\sum_i\alpha_iM_i-\lambda_{\min}=\widetilde L+\sum_i(|\alpha_i|-\alpha_i)$, and $|\alpha_i|-\alpha_i$ is
$2|\alpha_i|$ when $\alpha_i<0$ and $0$ otherwise, so every value of $\Sigma_N(\alpha)$ moves up by
$2\sum_{\alpha_i<0}|\alpha_i|$; this vanishes exactly when no $\alpha_i$ is negative.
\end{proof}

\begin{proposition}\label{prop:one-dim-sharp}
Let $Z_i$ be the family of Proposition~\ref{prop:signfamily-inhabited} over $\Rr$, on $E$. On
$E\otimes\Rr^2$ the operators $Z_i\otimes1$ are self-adjoint, commute, square to the identity and
attain both values $\pm1$, and for every $\alpha$ and $\mu$ the multiplicity of $\mu$ as an eigenvalue
of $\sum_i\alpha_i(Z_i\otimes1)+\sum_i\alpha_i$ is twice the number of subsets $S$ with
$2\sum_{i\in S}\alpha_i=\mu$. Its multiset of levels is not $\Sigma_N(\alpha)$.
\end{proposition}

\begin{proof}
In the basis $e_s\otimes e_b$, $b\in\{1,2\}$, each $Z_i\otimes1$ is diagonal with the real entry $s_i$
at $e_s\otimes e_b$, so it is self-adjoint, the family commutes, $(Z_i\otimes1)^2=Z_i^2\otimes1=1$,
and both values occur. The operator $\sum_i\alpha_i(Z_i\otimes1)+\sum_i\alpha_i$ is
$\widetilde L\otimes1$, with $\widetilde L$ as in Theorem~\ref{thm:signfamily} for the family $Z_i$,
diagonal with the entry $2\sum_{i:s_i=+1}\alpha_i$ at $e_s\otimes e_b$, so each value occurs once for
each $b$, that is twice the number of subsets realising it. At $\mu=0$, which the empty subset
realises, the multiplicity is therefore twice and not once the number of subsets realising it, so
the multiset differs from $\Sigma_N(\alpha)$.
\end{proof}

In the notation of \cite{FieldsGlazebrookMarciano2022} the $M^k_i$ are $z$-spin operators on individual qubits. There the classifier attached to each acts as its local $z$ axis, a one-bit reference frame whose choice is the choice of basis. The $\phi$-spin operators
with one reference angle per qubit collect such choices into one family, and writing each observable as
one of them leaves the spectrum unchanged (Proposition~\ref{prop:phi}). Spin operators along arbitrary axes
give the same spectrum. For a unit vector $n\in\Rr^3$ the spin operator $n\cdot\sigma$ is self-adjoint, as a real
combination of the self-adjoint Pauli matrices, squares to $\sum_{a,b}n_an_b\sigma_a\sigma_b=|n|^2\,\mathbf 1=\mathbf 1$
by $\sigma_a\sigma_b+\sigma_b\sigma_a=2\delta_{ab}\mathbf 1$, and has trace zero as each $\sigma_a$ does, so its eigenvalues are $+1$ and $-1$, and a unitary $V_n$ carrying
an orthonormal eigenbasis of it, the eigenvector of $+1$ first, to the standard basis has $V_n(n\cdot\sigma)V_n^{-1}=Z$; the tensor
product of the $V_{n_i}$ therefore conjugates $\sum_i\alpha_i\,n_i\cdot\sigma^{(i)}$, with
$\sigma^{(i)}$ acting on the $i$-th qubit, to $\sum_i\alpha_iZ_i$, and by Lemma~\ref{lem:freedoms}
the two have the same eigenvalues with their multiplicities.

\begin{proposition}\label{prop:phi}
Index the screen by sign vectors $s \in \{\pm1\}^{N}$. Let $Z_i$ be the diagonal operator reading
the $i$-th sign, $F_i$ the flip of the $i$-th coordinate, and for $\phi \in \Rr^{N}$ put
$\sigma_i(\phi) := \cos(\phi_i)\, Z_i + \sin(\phi_i)\, F_i$. Let
$r(\theta)=\bigl(\begin{smallmatrix}\cos(\theta/2)&-\sin(\theta/2)\\ \sin(\theta/2)&\cos(\theta/2)\end{smallmatrix}\bigr)$
be the half-angle rotation, indexed by the signs $+,-$, and
$R(\phi)_{s,t} := \prod_{j} r(\phi_j)_{s_j,t_j}$. Then
\[
R(\phi)\,R(\phi)^{\mathsf{T}} = 1,
\qquad
R(\phi)\, Z_i \;=\; \sigma_i(\phi)\, R(\phi),
\qquad
\sigma_i(\phi) \;=\; R(\phi)\, Z_i\, R(\phi)^{-1} .
\]
Consequently $H_\phi(\alpha):=\sum_i\alpha_i\sigma_i(\phi)$ is conjugate to
$H(\alpha):=\sum_i\alpha_iZ_i$, the two have the same characteristic polynomial, whose
roots after the shift of Theorem~\ref{thm:signfamily} are $\Sigma_N(\alpha)$ with multiplicity, and
for every assignment $\mathrm{cl}$ of a class to a spectrum with multiplicity,
$\mathrm{cl}(\operatorname{spec}H_\phi(\alpha)) = \mathrm{cl}(\operatorname{spec}H_{\phi'}(\alpha))$
for all $\phi, \phi' \in \Rr^N$.
\end{proposition}

\begin{proof}
Identify the space indexed by $\{\pm1\}^N$ with the $N$-fold tensor product of the two-dimensional
space indexed by $\{+,-\}$, the basis vector at $s$ being $e_{s_0}\otimes\cdots\otimes e_{s_{N-1}}$.
For matrices $a_0,\dots,a_{N-1}$ on the factor, the Kronecker product $a_0\otimes\cdots\otimes a_{N-1}$ has
entry $\prod_ja_j(s_j,t_j)$ at $(s,t)$, the factors taken in the order of $[N]$, so
$R(\phi)=r(\phi_0)\otimes\cdots\otimes r(\phi_{N-1})$. With
$Z=\mathrm{diag}(1,-1)$ and $F=\bigl(\begin{smallmatrix}0&1\\1&0\end{smallmatrix}\bigr)$ on the
factor, $Z_i$ and $F_i$ are the Kronecker products with $Z$, respectively $F$, in the $i$-th place and
the identity elsewhere. The mixed-product rule
$(a_0\otimes\cdots\otimes a_{N-1})(b_0\otimes\cdots\otimes b_{N-1})=a_0b_0\otimes\cdots\otimes
a_{N-1}b_{N-1}$ and the rule $(a_0\otimes\cdots\otimes a_{N-1})^{\mathsf T}=a_0^{\mathsf
T}\otimes\cdots\otimes a_{N-1}^{\mathsf T}$
reduce every claim to one factor.

Write $c=\cos(\theta/2)$ and $d=\sin(\theta/2)$. Then
$r(\theta)r(\theta)^{\mathsf T}=\bigl(\begin{smallmatrix}c^2+d^2&cd-dc\\dc-cd&d^2+c^2\end{smallmatrix}\bigr)=1$,
and
\[
r(\theta)\,Z\,r(\theta)^{\mathsf T}
=\begin{pmatrix}c&d\\d&-c\end{pmatrix}\begin{pmatrix}c&d\\-d&c\end{pmatrix}
=\begin{pmatrix}c^2-d^2&2cd\\2cd&d^2-c^2\end{pmatrix}
=\cos\theta\,Z+\sin\theta\,F,
\]
by the double-angle formulas $\cos\theta=c^2-d^2$ and $\sin\theta=2cd$. Hence
$R(\phi)R(\phi)^{\mathsf T}=\bigotimes_j r(\phi_j)r(\phi_j)^{\mathsf T}=1$, so $R(\phi)$ is
orthogonal with $R(\phi)^{-1}=R(\phi)^{\mathsf T}$; and
$R(\phi)Z_iR(\phi)^{\mathsf T}$ is the Kronecker product with
$r(\phi_i)Zr(\phi_i)^{\mathsf T}=\cos\phi_i\,Z+\sin\phi_i\,F$ in the $i$-th place and
$r(\phi_j)r(\phi_j)^{\mathsf T}=1$ elsewhere, which is $\cos\phi_i\,Z_i+\sin\phi_i\,F_i
=\sigma_i(\phi)$. Multiplying $R(\phi)Z_iR(\phi)^{-1}=\sigma_i(\phi)$ on the right by $R(\phi)$
gives the intertwining relation.

By linearity of conjugation,
$H_\phi(\alpha)=\sum_i\alpha_iR(\phi)Z_iR(\phi)^{-1}=R(\phi)\,H(\alpha)\,R(\phi)^{-1}$,
and conjugate matrices have the same characteristic polynomial,
$\det(X-RAR^{-1})=\det\bigl(R(X-A)R^{-1}\bigr)=\det(X-A)$. The $Z_i$ form a sign family, as shown
after Definition~\ref{def:signfamily}, so by Theorem~\ref{thm:signfamily} the roots of the
characteristic polynomial of $H(\alpha)+\sum_i\alpha_i$ are $\Sigma_N(\alpha)$ with
multiplicity, and the same shift applied to $H_\phi(\alpha)$ gives the same roots, the shift by a
scalar commuting with the conjugation. So $H_\phi(\alpha)$ and $H_{\phi'}(\alpha)$ have the same spectrum
with multiplicity, and $\mathrm{cl}$ takes the same value on it.
\end{proof}

The choice of basis fixes which of the $2^N$
eigenvalues is currently encoded \cite{FieldsGlazebrookMarciano2022}, not the spectrum, so by
Proposition~\ref{prop:phi} a system recomputing its reference angles on every interaction cycle runs through isospectral
screens. Once the prefactor is set to one and the joint eigenspaces are one-dimensional, the profile
is therefore the only datum of \eqref{eq:ham} on which a reading of the spectrum can depend. In particular a quantum
reference frame that is a choice of reference angle per qubit selects no dimension, and a
rescaling of the profile only translates the running dimensions along $\log t$
(Lemma~\ref{lem:rescale}), moving the position of a peak or a plateau and not its height; the plateau
number of Convention~\ref{conv:dial} below is not moved either, since multiplying every level by $c$
turns the counting function $\mathcal N(\lambda)$, the number of levels at most $\lambda$ counted
with multiplicity, into $\mathcal N(\lambda/c)$, whose exponent is the same.

For example, take a quantum reference frame that attends to a single sector of the screen, as the
frame reading the pointer state of an identified system attends to its pointer sector
\cite{FieldsGlazebrookMarciano2022}. We model it by the mask $\alpha=\mathbf{1}_{S_0}/k$, which
spreads equal weight over a set $S_0$ of $k\ge1$ qubits and gives weight zero to the other $N-k$; its
weights lie in $[0,1]$ and sum to $k\cdot(1/k)=1$, as \eqref{eq:ham} requires. In the product of Corollary~\ref{cor:heat} each ignored qubit
contributes the factor $1+e^{0}=2$, and in the sum \eqref{eq:ds} it contributes $0$. Hence
$\Theta_{\mathrm{full}}(t)=2^{N-k}(1+e^{-2t/k})^{k}$ and $D_S^{\mathrm{full}}(t)=4t/(1+e^{2t/k})$,
the running dimension of the equal-weight screen of $k$ qubits (Proposition~\ref{prop:uniform} with
$k$ in place of $N$). The factor $2^{N-k}$ drops out of the logarithmic derivative that defines $D_S$, since it adds only the constant $(N-k)\log2$ to $\log\Theta_{\mathrm{full}}$.
In this model an observer that reads the pointer state of an instrument from $k$ qubits and ignores
the rest of its environment therefore finds the running dimension $D_S^{\mathrm{full}}$ of a screen of $k$ qubits, whatever the size of that
environment. Weight one on each qubit of the sector only rescales the clock by $k$
(Lemma~\ref{lem:rescale}).

\section{The profile decides the reading}\label{sec:profiles}

\begin{proposition}\label{prop:uniform}
Under the hypothesis of Proposition~\ref{prop:cube}, for $\alpha_i = 1/N$ we have
$\Theta_{\mathrm{full}}(t) = (1+e^{-2t/N})^{N}$ and
\[
D_S^{\mathrm{full}}(t) \;=\; \frac{4t}{1+e^{2t/N}} ,
\]
which attains a single maximum at $t = Nu^{*}/2$, $u^{*}$ the unique positive root of
$e^{u}(u-1) = 1$, with value
\[
D_S^{\max} \;=\; \frac{2u^{*}}{1+e^{u^{*}}}\,N \;=\; 2\bigl(u^{*}-1\bigr)N \;=\; 2W\!\left(e^{-1}\right)N ,
\]
$W$ the Lambert function. The reading grows linearly in $N$ and admits no bound uniform in the size
of the screen.
\end{proposition}

\begin{proof}
\emph{The formulas.} With every $\alpha_i=1/N$ the $N$ factors of Corollary~\ref{cor:heat} are
equal to $1+e^{-2t/N}$, so $\Theta_{\mathrm{full}}(t)=(1+e^{-2t/N})^N$; and the $N$ summands of
\eqref{eq:ds} are equal to $4t(1/N)/(1+e^{2t/N})$, whose sum is $4t/(1+e^{2t/N})$.

\emph{Reduction to one variable.} Put $u=2t/N$, a strictly increasing bijection of $(0,\infty)$ onto
itself. Then $4t=2Nu$ and $D_S^{\mathrm{full}}=2N\,h(u)$ with $h(u):=u/(1+e^u)$, so the maxima of
$D_S^{\mathrm{full}}$ on $t>0$ are the images of the maxima of $h$ on $u>0$, with values multiplied
by $2N$. By the quotient rule,
\[
h'(u)=\frac{(1+e^u)-ue^u}{(1+e^u)^2},
\]
whose sign is that of $n(u):=1+e^u-ue^u=1-e^u(u-1)$.

\emph{The critical point.} The function $k(u):=e^u(u-1)$ has $k'(u)=ue^u>0$ for $u>0$, so it is
strictly increasing on $[0,\infty)$; $k(0)=-1$ and $k(u)\ge u-1\to\infty$. By the intermediate value
theorem $k$ takes the value $1$, and by strict monotonicity at exactly one point $u^*>0$; since
$k(1)=0<1$, $u^*>1$. Therefore $n(u)=1-k(u)$ is positive on $(0,u^*)$ and negative on
$(u^*,\infty)$, so $h$ is strictly increasing on $(0,u^*]$ and strictly decreasing on
$[u^*,\infty)$. The point $u^*$ is the unique maximiser of $h$ on $(0,\infty)$, and
$t=Nu^*/2$ the unique maximiser of $D_S^{\mathrm{full}}$.

\emph{The value.} The maximum is $2N\,u^*/(1+e^{u^*})$. From $e^{u^*}(u^*-1)=1$ we get
$e^{u^*}=1/(u^*-1)$, so $1+e^{u^*}=u^*/(u^*-1)$ and $u^*/(1+e^{u^*})=u^*-1$. Writing $v=u^*-1>0$,
the defining equation becomes $e^{v+1}v=1$, that is $ve^v=e^{-1}$. The map $v\mapsto ve^v$ is
strictly increasing on $[-1,\infty)$, its derivative being $(1+v)e^v$, so $v$ is the value at
$e^{-1}$ of its inverse, the principal branch $W$ of the Lambert function; hence
$u^*-1=W(e^{-1})$ and $D_S^{\max}=2W(e^{-1})N$.

\emph{Growth.} The coefficient $2W(e^{-1})$ is a positive constant, so the maximum exceeds any given
bound once $N$ is large enough.
\end{proof}

\begin{corollary}\label{cor:uniform-line}
Under the hypothesis of Proposition~\ref{prop:cube}, at $\alpha_i=1/N$ and for every $t>0$,
\[
0\;<\;2t-D_S^{\mathrm{full}}(t)\;=\;\frac{2t\,\bigl(e^{2t/N}-1\bigr)}{e^{2t/N}+1}\;\le\;\frac{2t^2}{N} .
\]
Hence $1-t/N\le D_S^{\mathrm{full}}(t)/(2t)<1$, so for every $T>0$ and $0<t\le T$ the relative
defect satisfies $0<1-D_S^{\mathrm{full}}(t)/(2t)\le T/N$; at $t=1$, $2-2/N\le D_S^{\mathrm{full}}(1)<2$,
and $D_S^{\mathrm{full}}(1)\to2$ as $N\to\infty$.
\end{corollary}

\begin{proof}
Put $x=2t/N>0$. By Proposition~\ref{prop:uniform}, $D_S^{\mathrm{full}}(t)=4t/(1+e^{x})$, so
\[
2t-D_S^{\mathrm{full}}(t)=\frac{2t(1+e^x)-4t}{1+e^x}=\frac{2t\,(e^x-1)}{e^x+1}.
\]
Since $e^x>1+x>1$, this is positive. For the upper bound we show $(2-x)e^x\le2+x$ for $x\ge0$. The
function $f(x)=2+x-(2-x)e^x$ has $f(0)=0$ and $f'(x)=1-(1-x)e^x$; from $e^{-x}\ge1-x$, valid for all
real $x$, we get $(1-x)e^x\le1$, so $f'\ge0$ and $f\ge0$ on $[0,\infty)$ by the mean value theorem.
Rearranged, $(2-x)e^x\le2+x$ reads $2(e^x-1)\le x(e^x+1)$, so
\[
\frac{2t\,(e^x-1)}{e^x+1}\;\le\;t\,x\;=\;\frac{2t^2}{N}.
\]
Dividing by $2t>0$ gives $1-t/N\le D_S^{\mathrm{full}}(t)/(2t)<1$, so the relative defect
$1-D_S^{\mathrm{full}}(t)/(2t)$ is positive and at most $t/N\le T/N$ for $0<t\le T$. At $t=1$ the two bounds read
$2-2/N\le D_S^{\mathrm{full}}(1)<2$; both ends tend to $2$ as $N\to\infty$, so $D_S^{\mathrm{full}}(1)$
does too.
\end{proof}

This is the line $2t$ that \cite{Oggad2026} reads off a single scale $\{0\}\cup\{1\}^{(p)}$ as $p$
grows, crossing two at its relaxation time $t=1$; the equal-weight screen reads within $2/N$ of two at
$t=1$ by following that line.

The trichotomy of \cite{Oggad2026} sorts a spectrum by its density near zero into a power-law tail,
a single scale $\{0\}\cup\{a\}^{(p)}$ and a gapped two-scale band
$\{0\}\cup\{a\}^{(p)}\cup\{b\}^{(q)}$. The equal-weight spectrum is evenly spaced.

\begin{proposition}\label{prop:comb}
Under the hypothesis of Proposition~\ref{prop:cube}, at $\alpha_i=1/N$ the spectrum of $L$ is
$\{2k/N:0\le k\le N\}$, the value $2k/N$ occurring $\binom Nk$ times; these are $N+1$ distinct values.
\end{proposition}

\begin{proof}
The level of a subset $S$ is $2|S|/N$, which equals $2k/N$ exactly when $|S|=k$, so by
Proposition~\ref{prop:cube} the value $2k/N$ occurs once for each subset of size $k$, that is
$\binom Nk$ times, which is positive for $0\le k\le N$; and $2k/N=2k'/N$ forces $k=k'$.
\end{proof}

Being finite, this evenly spaced spectrum carries no power-law tail, by the computation for finite
spectra that follows Convention~\ref{conv:dial}.

\begin{convention}\label{conv:dial}
If a spectrum on $(0,\infty)$ has density exactly $c\lambda^{\theta}$ with $\theta > -1$, the
counting function is $\mathcal{N}(\lambda) = c\lambda^{\theta+1}/(\theta+1)$, the heat trace is
$c\,\Gamma(\theta+1)\,t^{-(\theta+1)}$, and the running dimension is exactly $D_S = 2(\theta+1)$ at
every $t>0$. Accordingly a counting function behaving as $\lambda^{\theta+1}$ is read as the
\emph{plateau} $D_S = 2(\theta+1)$. This is the reading of \cite{Oggad2026} under the substitution
$\theta = d/2 - 1$; that paper states the plateau under regular variation at zero of index
$d/2>0$, which no finite spectrum has, so at a finite stage the plateau number is the reading of
the counting exponent over the thresholds at which that exponent is exhibited.
\end{convention}

The three computations of the exact case are short. Since $\theta>-1$ the density is integrable at
zero and $\mathcal{N}(\lambda)=\int_0^\lambda c\mu^\theta\,\mathrm{d}\mu=c\lambda^{\theta+1}/(\theta+1)$.
The substitution $x=t\lambda$ gives
$\int_0^\infty e^{-t\lambda}c\lambda^\theta\,\mathrm{d}\lambda=ct^{-(\theta+1)}\int_0^\infty
e^{-x}x^\theta\,\mathrm{d}x=c\,\Gamma(\theta+1)t^{-(\theta+1)}$, the Euler integral converging for
$\theta>-1$. Then $\log\Theta=\mathrm{const}-(\theta+1)\log t$, and
$-2\,\mathrm{d}\log\Theta/\mathrm{d}\log t=2(\theta+1)$ at every $t$. The counting exponent
$\theta+1$ and the running dimension $2(\theta+1)$ are thus one number read twice, and the
convention transports that identity to spectra whose counting function obeys the power law only
along a sequence of thresholds. A finite spectrum in $[0,\infty)$ has a counting function constant
on $[0,\epsilon)$, $\epsilon$ its least positive level (any $\epsilon>0$ if there is none), so for
$c>0$ and $0<\lambda<\epsilon/\max(1,c)$ both $\mathcal{N}(c\lambda)$ and $\mathcal{N}(\lambda)$ equal
the multiplicity of the level $0$; the ratio $\mathcal{N}(c\lambda)/\mathcal{N}(\lambda)$ is then
undefined or equal to $1$, while $c^{d/2}\ne1$ for $c\ne1$ and $d>0$.

\begin{theorem}\label{thm:plateau}
Under the hypothesis of Proposition~\ref{prop:cube}, let $\alpha_i = r^{i}(1-r)/(1-r^N)$ for
$i\in[N]$ and a fixed $0<r<1$. At the thresholds $\lambda_k = 2r^{k}/(1-r^{N})$,
$0\le k<N$, the counting function of $\operatorname{spec}(L)$ obeys
\[
\mathcal{N}(\lambda_k) \;\asymp\; 2^{N}\lambda_k^{\,\log 2/\log(1/r)},
\]
with constants depending on $r$ alone. Convention~\ref{conv:dial} therefore reads the plateau
\[
D_S \;=\; \frac{2\log 2}{\log(1/r)},
\]
which carries no cap: it exceeds $2$ for every $r > 1/2$ and diverges as $r \uparrow 1$,
and every $D>0$ is this plateau for exactly one ratio, $r=2^{-2/D}$.
\end{theorem}

\begin{proof}
Here $\mathcal{N}(\lambda)$ is the number of eigenvalues of $L$ at most $\lambda$, counted with
multiplicity, which by Proposition~\ref{prop:cube} is the number of subsets
$S\subseteq[N]$ with $2\sum_{i\in S}\alpha_i\le\lambda$. Write $A=(1-r)/(1-r^N)>0$, so
that $\alpha_i=Ar^i$, $\sum_i\alpha_i=A(1-r^N)/(1-r)=1$, and $\lambda_k=2Ar^k/(1-r)$.

\emph{Lower bound.} Let $S$ be a subset of the tail $\{k,\dots,N-1\}$. Its level is at most the level
of the whole tail,
\[
2\sum_{i\in S}\alpha_i\;\le\;2A\sum_{i=k}^{N-1}r^i\;=\;\frac{2Ar^k(1-r^{N-k})}{1-r}\;\le\;\lambda_k .
\]
The tail has $N-k$ elements and so $2^{N-k}$ subsets, all counted at $\lambda_k$; hence
$\mathcal{N}(\lambda_k)\ge2^{N-k}$.

\emph{Upper bound.} Since $r^{n}\to0$, there is a least integer $j\ge0$ with $r^{j+1}<1-r$; it
depends on $r$ alone. Let $S$ contain an index $i\le k-j-1$. Its level is at least
$2\alpha_i=2Ar^i\ge2Ar^{k-j-1}$, the weights being nonnegative and $r^i$ decreasing in $i$; and
$2Ar^{k-j-1}>\lambda_k$ is equivalent, after division by $2Ar^{k}>0$, to $r^{-(j+1)}>1/(1-r)$, that
is to $r^{j+1}<1-r$, which holds by the choice of $j$. So no such $S$ is counted at $\lambda_k$, and
every counted subset lies in $\{\max(0,k-j),\dots,N-1\}$, a set of $N-\max(0,k-j)\le N-k+j$
elements. Hence $\mathcal{N}(\lambda_k)\le2^{N-k+j}$.

\emph{The exponent.} Put $\gamma=\log2/\log(1/r)>0$. Then
$(r^k)^\gamma=e^{-k\gamma\log(1/r)}=e^{-k\log2}=2^{-k}$. From $\lambda_k=2r^k/(1-r^N)$ we have
$r^k=\lambda_k(1-r^N)/2$, and $1-r\le1-r^N\le1$ for $N\ge1$, so
\[
\Bigl(\frac{1-r}{2}\Bigr)^{\!\gamma}\lambda_k^\gamma\;\le\;2^{-k}\;\le\;2^{-\gamma}\lambda_k^\gamma ,
\]
$x\mapsto x^\gamma$ being increasing on $(0,\infty)$. Combining with the two bounds,
\[
\Bigl(\frac{1-r}{2}\Bigr)^{\!\gamma}\,2^N\lambda_k^\gamma\;\le\;2^{N-k}\;\le\;\mathcal{N}(\lambda_k)
\;\le\;2^{j}\,2^{N-k}\;\le\;2^{j-\gamma}\,2^N\lambda_k^\gamma ,
\]
and both constants, $((1-r)/2)^\gamma$ and $2^{j-\gamma}$, depend on $r$ alone. This is the
asserted comparison.

\emph{The reading.} Convention~\ref{conv:dial} with $\theta+1=\gamma$ reads $D_S=2\gamma=2\log2/\log(1/r)$.
As a function of $r\in(0,1)$, $\log(1/r)$ decreases strictly and continuously from $+\infty$ to $0$,
so $2\gamma$ increases strictly and continuously from $0$ to $+\infty$: it exceeds $2$ exactly when
$\log(1/r)<\log2$, that is when $r>1/2$, and it diverges as $r\uparrow1$. For $D>0$ and
$r=2^{-2/D}\in(0,1)$ one has $\log(1/r)=(2/D)\log2$, so $2\gamma=D$; strict monotonicity makes this
ratio the only one.
\end{proof}

\begin{definition}\label{def:bump}
The \emph{bump} is the function $\varpi(x) := x/(1+e^{x})$ on $(0,\infty)$.
\end{definition}

\begin{proposition}\label{prop:bumpsum}
The running dimension is a sum of bumps, one per weight,
$D_S^{\mathrm{full}}(t) = 2\sum_{i\in[N]} \varpi(2t\alpha_i)$, and on a logarithmic time axis,
with $t = e^{s}$ and $\alpha_i = e^{-a_i}$,
\[
D_S^{\mathrm{full}}(e^{s}) \;=\; 2\sum_{i\in[N]} \check\varpi(s - a_i), \qquad \check\varpi(y) \;:=\; \varpi(2e^{y}) .
\]
So $D_S$ against $\log t$ is one fixed bump convolved with the counting measure of the log-weights.
\end{proposition}

\begin{proof}
The $i$-th summand of \eqref{eq:ds} is $4t\alpha_i/(1+e^{2t\alpha_i})=2\cdot\bigl(x/(1+e^x)\bigr)$
at $x=2t\alpha_i$, which is $2\,\varpi(2t\alpha_i)$. For the second form, which presupposes
$\alpha_i>0$ so that $a_i=-\log\alpha_i$ is defined, $2t\alpha_i=2e^se^{-a_i}=2e^{s-a_i}$, so the
$i$-th summand is $2\check\varpi(s-a_i)$. Writing $\nu=\sum_i\delta_{a_i}$ for the counting measure of
the log-weights, $\sum_i\check\varpi(s-a_i)=\int\check\varpi(s-a)\,\mathrm{d}\nu(a)=(\check\varpi*\nu)(s)$.
\end{proof}

\begin{proposition}\label{prop:bumpmass}
The bump carries mass $\log 2$:
$\int_{\Rr} \varpi(2e^{y})\,\mathrm{d}y = \int_{0}^{\infty} \varpi(x)\,\mathrm{d}x/x
= \int_{0}^{\infty} \mathrm{d}x/(1+e^{x}) = \log 2$.
\end{proposition}

\begin{proof}
The map $y\mapsto x=2e^y$ is a smooth increasing bijection of $\Rr$ onto $(0,\infty)$ with
$\mathrm{d}x=x\,\mathrm{d}y$, so the change of variables gives the first equality, the integrands
being nonnegative. The second is $\varpi(x)/x=1/(1+e^x)$. For the third, the function
$F(x)=-\log(1+e^{-x})$ has
$F'(x)=e^{-x}/(1+e^{-x})=1/(e^x+1)$, multiplying numerator and denominator by $e^x$. The integrand is
positive and bounded by $e^{-x}$, which is integrable on $(0,\infty)$, so the improper integral
converges and equals $\lim_{X\to\infty}F(X)-F(0)=0-(-\log2)=\log2$.
\end{proof}

\begin{corollary}\label{cor:plateaumass}
A geometric profile $\alpha_i \propto r^{i}$ has log-weights in arithmetic progression with spacing
$\log(1/r)$, so its comb carries mass $\log2$ per spacing, and
\[
\frac{2\log 2}{\log(1/r)} \;=\; 2 \cdot \frac{\text{mass of the bump}}{\log(1/r)} .
\]
At $r = 1/2$ the spacing equals the mass of the bump and the plateau is exactly $2$.
\end{corollary}

\begin{proof}
If $\alpha_i=Ar^i$ with $A>0$ then $a_i=-\log\alpha_i=-\log A+i\log(1/r)$, an arithmetic progression
with spacing $h:=\log(1/r)>0$. Each copy $\check\varpi(\cdot-a_i)$ of the bump has mass $\log2$ by
Proposition~\ref{prop:bumpmass}, and consecutive copies are translated by $h$, so the comb carries
mass $\log2$ per length $h$ of the logarithmic axis. In the sense of a mean height: for the
two-sided comb $\sum_{i\in\mathbb{Z}}\check\varpi(s-a_0-ih)$, Tonelli's theorem applied to the nonnegative
terms gives
\[
\frac1h\int_0^h\sum_{i\in\mathbb{Z}}\check\varpi(s-a_0-ih)\,\mathrm{d}s
=\frac1h\sum_{i\in\mathbb{Z}}\int_{-ih}^{-(i-1)h}\check\varpi(y-a_0)\,\mathrm{d}y
=\frac1h\int_{\Rr}\check\varpi\,\mathrm{d}y=\frac{\log2}{h},
\]
the intervals $[-ih,-(i-1)h)$ tiling $\Rr$; doubling, as in Proposition~\ref{prop:bumpsum}, gives
the mean height $2\log2/\log(1/r)$, the displayed number. At $r=1/2$ the spacing
$\log(1/r)=\log2$ equals the mass, and the number is $2$.
\end{proof}

\begin{corollary}\label{cor:bumppeak}
The bump has a single maximum on $(0,\infty)$, at the $u^*$ of Proposition~\ref{prop:uniform}, with
value $u^*-1=W(e^{-1})$; and at $\alpha_i=1/N$, $D_S^{\mathrm{full}}(t)=2N\,\varpi(2t/N)$, so the
maximum of Proposition~\ref{prop:uniform} is $2N$ times the peak of the bump.
\end{corollary}

\begin{proof}
At $\alpha_i=1/N$ the sum of Proposition~\ref{prop:bumpsum} has $N$ equal terms
$2\,\varpi(2t/N)$. The function $h$ of the proof of Proposition~\ref{prop:uniform} is the bump,
and that proof shows that $h$ has the single maximiser $u^*$ on $(0,\infty)$, with
$h(u^*)=u^*-1=W(e^{-1})$.
\end{proof}

\begin{theorem}\label{thm:dyadic}
For the geometric profile at $r = 1/2$ the running dimension $D_S=D_S^{\mathrm{full}}$ of
Corollary~\ref{cor:heat} is a
difference of two values taken at the two ends of the ladder of weights, every interior rung
cancelling; consequently $D_S < 2$ at every finite stage and every scale, and
\[
0 \;<\; 2 - D_S(t) \;\le\; \frac{2c}{2^{N}} + 2\,b(2c),
\qquad b(u) = \frac{u}{e^{u}-1},\quad c = \frac{t}{1 - 2^{-N}} .
\]
The first term recedes like $2^{-N}$, and for every $\varepsilon > 0$ the bound is at most
$\varepsilon$ wherever $4/\varepsilon \le t \le \varepsilon(2^{N}-1)/4$, a stretch whose ultraviolet
end does not depend on $N$ and whose infrared end grows by a factor tending to $2$ per stage.
\end{theorem}

\begin{proof}
Throughout $t>0$, $N\ge1$, and $D_S$ is $D_S^{\mathrm{full}}$.

\emph{A partial-fraction identity.} For $u>0$,
\[
\frac{1}{e^u-1}-\frac{2}{e^{2u}-1}=\frac{(e^u+1)-2}{(e^u-1)(e^u+1)}=\frac{1}{e^u+1},
\]
using $e^{2u}-1=(e^u-1)(e^u+1)$. Multiplying by $u$ gives $u/(1+e^u)=b(u)-b(2u)$, since
$2u/(e^{2u}-1)=b(2u)$. So the summand $4t\alpha/(1+e^{2t\alpha})$ of \eqref{eq:ds}, which is
$2u/(1+e^u)$ at $u=2t\alpha$, equals $2\bigl(b(u)-b(2u)\bigr)$.

\emph{Telescoping.} At $r=1/2$ the weights of Theorem~\ref{thm:plateau} are
$\alpha_i=2^{-i}(1/2)/(1-2^{-N})$ for $i=0,\dots,N-1$, so $u_i:=2t\alpha_i=c\,2^{-i}$ and
$2u_{i+1}=u_i$. Hence
\[
D_S(t)=2\sum_{i=0}^{N-1}\bigl(b(u_i)-b(2u_i)\bigr)
=2\Bigl(\sum_{i=0}^{N-1}b(u_i)-\sum_{i=-1}^{N-2}b(u_{i})\Bigr)
=2\bigl(b(u_{N-1})-b(2u_0)\bigr),
\]
writing $u_{-1}:=2u_0$ and using $2u_i=u_{i-1}$; every interior term cancels. With
$u_{N-1}=c\,2^{1-N}=2c\,2^{-N}$ and $2u_0=2c$ this is $D_S(t)=2\bigl(b(2c\,2^{-N})-b(2c)\bigr)$.

\emph{Strict bounds.} For $u>0$ the inequality $e^u>1+u$ gives $e^u-1>u>0$, so $0<b(u)<1$.
Therefore $D_S(t)<2\,b(2c\,2^{-N})<2$, and
$2-D_S(t)=2\bigl(1-b(2c\,2^{-N})\bigr)+2\,b(2c)$ is a sum of two positive numbers.

\emph{The upper bound.} We claim $b(u)\ge1-u/2$ for $u>0$. Since $e^u-1>0$ this is equivalent to
$2u\ge(2-u)(e^u-1)$, that is to $(2-u)e^u\le2+u$. Put $f(u)=2+u-(2-u)e^u$; then $f(0)=0$ and
$f'(u)=1-(1-u)e^u$. From $e^{-u}\ge1-u$, valid for all real $u$, we get $(1-u)e^u\le1$, so
$f'\ge0$ and $f(u)\ge0$ for $u\ge0$. Applied at $u=2c\,2^{-N}$ this gives
$2\bigl(1-b(2c\,2^{-N})\bigr)\le2c\,2^{-N}$, whence $2-D_S(t)\le2c/2^N+2b(2c)$.

\emph{The window.} First, $2c/2^N=2t/\bigl((1-2^{-N})2^N\bigr)=2t/(2^N-1)$, which is at most
$\varepsilon/2$ exactly when $t\le\varepsilon(2^N-1)/4$; for fixed $t$ it decays like $2^{-N}$.
Second, every term of the exponential series of $e^{2c}$ is nonnegative, so
$e^{2c}-1\ge(2c)^2/2=2c^2$ and $2b(2c)=4c/(e^{2c}-1)\le2/c$; and $c=t/(1-2^{-N})\ge t$, so
$2b(2c)\le2/t$, which is at most $\varepsilon/2$ when $t\ge4/\varepsilon$. On the stretch where both
hold, $2-D_S(t)\le\varepsilon$. Its lower end $4/\varepsilon$ does not involve $N$, and its upper
end at stage $N+1$ divided by that at stage $N$ is $(2^{N+1}-1)/(2^N-1)=2+1/(2^N-1)\to2$.
\end{proof}

\begin{corollary}\label{cor:split}
Under the hypothesis of Proposition~\ref{prop:cube}, every class read off the spectrum is a function
of the weight profile and of nothing else: for every assignment $\mathrm{cl}$ of a class to a
spectrum with multiplicity there is a map $F$ with $\mathrm{cl}(\operatorname{spec}(L)) = F(\alpha)$
at every realisation meeting that hypothesis. The constraint $\sum_i \alpha_i = 1$ does not
determine the reading: at total weight one the number $2\log2/\log(1/r)$ that
Convention~\ref{conv:dial} reads off a geometric profile does not depend on $N$, while the
equal-weight maximum $2W(e^{-1})N$ exceeds every bound as $N$ grows. The two running dimensions also differ in
shape: at equal weights $D_S^{\mathrm{full}}$ is strictly increasing on $(0,Nu^*/2]$ and strictly
decreasing on $[Nu^*/2,\infty)$, while at $r=1/2$ the geometric profile has
$0<2-D_S^{\mathrm{full}}(t)\le\varepsilon$ for every $\varepsilon>0$ and every
$t\in[4/\varepsilon,\,\varepsilon(2^N-1)/4]$.
\end{corollary}

\begin{proof}
\emph{Factorisation.} Define $F(\alpha):=\mathrm{cl}(\Sigma_N(\alpha))$. At any realisation meeting
the hypothesis of Proposition~\ref{prop:cube}, that proposition gives
$\operatorname{spec}(L)=\Sigma_N(\alpha)$ as multisets, so
$\mathrm{cl}(\operatorname{spec}(L))=\mathrm{cl}(\Sigma_N(\alpha))=F(\alpha)$, whatever rule
$\mathrm{cl}$ is. Without one-dimensionality of the joint eigenspaces this fails: doubling every
joint eigenspace, as in Proposition~\ref{prop:one-dim-sharp}, meets every other clause and doubles every multiplicity, presenting
another multiset at the same profile, on which $\mathrm{cl}$ may take another value.

\emph{The constraint does not fix the reading.} Both profiles have total weight one: $N\cdot(1/N)=1$,
and $\sum_iAr^i=A(1-r^N)/(1-r)=1$ for $A=(1-r)/(1-r^N)$. By Theorem~\ref{thm:plateau} the geometric
profile reads $2\log2/\log(1/r)$, a number not involving $N$; by Proposition~\ref{prop:uniform}
the equal-weight profile has maximum $2W(e^{-1})N$, which exceeds $2\log2/\log(1/r)$, and every
other bound, for all large $N$. So two profiles subject to the same constraint read different
numbers.

\emph{The two shapes.} By the proof of Proposition~\ref{prop:uniform},
$D_S^{\mathrm{full}}(t)=2Nh(2t/N)$ with $h$ strictly increasing on $(0,u^*]$ and strictly
decreasing on $[u^*,\infty)$, and $t\mapsto2t/N$ is strictly increasing, so $D_S^{\mathrm{full}}$ is
strictly increasing on $(0,Nu^*/2]$ and strictly decreasing on $[Nu^*/2,\infty)$. The clause at
$r=1/2$ is Theorem~\ref{thm:dyadic}: its strict bound gives $2-D_S^{\mathrm{full}}(t)>0$, and its
window gives $2-D_S^{\mathrm{full}}(t)\le\varepsilon$ on $[4/\varepsilon,\varepsilon(2^N-1)/4]$.
\end{proof}

The weighted form asks of the profile $\alpha_i\in[0,1]$ and $\sum_i\alpha_i=1$
\cite{FieldsGlazebrookMarciano2022}. Among profiles meeting these constraints,
Corollary~\ref{cor:split} exhibits a reading that does not depend on $N$, the geometric number
$2\log2/\log(1/r)$, and one that exceeds every bound as $N$ grows, the equal-weight maximum; and a
running dimension with a single strict peak beside one held within any $\varepsilon$ of two on a
stretch whose infrared end $\varepsilon(2^N-1)/4$ grows with $N$. By Proposition~\ref{prop:phi}
a choice of reference angles changes none of these readings, which depend on the profile
alone. By Lemma~\ref{lem:rational} a grouping of an
equal-weight screen collects the dyadic profile of Theorem~\ref{thm:dyadic} at every finite $N$.

On the joint eigenspace of a sign vector $s$ each $M_i$ acts by $s_i$, so the operator $\widetilde L$
of Theorem~\ref{thm:signfamily} acts there by $\sum_i(s_i+1)\alpha_i=2\sum_{i:s_i=+1}\alpha_i$, twice the
weight of the qubits that flash $+1$. At equal weights this is $2/N$ times their number, so a frame
with equal weights reads a count, like a tally of how many of $N$ detectors have fired. The mask of
Section~\ref{sec:spectrum}, with weight $1/k$ on each of the $k$ qubits of its sector and $0$ elsewhere,
gives $2/k$ times the number of those qubits that flash $+1$. At $r=\tfrac12$ the profile of Theorem~\ref{thm:plateau} gives the qubits the weights
$2^{-i}\cdot\tfrac12/(1-2^{-N})=2^{N-1-i}/(2^N-1)$, which up to a common factor are the place values $2^{N-1},\dots,2,1$ of an $N$-digit binary number.
The qubits that flash $+1$ then spell a number in binary, and the level is $2/(2^N-1)$ times that
number. Binary expansions are unique, so the level tells exactly which qubits flash $+1$. In
Section~\ref{sec:relabel} we read other profiles as stages in learning a mask.

\section{Relabelling and regrouping}\label{sec:relabel}

In the model of \cite{FieldsGlazebrookMarciano2021,FieldsGlazebrookMarciano2022}, $H_{\Omega\mho}$ is swap-symmetric under the permutation group $S_N$, the $M_i$
commuting \cite{FieldsGlazebrookMarciano2022}. The terms $\alpha_iM_i$ can be
re-arranged in any order, so that the qubits of the screen can be permuted arbitrarily and its
state is invariant under $S_N$ \cite{FieldsGlazebrookMarciano2021}. Lemma~3 of
\cite{FieldsGlazebrookMarciano2021} states that fixing a reference state over a macroscopic time
interval locally breaks this symmetry, and the remark after its proof \cite{FieldsGlazebrookMarciano2021} reads the classifier network
that computes the reference state as dividing the screen into two regions, one in which the qubits are classically phase-correlated and
one in which they are not. The lemma takes the reference state as already fixed
\cite{FieldsGlazebrookMarciano2021} and leaves open how a system learns a reference state that
identifies a component of its environment. As a profile, the division into two regions is a step
function, a mask on the phase-correlated qubits. Any other profile can then be read as a stage in
learning such a mask, before the weights have settled on the qubits that the reference state fixes.
In this reading the settled mask identifies the system, as the reference state does in
\cite{FieldsGlazebrookMarciano2021}, and its pointer state is read as in the example of
Section~\ref{sec:spectrum}. Read as relabelling, the symmetry says that renaming the qubits changes
nothing an observer reads, and Proposition~\ref{prop:relabel} below proves it at every profile. A stricter invariance, of the
operator itself under permuting its qubits with each weight held at its index, holds only at equal
weights.

\begin{proposition}\label{prop:swap}
Take the $M_k$ of \eqref{eq:ham} to be the $z$-spin operators $Z_ke_s=s_ke_s$ of the source, and for a
permutation $\sigma$ of $[N]$ let $P_\sigma e_s:=e_{s\circ\sigma}$. Then
$P_\sigma H_{\Omega\mho}P_\sigma^{-1}=\sum_k\alpha_{\sigma(k)}Z_k$, and the $Z_k$ are linearly independent. Hence
$H_{\Omega\mho}$ is invariant under conjugation by the swap of the qubits $i$ and $j$ if and only if
$\alpha_i=\alpha_j$, and under conjugation by every swap if and only if all weights are equal.
\end{proposition}

\begin{proof}
Since $P_\sigma^{-1}e_s=e_{s\circ\sigma^{-1}}$, $P_\sigma Z_kP_\sigma^{-1}e_s=s_{\sigma^{-1}(k)}e_s$, so
$P_\sigma Z_kP_\sigma^{-1}=Z_{\sigma^{-1}(k)}$ and
$P_\sigma H_{\Omega\mho}P_\sigma^{-1}=\sum_k\alpha_kZ_{\sigma^{-1}(k)}=\sum_k\alpha_{\sigma(k)}Z_k$. If
$\sum_kc_kZ_k=0$, applying it to $e_s$ and to $e_{s^{(k)}}$, which differ only in the $k$-th sign,
and subtracting the two eigenvalues gives $2c_ks_k=0$, so $c_k=0$. Hence $\sum_k\gamma_kZ_k=\sum_k\alpha_kZ_k$
forces $\gamma=\alpha$, and $P_\sigma H_{\Omega\mho}P_\sigma^{-1}=H_{\Omega\mho}$ if and only if $\alpha\circ\sigma=\alpha$. For
the transposition of $i$ and $j$ this is $\alpha_i=\alpha_j$, and it holds for every transposition
exactly when all weights are equal.
\end{proof}

\begin{definition}\label{def:push}
For $g : [N] \to [m]$ the \emph{block} of $j \in [m]$ is $g^{-1}(j)$, and the \emph{pushforward} of
$\alpha : [N] \to \Rr$ along $g$ is $(g_{*}\alpha)(j) := \sum_{i \in g^{-1}(j)} \alpha(i)$.
\end{definition}

On a sign vector constant on every block, every site of the block $g^{-1}(j)$ carrying the same sign
$\varsigma_j$, so that $s_i=\varsigma_{g(i)}$, summing block by block over the blocks $g^{-1}(j)$,
which partition $[N]$ (Proposition~\ref{prop:push-basic} below), gives
\[
\sum_{i\in[N]}\alpha(i)s_i=\sum_{j\in[m]}(g_*\alpha)(j)\,\varsigma_j .
\]
On these sign vectors the screen therefore takes the values of the screen of $m$ sign observables
with profile $g_*\alpha$, each block contributing the total weight of its sites.

\begin{proposition}\label{prop:push-basic}
For every $g : [N] \to [m]$ and $\alpha, f : [N] \to \Rr$,
$\sum_{j \in [m]} \sum_{i \in g^{-1}(j)} f(i) = \sum_{i \in [N]} f(i)$, hence
$\sum_{j} (g_{*}\alpha)(j) = \sum_{i} \alpha(i)$. If $\alpha \geq 0$ then $g_{*}\alpha \geq 0$, and
if $\sum_i \alpha(i) = 1$ then $\sum_j (g_{*}\alpha)(j) = 1$.
\end{proposition}

\begin{proof}
The blocks partition $[N]$: each $i$ lies in $g^{-1}(g(i))$, and if $i\in g^{-1}(j)$ then $j=g(i)$,
so $i$ lies in no other block. The double sum is therefore a sum over the disjoint union
$\bigsqcup_jg^{-1}(j)=[N]$ and enumerates each $i$ exactly once, which is the first identity. Taking
$f=\alpha$ and recalling Definition~\ref{def:push} gives the second. If $\alpha\ge0$ then each
$(g_*\alpha)(j)$ is a sum of nonnegative numbers, an empty block giving the empty sum $0$; and if
$\sum_i\alpha(i)=1$ the second identity gives $\sum_j(g_*\alpha)(j)=1$.
\end{proof}

\begin{proposition}\label{prop:push-perm}
For every $g:[N]\to[m]$, $\alpha:[N]\to\Rr$ and permutation $\sigma$ of $[N]$,
$(g\circ\sigma)_*(\alpha\circ\sigma)=g_*\alpha$.
\end{proposition}

\begin{proof}
$\bigl((g\circ\sigma)_*(\alpha\circ\sigma)\bigr)(j)
=\sum_{i:g(\sigma(i))=j}\alpha(\sigma(i))=\sum_{i':g(i')=j}\alpha(i')=(g_*\alpha)(j)$, by the
substitution $i'=\sigma(i)$, which is a bijection of $\{i:g(\sigma(i))=j\}$ onto $g^{-1}(j)$.
\end{proof}

Recall that $u_N$ is the equal-weight profile, $u_N(i)=1/N$ for $i\in[N]$.

\begin{lemma}\label{lem:rational}
For every $\alpha:[m]\to\Qq$ with $\alpha\ge0$ and $\sum_j\alpha(j)=1$ there are $N\ge1$ and
$g:[N]\to[m]$ with $g_*u_N=\alpha$, and $g$ is onto when every $\alpha(j)>0$. Conversely, for every
$N\ge1$ and $g:[N]\to[m]$ the profile $g_*u_N$ takes rational values. The profile of
Theorem~\ref{thm:plateau} at $r=1/2$ is rational and strictly positive at every $N\ge1$, so it is
$g_*u_{N'}$ for some $N'$ and some onto $g$.
\end{lemma}

\begin{proof}
Let $N\ge1$ be the product of the denominators of the $\alpha(j)$ in lowest terms, so that
$\alpha(j)=k_j/N$ with $k_j\ge0$ an integer, and $\sum_jk_j=N\sum_j\alpha(j)=N$. Let $g$ send the
sites $i$ with $k_0+\dots+k_{j-1}\le i<k_0+\dots+k_j$ to $j$; these $m$ intervals have lengths
$k_0,\dots,k_{m-1}$ summing to $N$, so they partition $[N]$ and $g$ is a map $[N]\to[m]$ whose block
$g^{-1}(j)$ has $k_j$ sites. By Definition~\ref{def:push},
$(g_*u_N)(j)=\sum_{i\in g^{-1}(j)}1/N=k_j/N=\alpha(j)$. If $\alpha(j)>0$ then $k_j>0$, the block of $j$
is nonempty, and $j$ lies in the image of $g$. Conversely $(g_*u_N)(j)=|g^{-1}(j)|/N$ is a quotient of
integers. At $r=1/2$ the weights of Theorem~\ref{thm:plateau} are
$\alpha_i=2^{-i}\cdot\tfrac12/(1-2^{-N})=2^{N-1-i}/(2^N-1)$, rational and positive, so the first
part applies; with $N'=2^N-1$ the block of $i$ has $2^{N-1-i}$ sites.
\end{proof}

\begin{definition}\label{def:two-groupings}
Two groupings $[4] \to [2]$:
$g_{22}(0) = g_{22}(1) = 0$, $g_{22}(2) = g_{22}(3) = 1$; and $g_{13}(0) = 0$,
$g_{13}(1) = g_{13}(2) = g_{13}(3) = 1$.
\end{definition}

\begin{proposition}\label{prop:groupings-separate}
With $\alpha$ uniform on $[4]$, $(g_{22})_{*}\tfrac14 = (\tfrac12, \tfrac12)$ and
$(g_{13})_{*}\tfrac14 = (\tfrac14, \tfrac34)$, and the screen spectra $\Sigma_2$ of these two
profiles differ, the level $1$ occurring twice for $g_{22}$ and not at all for $g_{13}$. The profile is a functional
of the grouping and not of the underlying profile alone.
\end{proposition}

\begin{proof}
The blocks of $g_{22}$ are $\{0,1\}$ and $\{2,3\}$, of sizes $2,2$; those of $g_{13}$ are $\{0\}$ and
$\{1,2,3\}$, of sizes $1,3$. Each site carries $\tfrac14$,
so a block of size $n$ collects $n/4$, which gives the two pushforwards. As $S$ runs over
$\varnothing,\{0\},\{1\},\{0,1\}$, the level $2\sum_{j\in S}((g_{22})_*\tfrac14)(j)$ runs over
$0,1,1,2$, and $2\sum_{j\in S}((g_{13})_*\tfrac14)(j)$ over $0,\tfrac12,\tfrac32,2$. The value $1$ has
multiplicity two in the first multiset and zero in the second, so the multisets differ. The
underlying profile is the same in both cases, so the difference is carried by the grouping alone.
\end{proof}

\begin{proposition}\label{prop:relabel}
For every permutation $\sigma$ of $[N]$, $\Sigma_{N}(\alpha \circ \sigma) = \Sigma_{N}(\alpha)$, so
every class read off the spectrum is invariant and a relabelling moves no reading. A regrouping does:
with $g_{22}, g_{13}$ as in Definition~\ref{def:two-groupings} and $\alpha$ uniform,
$\Sigma_2((g_{22})_{*}\alpha)$ contains the level $1$ twice and $\Sigma_2((g_{13})_{*}\alpha)$ not at
all.
\end{proposition}

\begin{proof}
The map $S\mapsto\sigma(S)$ is a bijection of the subsets of $[N]$, with inverse
$S'\mapsto\sigma^{-1}(S')$, and since $\sigma$ restricts to a bijection of $S$ onto $\sigma(S)$,
\[
\sum_{i \in S}\alpha(\sigma(i)) \;=\; \sum_{i' \in \sigma(S)}\alpha(i') .
\]
So for every real $\mu$ the subsets $S$ with $2\sum_{i\in S}\alpha(\sigma(i))=\mu$ correspond
bijectively to the subsets $S'$ with $2\sum_{i'\in S'}\alpha(i')=\mu$, and $\mu$ has the same
multiplicity in $\Sigma_N(\alpha\circ\sigma)$ as in $\Sigma_N(\alpha)$; two multisets with the same
multiplicity function are equal. A class read off the spectrum is a function of this multiset and
is therefore unchanged. The contrast is Proposition~\ref{prop:groupings-separate}, the uniform
profile being fixed by every permutation, $\alpha\circ\sigma=\alpha$, so that no relabelling of the
sites can account for the difference between $\Sigma_2((g_{22})_*\alpha)$ and
$\Sigma_2((g_{13})_*\alpha)$.
\end{proof}

So the swap symmetry, read as relabelling, is exactly the statement that relabelling is invisible,
while a regrouping is not (Proposition~\ref{prop:groupings-separate}). The nonempty blocks of a grouping $g$ in the sense of
Definition~\ref{def:push} partition the sites, each site $i$ lying in the block $g^{-1}(g(i))$ and
in no other, $g$ being a function. A grouping models the identification of systems, which a
reference frame carries out \cite{FieldsGlazebrookMarciano2021,FieldsGlazebrookMarciano2022}. Each
nonempty block is then an identified system, and two distinct systems never share a site. A system
may still have parts, which we take to be the blocks of a finer grouping that lie inside its block. We
read such a system as classical, its block playing the part of the region of classically
phase-correlated qubits in \cite{FieldsGlazebrookMarciano2021}. Call the family of sets an observer correlates a
\emph{readout}; unlike the blocks of a grouping, its sets may overlap.

\section{What a readout can read}\label{sec:grouping}

Let the observer correlate the sets $A_0,\dots,A_{m-1}\subseteq[N]$, each correlation taken as the
single-shot parity of the $\pm1$ outputs of a set. On a sign vector $s$ the correlation of $A_j$ is
then $\prod_{i\in A_j}s_i=\chi_{A_j}(s)$, where $\chi_S(s):=\prod_{i\in S}s_i$ is the Walsh character
defined in the proof of Proposition~\ref{prop:cube}; flipping the $i$-th sign negates $\chi_S$ when
$i\in S$ and fixes it otherwise, so $(\Delta\chi_S)(s)=\sum_i\alpha_i\bigl(\chi_S(s)-\chi_S(s^{(i)})\bigr)$ gives
$\Delta\chi_S=\bigl(2\sum_{i\in S}\alpha_i\bigr)\chi_S$ for the weighted cube Laplacian $\Delta$. A
\emph{choice of sets}, a subset $J\subseteq[m]$ of indices, combines
the correlations of the chosen sets into a product.
Write $\tau_A(J)$ for the set of sites lying in an odd number of the chosen sets, the parity
transpose of Definition~\ref{def:readout} below.

\begin{lemma}\label{lem:parity}
$\prod_{j\in J}\chi_{A_j}=\chi_{\tau_A(J)}$, and $\chi_{\tau_A(J)}$ is an eigenvector of the weighted
cube Laplacian at the level $2\sum_{i\in\tau_A(J)}\alpha_i$.
\end{lemma}

\begin{proof}
$\prod_{j\in J}\chi_{A_j}(s)=\prod_{i\in[N]}s_i^{\,|\{j\in J:\,i\in A_j\}|}$, and $s_i^2=1$ removes every
even power, leaving $\prod_{i\in\tau_A(J)}s_i$. The second clause is
$\Delta\chi_S=\bigl(2\sum_{i\in S}\alpha_i\bigr)\chi_S$, from the proof of Proposition~\ref{prop:cube}.
\end{proof}

\begin{definition}\label{def:readout}
For $\alpha : [N] \to \Rr$ the \emph{screen spectrum} is $\Sigma_N(\alpha)$, one level
$2\sum_{i\in S}\alpha(i)$ per subset $S\subseteq[N]$. Let the readout $A_0,\dots,A_{m-1} \subseteq
[N]$ be the sets an observer correlates. Its \emph{parity transpose} at a choice of sets
$J \subseteq [m]$ is
\[
\tau_A(J) \;:=\; \bigl\{\, i \in [N] \;:\; |\{\, j \in J : i \in A_j \,\}| \ \text{is odd} \,\bigr\},
\]
and the spectrum it reads is the multiset $R_A(\alpha)$ of the $2\sum_{i \in \tau_A(J)}\alpha(i)$,
one per $J \subseteq [m]$. Say $A$ \emph{carries a profile at} $\alpha$ when some
$\eta : [m] \to \Rr$ satisfies $\Sigma_m(\eta) = R_A(\alpha)$.
\end{definition}

Both multisets in the last clause have $2^m$ entries, $R_A(\alpha)$ one per choice $J\subseteq[m]$
and $\Sigma_m(\eta)$ one per subset $S\subseteq[m]$, so the question is
whether the $2^m$ levels an observer reads can be rearranged into the subset sums of $m$ numbers.

\begin{lemma}[Reindexing]\label{lem:reindex}
If $\pi$ is a bijection of $2^{[m]}$ and $\ell,\ell'$ are functions on $2^{[m]}$ with
$\ell(J)=\ell'(\pi J)$ for all $J$, then $\ms{\ell(J)}_J=\ms{\ell'(S)}_S$.
\end{lemma}

\begin{proof}
For each $\mu$ the bijection $\pi$ carries $\{J:\ell(J)=\mu\}$ onto $\{S:\ell'(S)=\mu\}$, so $\mu$ has
the same multiplicity in both multisets.
\end{proof}

\begin{proposition}\label{prop:readout}
Let $g : [N] \to [m]$ and $A_j := g^{-1}(j)$. Then $A$ carries a profile at every $\alpha$, namely the
pushforward: $R_{A}(\alpha) = \Sigma_{m}(g_{*}\alpha)$.
\end{proposition}

\begin{proof}
A site $i$ lies in exactly one block, $A_{g(i)}$, so $|\{j\in J:i\in A_j\}|$ is $1$ if $g(i)\in J$
and $0$ otherwise; hence $\tau_A(J)=\{i:g(i)\in J\}=g^{-1}(J)$. The set $g^{-1}(J)$ is the disjoint
union of the blocks $g^{-1}(j)$, $j\in J$, so
\[
2\sum_{i\in\tau_A(J)}\alpha(i)=2\sum_{j\in J}\sum_{i\in g^{-1}(j)}\alpha(i)=2\sum_{j\in J}(g_*\alpha)(j)
\]
by Definition~\ref{def:push}. The level of $R_A(\alpha)$ at $J$ is thus the level of
$\Sigma_m(g_*\alpha)$ at the same $J$, and reindexing by the identity gives equality of the
multisets.
\end{proof}

A readout whose sets overlap can still carry a profile at every weighting
(Proposition~\ref{prop:sectors}), and the interpretation of \cite{FieldsGlazebrookMarciano2022}
depends on such overlap. In that interpretation the reference and
pointer sectors of an identified system are proper subsets of the observer's environment sector,
of which the identified system is itself part.

\begin{proposition}\label{prop:sectors}
Take $N = 3$, $A_0 = \{0,1,2\}$, $A_1 = \{1\}$, $A_2 = \{2\}$, an environment sector with a reference
and a pointer component as proper subsets. It is no partition, since $1 \in A_0 \cap A_1$, and no
chain, since $A_1$ and $A_2$ are disjoint. Nevertheless $\tau_A$ is a bijection of $2^{[3]}$ and
$R_{A}(\alpha) = \Sigma_{3}(\alpha)$ for every $\alpha$.
\end{proposition}

\begin{proof}
Site $0$ lies in $A_0$ only, site $1$ in $A_0$ and $A_1$, site $2$ in $A_0$ and $A_2$. Hence, for
$J\subseteq[3]$ and $U:=\tau_A(J)$,
\[
[0\in U]=[0\in J],\qquad [1\in U]\equiv[0\in J]+[1\in J],\qquad [2\in U]\equiv[0\in J]+[2\in J]
\pmod 2 .
\]
These relations can be solved for $J$: $[0\in J]=[0\in U]$, $[1\in J]\equiv[1\in U]+[0\in U]$ and
$[2\in J]\equiv[2\in U]+[0\in U]$ modulo $2$. So $J$ is determined by $\tau_A(J)$, and $\tau_A$ is an
injective map of the eight-element set $2^{[3]}$ to itself, hence a bijection. The level of
$R_A(\alpha)$ at $J$ is the level of $\Sigma_3(\alpha)$ at $\tau_A(J)$, and reindexing by $\tau_A$
gives $R_A(\alpha)=\Sigma_3(\alpha)$. The family is no partition because $1\in A_0\cap A_1$, and no
chain because $A_1\cap A_2=\varnothing$ while both are nonempty, so neither contains the other.
\end{proof}

Two sets \emph{cross} when they share a site and neither contains the other.

\begin{theorem}\label{thm:overlap}
Take $N = 3$, $\alpha \equiv \tfrac13$, $A_0 = \{0,1\}$ and $A_1 = \{1,2\}$, which share the site
$1$ while neither contains the other. Then no $\eta : [2] \to \Rr$ satisfies
$\Sigma_{2}(\eta) = R_{A}(\alpha)$.
\end{theorem}

\begin{proof}
The parity transposes are $\tau_A(\varnothing)=\varnothing$, $\tau_A(\{0\})=A_0=\{0,1\}$,
$\tau_A(\{1\})=A_1=\{1,2\}$, and $\tau_A(\{0,1\})=\{0,2\}$, site $1$ lying in both chosen sets and
so in an even number of them. Each nonempty transpose has two sites of weight $\tfrac13$, so
$R_A(\alpha)=\ms{0,\tfrac43,\tfrac43,\tfrac43}$, with total $4$. Let $\eta:[2]\to\Rr$ be arbitrary.
The four levels of $\Sigma_2(\eta)$ are $0$, $2\eta(0)$, $2\eta(1)$ and $2(\eta(0)+\eta(1))$,
with total $4(\eta(0)+\eta(1))$. Equal multisets have equal totals, so
$\Sigma_2(\eta)=R_A(\alpha)$ would force $\eta(0)+\eta(1)=1$; but then $\Sigma_2(\eta)$ contains
the level $2(\eta(0)+\eta(1))=2$, which is neither $0$ nor $\tfrac43$ and so not an element of
$R_A(\alpha)$. Hence no $\eta$ works.
\end{proof}

\begin{definition}\label{def:trace}
The \emph{trace} of a site $i \in [N]$ is
$t_A(i) := ([ i \in A_0 ], \dots, [ i \in A_{m-1} ]) \in \Ff^{m}$, and $\mathbf{1}_J\in\Ff^m$ is the
indicator of $J \subseteq [m]$, so that $\mathbf{e}_j:=\mathbf{1}_{\{j\}}$ is the $j$-th standard
basis vector, and $\langle u,v\rangle:=\sum_ju_jv_j\in\Ff$ is the standard pairing of
$u,v\in\Ff^m$. The \emph{trace set}
$\mathcal{T}(A) \subseteq \Ff^m$ is the set of nonzero traces occurring among the sites; the
\emph{character support} $\mathcal{T}_\alpha(A)$ is the set of nonzero traces $v$ whose class carries
nonzero total weight $\omega_v := \sum_{t_A(i) = v} \alpha(i)$.
\end{definition}

\begin{lemma}\label{lem:levelchar}
For $J\subseteq[m]$ and $i\in[N]$, $i\in\tau_A(J)$ if and only if $\langle t_A(i),\mathbf{1}_J\rangle=1$
in $\Ff$. Moreover $\mathcal{T}_\alpha(A)\subseteq\mathcal{T}(A)$, and
\begin{equation}\label{eq:levelchar}
2\!\!\sum_{i \in \tau_A(J)}\!\!\alpha(i) \;=\; 2\!\!\sum_{v \in \mathcal{T}_\alpha(A)}\!\! \omega_v\,
\bigl[\, \langle v, \mathbf{1}_J \rangle = 1 \,\bigr] .
\end{equation}
\end{lemma}

\begin{proof}
Over the integers $\sum_{j}t_A(i)_j(\mathbf{1}_J)_j=\sum_{j\in J}[i\in A_j]=|\{j\in J:i\in A_j\}|$, and
the pairing $\langle t_A(i),\mathbf{1}_J\rangle$ is this sum reduced modulo $2$, which is $1$ exactly
when the count is odd. For \eqref{eq:levelchar}, write the left side as
$2\sum_i\alpha(i)[\langle t_A(i),\mathbf{1}_J\rangle=1]$ and group the sites by their trace: the sites
of trace $v$ contribute $2\omega_v[\langle v,\mathbf{1}_J\rangle=1]$. The trace $v=0$ pairs to zero
with every choice of sets and contributes nothing, and a nonzero trace with $\omega_v=0$ contributes nothing,
so only the traces of $\mathcal{T}_\alpha(A)$ remain. The inclusion
$\mathcal{T}_\alpha(A)\subseteq\mathcal{T}(A)$ holds because $\omega_v\ne0$ requires a site of trace
$v$.
\end{proof}

\begin{proposition}\label{prop:indep-sufficient}
If $\mathcal{T}_\alpha(A)$ is linearly independent over $\Ff$ then $A$ carries a profile at
$\alpha$, namely the class weights: $\eta(j) := \omega_{v_j}$ on an enumeration $v_0,\dots,v_{k-1}$
of the support and $0$ beyond it.
\end{proposition}

\begin{proof}
An independent set in $\Ff^m$ has at most $m$ elements, so $k\le m$, and by the basis extension
theorem of linear algebra $v_0,\dots,v_{k-1}$ extends to a basis $v_0,\dots,v_{m-1}$ of $\Ff^m$.
Consider the linear map $\Xi:\Ff^m\to\Ff^m$, $x\mapsto(\langle v_j,x\rangle)_{j\in[m]}$. If
$\Xi(x)=0$ then $\langle v_j,x\rangle=0$ for all $j$, hence $\langle y,x\rangle=0$ for every $y$ in
the span of the $v_j$, which is $\Ff^m$; taking $y$ the $l$-th standard basis vector gives $x_l=0$
for every $l$. So $\Xi$ is injective, hence bijective, $\Ff^m$ being finite. Under the
identification of subsets with indicators, $\Xi$ becomes the bijection
$J\mapsto S(J):=\{j:\langle v_j,\mathbf{1}_J\rangle=1\}$ of $2^{[m]}$. By \eqref{eq:levelchar}, and
because $\eta(j)=0$ for $j\ge k$,
\[
2\sum_{i\in\tau_A(J)}\alpha(i)=2\sum_{j\in[k]} \omega_{v_j}[\langle v_j,\mathbf{1}_J\rangle=1]
=2\sum_{j\in S(J)}\eta(j),
\]
which is the level of $\Sigma_m(\eta)$ at $S(J)$. Reindexing by $S$ gives $R_A(\alpha)=\Sigma_m(\eta)$.
\end{proof}

\begin{theorem}\label{thm:dichotomy}
$A$ carries a profile at every $\alpha$ if and only if $\mathcal{T}(A)$ is linearly independent over
$\Ff$.
\end{theorem}

\begin{proof}
\emph{Sufficiency.} If $\mathcal{T}(A)$ is independent, so is its subset $\mathcal{T}_\alpha(A)$ for
every $\alpha$, a subset of an independent set being independent, and
Proposition~\ref{prop:indep-sufficient} applies at each $\alpha$.

\emph{From multisets to a permutation.} Suppose $\Sigma_m(\eta)=R_A(\alpha)$. Write $\ell(J)$ for
the level of $R_A(\alpha)$ at $J$ and $\ell'(S)$ for that of $\Sigma_m(\eta)$ at $S$. For each value
$\mu$ the fibres $\{J:\ell(J)=\mu\}$ and $\{S:\ell'(S)=\mu\}$ have the same cardinality, the
multiplicity of $\mu$; choosing a bijection between them for each $\mu$ and gluing these over the
finitely many values gives a permutation $\sigma$ of $2^{[m]}$ with $\ell'(\sigma J)=\ell(J)$, that is
\[
\sum_{i \in \tau_A(J)}\alpha(i) \;=\; \sum_{j \in \sigma J}\eta(j) \qquad (J \subseteq [m]) .
\]

\emph{The admissible set is a finite union of subspaces.} At $J=\sigma^{-1}\{j\}$ the identity
reads $\eta(j)=\sum_{i\in\tau_A(\sigma^{-1}\{j\})}\alpha(i)$, so $\eta$ is determined by $\alpha$
and $\sigma$, linearly in $\alpha$. For a permutation $\sigma$ of $2^{[m]}$ let $V_\sigma$ be the set
of $\alpha\in\Rr^N$ satisfying, for every $J$,
\[
\sum_{i\in\tau_A(J)}\alpha(i)\;=\;\sum_{j\in\sigma J}\;\sum_{i\in\tau_A(\sigma^{-1}\{j\})}\alpha(i).
\]
Each of these $2^m$ conditions says that a linear functional of $\alpha$ vanishes, so $V_\sigma$ is
an intersection of kernels of linear functionals: a linear subspace, closed in $\Rr^N$. If $A$
carries a profile at $\alpha$, the previous paragraph produces a $\sigma$ with $\alpha\in V_\sigma$.
Conversely, if $\alpha\in V_\sigma$, define $\eta(j):=\sum_{i\in\tau_A(\sigma^{-1}\{j\})}\alpha(i)$;
the defining conditions say $\ell(J)=\ell'(\sigma J)$ for all $J$, and reindexing by $\sigma$ gives
$R_A(\alpha)=\Sigma_m(\eta)$. Therefore
\begin{equation}\label{eq:admissible}
\{\, \alpha \;:\; A \text{ carries a profile at } \alpha \,\} \;=\; \bigcup_{\sigma} V_\sigma ,
\end{equation}
a union over the finitely many permutations of $2^{[m]}$.

\emph{One permutation serves every weighting.} Suppose now that $A$ carries a profile at every
$\alpha$, so the union \eqref{eq:admissible} is $\Rr^N$. By the Baire category theorem, applied to
the complete metric space $\Rr^N$ covered by finitely many closed sets, some $V_\sigma$ has nonempty
interior, containing the open ball of some radius $\varepsilon>0$ about a point $x$. For any
$y\ne0$ the vectors $x$ and $x+\tfrac{\varepsilon}{2|y|}y$ lie in $V_\sigma$, hence so does $y$, their difference rescaled. So
$V_\sigma=\Rr^N$, and this single $\sigma$ satisfies the identity at every $\alpha$, with the $\eta$
it determines.

\emph{The blocks are disjoint.} Fix that $\sigma$ and let $D_j:=\tau_A(\sigma^{-1}\{j\})$. Evaluate
the identity at the weighting $\delta_i$ concentrated on a site $i$, $\delta_i(i)=1$ and
$\delta_i(i')=0$ for $i'\ne i$:
\[
[\, i\in\tau_A(J)\,]\;=\;\sum_{j\in\sigma J}[\, i\in D_j\,]\qquad(J\subseteq[m]),
\]
an equality of integers whose left side is $0$ or $1$. If $i\in D_j\cap D_{j'}$ with $j\ne j'$, take
$J=\sigma^{-1}\{j,j'\}$: the right side is $2$, the left at most $1$, a contradiction. So the $D_j$
are pairwise disjoint.

\emph{Traces.} Let $i\in D_j$. By disjointness the only nonzero term on the right is the one for $j$,
so $[i\in\tau_A(J)]=[j\in\sigma J]$, that is $\langle t_A(i),\mathbf{1}_J\rangle=[j\in\sigma J]$
for every $J$. If $i'\in D_j$ as well, then $\langle t_A(i)-t_A(i'),\mathbf{1}_J\rangle=0$ for every
$J$; taking $J=\{l\}$ gives equality of the $l$-th coordinates, so $t_A(i)=t_A(i')$. Write $v_j$ for
this common trace, defined when $D_j\ne\varnothing$. A site lying in no $D_j$ has
$\langle t_A(i),\mathbf{1}_J\rangle=0$ for every $J$, so the same test gives $t_A(i)=0$. Hence
$\mathcal{T}(A)\subseteq\{v_j:D_j\ne\varnothing\}$.

\emph{Independence.} The $v_j$ are distinct: for $j\ne j'$, at $J=\sigma^{-1}\{j\}$ we have
$\langle v_j,\mathbf{1}_J\rangle=1$ and $\langle v_{j'},\mathbf{1}_J\rangle=[j'\in\{j\}]=0$. Over
$\Ff$ the only coefficients are $0$ and $1$, so a set is linearly dependent exactly when some
nonempty subset of it sums to zero. Let $I$ be a nonempty set of indices with $D_j\ne\varnothing$ for
$j\in I$, pick $j_0\in I$ and $J=\sigma^{-1}\{j_0\}$; then
$\langle\sum_{j\in I}v_j,\mathbf{1}_J\rangle=\sum_{j\in I}[j\in\{j_0\}]=1$, so
$\sum_{j\in I}v_j\ne0$. Thus $\{v_j:D_j\ne\varnothing\}$ is independent, and so is its subset
$\mathcal{T}(A)$.
\end{proof}

\begin{corollary}\label{cor:partition}
If $A$ carries a profile at every $\alpha$, there is a permutation $\sigma$ of $2^{[m]}$ such that
the sets $D_j:=\tau_A(\sigma^{-1}\{j\})$ are pairwise disjoint, a site has nonzero trace exactly when
it lies in some $D_j$, and a site of $D_j$ lies in $\tau_A(J)$ exactly when $j\in\sigma J$.
\end{corollary}

\begin{proof}
The permutation $\sigma$ and the disjointness are those of the proof of Theorem~\ref{thm:dichotomy},
whose trace step gives $[i\in\tau_A(J)]=[j\in\sigma J]$ for $i\in D_j$ and $t_A(i)=0$ for a site lying
in no $D_j$. A site of $D_j$ lies in $\tau_A(\sigma^{-1}\{j\})$, so by Lemma~\ref{lem:levelchar} its
trace pairs to $1$ with a choice of sets and is nonzero.
\end{proof}

In the choices recoded by $\sigma$, which carries $\sigma^{-1}\{j\}$ to $\{j\}$, an observer serving
every weighting thus reads each block $D_j$ alone: a partition of the sites it reads at all.

\begin{proposition}\label{cor:quantifier}
Take $N = 4$, $m = 3$, $A_0 = \{0,3\}$, $A_1 = \{1,3\}$, $A_2 = \{2,3\}$ and $\alpha = (3,1,1,1)$. The
traces are $\mathbf{e}_0$, $\mathbf{e}_1$, $\mathbf{e}_2$, $\mathbf{e}_0 + \mathbf{e}_1 +
\mathbf{e}_2$, four vectors in $\Ff^3$, so $\mathcal{T}_\alpha(A)$ is dependent; and
\[
R_A(\alpha) \;=\; \ms{0,4,4,4,8,8,8,12} \;=\; \Sigma_3(2,2,2),
\]
so $A$ carries a profile at $\alpha$. Proposition~\ref{prop:indep-sufficient} is therefore not a
biconditional at a fixed weighting, while $\mathcal{T}(A)$ is dependent and
Theorem~\ref{thm:dichotomy} gives a weighting at which the same family carries no profile.
\end{proposition}

\begin{proof}
Site $k\in\{0,1,2\}$ lies in $A_k$ alone and has trace $\mathbf{e}_k$; site $3$ lies in all three
sets and has trace $\mathbf{e}_0+\mathbf{e}_1+\mathbf{e}_2$. The four traces are distinct and
each class is a single site of nonzero weight, so
$\mathcal{T}_\alpha(A)=\mathcal{T}(A)$ consists of four vectors, which sum to zero and are therefore
dependent; in any case four vectors in a space of dimension three are dependent.

A site $i$ lies in $\tau_A(J)$ when $\langle t_A(i),\mathbf{1}_J\rangle=1$: site $k\le2$ when $k\in J$,
and site $3$ when $|J|$ is odd. The eight transposes and the doubled weights are therefore
\[
\begin{array}{c|cccccccc}
J&\varnothing&\{0\}&\{1\}&\{2\}&\{0,1\}&\{0,2\}&\{1,2\}&\{0,1,2\}\\\hline
\tau_A(J)&\varnothing&\{0,3\}&\{1,3\}&\{2,3\}&\{0,1\}&\{0,2\}&\{1,2\}&\{0,1,2,3\}\\
2\sum\alpha&0&8&4&4&8&8&4&12
\end{array}
\]
so $R_A(\alpha)=\ms{0,4,4,4,8,8,8,12}$. The subset sums of $(2,2,2)$ are $0$ once, $2$ three times,
$4$ three times and $6$ once, and doubling gives the same multiset. The family carries a profile at
$\alpha$ although its character support is dependent. Since $\mathcal{T}(A)$ is dependent,
Theorem~\ref{thm:dichotomy} shows that $A$ does not carry a profile at every weighting.
\end{proof}

\begin{corollary}\label{cor:thin}
If $\mathcal{T}(A)$ is dependent then the weightings at which $A$ carries a profile form a finite
union of proper linear subspaces of $\Rr^{N}$.
\end{corollary}

\begin{proof}
By \eqref{eq:admissible} the set is the finite union of the subspaces $V_\sigma$. If some
$V_\sigma$ were all of $\Rr^N$, that single $\sigma$ would satisfy the identity of the proof of
Theorem~\ref{thm:dichotomy} at every weighting, and the steps of that proof from the disjointness of
the blocks onwards, which start from that identity alone, would make $\mathcal{T}(A)$ independent,
contrary to the hypothesis. So every $V_\sigma$ is proper.
\end{proof}

\begin{corollary}\label{cor:null}
If $\mathcal{T}(A)$ is dependent, the weightings at which $A$ carries a profile form a closed,
Lebesgue-null and nowhere dense subset of $\Rr^N$.
\end{corollary}

\begin{proof}
By Corollary~\ref{cor:thin} the set is a finite union of proper subspaces, closed as a finite union
of closed sets. A proper subspace $V$ lies in a hyperplane $\{x:\sum_ia_ix_i=0\}$ with $a\in\Rr^N$, $a\ne0$,
and after an orthogonal change of coordinates, which preserves Lebesgue measure, that hyperplane is
$\{x_1=0\}$, of measure zero by Fubini's theorem; a finite union of null sets is null. A null set
contains no ball, so a closed null set has empty interior and is nowhere dense.
\end{proof}

The equivalence of Theorem~\ref{thm:dichotomy} also holds with $\alpha$ restricted to the
nonnegative weightings. If $A$ carries a profile at every nonnegative $\alpha$, the set of weightings
where it carries one contains the nonempty open set $(0,\infty)^N$ and so is not nowhere dense. By
Corollary~\ref{cor:null} $\mathcal{T}(A)$ is then linearly independent, and
Theorem~\ref{thm:dichotomy} gives a profile at every $\alpha$.

\begin{lemma}\label{lem:signchar}
For $v\in\Ff^m$ write $\vartheta_v(J)=(-1)^{\langle v,\mathbf{1}_J\rangle}$. Then
$\vartheta_u\vartheta_v=\vartheta_{u+v}$,
\begin{equation}\label{eq:orth}
\sum_{J \subseteq [m]} \vartheta_z(J) \;=\; 2^{m}\,[\,z=0\,],
\end{equation}
and with $\bar{\ell}=\sum_{v\in\mathcal{T}_\alpha(A)}\omega_v$,
\begin{equation}\label{eq:levelsign}
2\!\!\sum_{i \in \tau_A(J)}\!\!\alpha(i) \;=\; \bar{\ell} \;-\!\! \sum_{v \in \mathcal{T}_\alpha(A)}\!\!
\omega_v\, \vartheta_v(J) ,
\end{equation}
so that $\bar{\ell}$ is the mean of the $2^m$ levels.
\end{lemma}

\begin{proof}
The product rule holds because the pairing is additive in its first argument and
$(-1)^a(-1)^b=(-1)^{a+b}$ depends only on $a+b$ modulo two. For \eqref{eq:orth},
$\vartheta_0\equiv1$ gives $2^m$; if $z\ne0$, choose $l$ with $z_l=1$, and the involution
$J\mapsto J\triangle\{l\}$ of $2^{[m]}$ changes $\langle z,\mathbf{1}_J\rangle$ by $z_l=1$, so it pairs
the choices of sets into couples on which $\vartheta_z$ takes opposite values, and the sum is zero.
Writing the indicator in \eqref{eq:levelchar} as $(1-\vartheta_v)/2$ gives \eqref{eq:levelsign}, and
summing \eqref{eq:levelsign} over $J$ gives $2^m\bar{\ell}-\sum_v\omega_v\sum_J\vartheta_v(J)=2^m\bar{\ell}$
by \eqref{eq:orth}, every $v$ in the support being nonzero.
\end{proof}

\begin{theorem}\label{thm:oddgeneral}
Suppose an odd number of distinct elements of $\mathcal{T}(A)$ sum to zero. Then $A$ carries no
profile at any $\alpha$ whose class weights are all strictly positive.
\end{theorem}

\begin{proof}
Let $\mathcal{U}\subseteq\mathcal{T}(A)$ be a set of $p$ distinct traces summing to zero, $p$ odd.
Strict positivity of the class weights means $\omega_v>0$ for every $v\in\mathcal{T}(A)$, so
$\mathcal{T}_\alpha(A)=\mathcal{T}(A)\supseteq\mathcal{U}$.

\emph{The $p$-th moment of the reading.} By \eqref{eq:levelsign} the deviation of the level at $J$
from the mean $\bar{\ell}$ is $-\sum_{v\in\mathcal{T}_\alpha(A)}\omega_v\vartheta_v(J)$. Its $p$-th power
expands, by distributivity, into a sum over ordered $p$-tuples $(u_1,\dots,u_p)$ of elements of
$\mathcal{T}_\alpha(A)$ of the terms
$(-1)^p\omega_{u_1}\cdots\omega_{u_p}\vartheta_{u_1}(J)\cdots\vartheta_{u_p}(J)$,
and $\vartheta_{u_1}\cdots\vartheta_{u_p}=\vartheta_{u_1+\cdots+u_p}$ by the product rule. Summing
over $J$ and applying \eqref{eq:orth} to each tuple,
\begin{equation}\label{eq:pmoment}
\sum_{J \subseteq [m]} \Bigl( 2\!\!\sum_{i \in \tau_A(J)}\!\!\alpha(i) - \bar{\ell} \Bigr)^{\!p}
\;=\; (-1)^p\,2^{m} \!\!\!\!\sum_{\substack{u_1,\dots,u_p \in \mathcal{T}_\alpha(A) \\
u_1 + \cdots + u_p = 0}} \!\!\!\! \omega_{u_1}\cdots \omega_{u_p} .
\end{equation}

\emph{The same moment of a profile vanishes.} Let $\eta:[m]\to\Rr$. By
Proposition~\ref{prop:readout} applied to the identity grouping of $[m]$, whose blocks are the
singletons $\{j\}$, $\Sigma_m(\eta)$ is the reading of that family at the weighting $\eta$. Its
traces are the standard basis vectors $\mathbf{e}_j$, the class of $\mathbf{e}_j$ is the single
site $j$ with weight $\eta(j)$, and its character support consists of the $\mathbf{e}_j$ with
$\eta(j)\ne0$; basis vectors of zero weight contribute nothing to the sums below. So
\eqref{eq:levelsign} and \eqref{eq:pmoment} hold for $\Sigma_m(\eta)$ with mean
$\bar{\ell}_\eta=\sum_j\eta(j)$ and with tuples of standard basis vectors. The linear functional
$x\mapsto\sum_lx_l$ on $\Ff^m$ takes the value $1$ on every $\mathbf{e}_j$, hence the value
$p\equiv1$ on a sum of $p$ of them, which is therefore never zero. The sum on the right of
\eqref{eq:pmoment} is empty, and the $p$-th moment of $\Sigma_m(\eta)$ about its mean is zero.

\emph{The moment of the reading does not vanish.} On the right of \eqref{eq:pmoment} for $A$, every
term is a product of $p$ strictly positive class weights, so every term is positive; and any
enumeration $(u_1,\dots,u_p)$ of $\mathcal{U}$ is one of the tuples, with sum zero. The sum is
therefore strictly positive and the right side is nonzero.

\emph{Conclusion.} Suppose $\Sigma_m(\eta)=R_A(\alpha)$. Equal multisets of $2^m$ numbers have equal
totals, hence equal means, $\bar{\ell}_\eta=\bar{\ell}$; and for any function $f$ the sum of $f$ over a
multiset depends only on the multiset, so with $f(x)=(x-\bar{\ell})^p$ the two $p$-th moments about the common mean agree. The
first is zero and the second is not, a contradiction.
\end{proof}

\begin{corollary}\label{cor:crossing-every}
The crossing cover of Theorem~\ref{thm:overlap} carries a profile at no strictly positive weighting.
\end{corollary}

\begin{proof}
Site $0$ lies in $A_0$ alone, site $1$ in both sets and site $2$ in $A_1$ alone, so the traces are
$\mathbf{e}_0$, $\mathbf{e}_0+\mathbf{e}_1$ and $\mathbf{e}_1$: three distinct vectors of
$\Ff^2$ whose sum is $2\mathbf{e}_0+2\mathbf{e}_1=0$. Each trace is carried by a single site, so
the class weights are the site weights and are strictly positive when $\alpha$ is.
Theorem~\ref{thm:oddgeneral} with $p=3$ applies.
\end{proof}

\begin{proposition}\label{prop:evenonly}
The four traces of the family of Proposition~\ref{cor:quantifier} sum to zero, and no subset of them of
odd size does.
\end{proposition}

\begin{proof}
The traces are $\mathbf{e}_0$, $\mathbf{e}_1$, $\mathbf{e}_2$ and
$\mathbf{e}_0+\mathbf{e}_1+\mathbf{e}_2$, whose sum is $2(\mathbf{e}_0+\mathbf{e}_1+\mathbf{e}_2)=0$.
A subset of odd size has one or three elements. A single trace is nonzero, and since the four sum to
zero, any three of them sum to the fourth, which is nonzero.
\end{proof}

A readout carries a profile when it reads the levels of a screen with one weight per set
(Definition~\ref{def:readout}), as a grouping does (Proposition~\ref{prop:readout}). When its traces
are independent this holds at every weighting (Theorem~\ref{thm:dichotomy}), as for the sectors of
Proposition~\ref{prop:sectors}. When an odd number of distinct traces sum to zero it fails at every
positive weighting (Theorem~\ref{thm:oddgeneral}), as for the crossing cover of
Corollary~\ref{cor:crossing-every}. In both cases the sets decide, at every weighting in the first and
at every positive one in the second (Theorems~\ref{thm:dichotomy} and~\ref{thm:oddgeneral}). Between these cases the weights can decide. The family of
Proposition~\ref{cor:quantifier} has dependent traces, and no odd number of them sums to zero
(Proposition~\ref{prop:evenonly}). It carries a profile at $(3,1,1,1)$ and only on finitely many
proper subspaces of weightings (Corollary~\ref{cor:thin}). At $(3,1,1,1)$
these sets thus read as three separate sites of weight $2$ do (Proposition~\ref{cor:quantifier}).

\section{The boundary is profinite}\label{sec:tower}

Groupings compose, and a composite of onto maps is onto. A coarse-graining of an instrument is a
grouping of its outcomes along a partition of the index set \cite{Chitambar2014}, and the
operational $\mathrm{ER}=\mathrm{EPR}$ setting takes its protocols to be LOCC, whose definition
proceeds through coarse-graining in this sense \cite{FieldsGlazebrookMarcianoZappala2025}; profiles push along a grouping, and the tower below reads
the same data by refinement. A stage of the tower is a list $[N]$ of sites, a refinement is a
grouping of a finer list onto a coarser one, and a point of the boundary is a choice of one site per
stage that every refinement respects. A refinement is onto, and a map of a finite set onto
$[\sz(n)]$ has at least $\sz(n)$ elements in its domain, its fibres over the $\sz(n)$ sites being
nonempty and pairwise disjoint, so the sizes of the stages never decrease;
each stage is a screen of $\sz(n)$ sites and each refinement a grouping of it onto the one before.

\begin{definition}\label{def:tower}
A \emph{tower} $T$ is a sequence of sizes $N_n>0$, $n\in\Nn$, together with refinements
$\rf_n : [\sz(n+1)] \to [\sz(n)]$, each onto. Stage $n$ is the list of sites $[\sz(n)]$.
\end{definition}

\begin{definition}\label{def:path}
The \emph{boundary} of $T$ is the set of compatible choices of one site per stage,
\[
\Pi_T \;:=\; \Bigl\{\, p \in \prod_{n \in \Nn} [\sz(n)] \;:\; \rf_n(p_{n+1}) = p_n \ \text{for all } n \,\Bigr\},
\]
topologised as a subspace of the product of the discrete finite stages, with projections
$\pi_n(p) = p_n$.
\end{definition}

Each stage carries the discrete topology, in which every subset is open, and so also closed, its
complement being open. A basic open set
of the product is a set $\mathcal{C}(F,B) := \{x : x_k \in B_k \text{ for } k \in F\}$ with
$F \subseteq \Nn$ finite and $B_k \subseteq [\sz(k)]$, and the open sets of $\Pi_T$ are the
intersections with $\Pi_T$ of unions of such sets. Every coordinate map of the product is
continuous, the preimage of $B\subseteq[\sz(n)]$ under the $n$-th one being the basic open set
$\mathcal{C}(\{n\},B)$, so each $\pi_n$, being the restriction of the $n$-th coordinate map, is
continuous into the discrete stage $[\sz(n)]$. For $i \in [\sz(n)]$ the \emph{cylinder} is
$C_{n,i} := \{p \in \Pi_T : p_n = i\} = \pi_n^{-1}(i)$.

\begin{lemma}\label{lem:coarser}
If $p, q \in \Pi_T$ and $p_n = q_n$, then $p_k = q_k$ for every $k \leq n$: the coordinate of a point
at a stage determines all its coarser coordinates. Moreover $C_{n,i}$ is the union over
$j \in (\rf_n)^{-1}(i)$ of the sets $C_{n+1,j}$.
\end{lemma}

\begin{proof}
By descending induction on $k$ from $k=n$: the case $k=n$ is the hypothesis, and if
$p_{k+1} = q_{k+1}$ with $k+1 \leq n$ then $p_k = \rf_k(p_{k+1}) = \rf_k(q_{k+1}) = q_k$ by the
compatibility of $p$ and $q$. For the second clause, $p_n = i$ holds exactly when
$\rf_n(p_{n+1}) = i$.
\end{proof}

A \emph{profinite set} is a compact Hausdorff totally disconnected space. Fix an uncountable strong
limit cardinal $\kappa$. A $\kappa$-condensed set is a sheaf of sets on the site of profinite sets of
cardinality below $\kappa$, covers being finite jointly surjective families, and condensed sets are
their colimit over all such $\kappa$, every transition functor being fully faithful
\cite{Scholze2026}; we work with $\kappa$-condensed sets throughout and call them condensed. Write
$\mathrm{Prof}$ for the category of profinite sets of cardinality below $\kappa$ with continuous
maps, $\mathrm{Cond}$ for the category of condensed sets with morphisms of sheaves, and
$\mathrm{Hom}_{\mathrm{Prof}}$, $\mathrm{Hom}_{\mathrm{Cond}}$ for their sets of morphisms. A
topological space $X$ gives the condensed set $\underline{X}$ whose value at a profinite set $S'$ is
the set $C(S',X)$ of continuous maps $S'\to X$, with restriction along a continuous $\psi:S''\to S'$
given by composition with $\psi$ \cite{Scholze2026}. For a profinite set $S$ in the site,
$\underline{S}(S')=C(S',S)=\mathrm{Hom}_{\mathrm{Prof}}(S',S)$, the morphisms of $\mathrm{Prof}$ being
the continuous maps, so $\underline{S}$ is the presheaf $S$ represents. A continuous
$h : S_1 \to S_2$ is sent to the morphism $\underline{h}$ whose component at $S'$ is
$\psi \mapsto h \circ \psi$, natural in $S'$ because $(h\circ\psi)\circ\varphi=h\circ(\psi\circ\varphi)$;
since $\mathrm{id}\circ\psi=\psi$ and $(h_2\circ h_1)\circ\psi=h_2\circ(h_1\circ\psi)$, this is a
functor $\underline{(-)}:\mathrm{Prof}\to\mathrm{Cond}$.

\begin{theorem}\label{thm:screen-profinite}
$\Pi_T$ is closed in $\prod_n [\sz(n)]$, hence compact, Hausdorff and totally disconnected: it is a
profinite set, of cardinality at most $2^{\aleph_0}<\kappa$, so it lies in the site, and the
$\kappa$-condensed set $\underline{\Pi_T}$ of \cite{Scholze2026} is the \emph{condensed boundary}
of $T$.
\end{theorem}

\begin{proof}
\emph{Closed.} Fix $n$ and let $Z_n := \{x \in \prod_k [\sz(k)] : \rf_n(x_{n+1}) = x_n\}$. The map
$x \mapsto (x_{n+1}, x_n)$ from the product to the finite discrete space
$[\sz(n+1)] \times [\sz(n)]$ is continuous, its two components being coordinate maps, and $Z_n$ is
the preimage under it of the graph $\{(j, \rf_n(j)) : j \in [\sz(n+1)]\}$. Every subset of a
discrete space is closed, so $Z_n$ is closed, and $\Pi_T = \bigcap_n Z_n$ is closed as an
intersection of closed sets.

\emph{Compact.} Let $\mathcal{U}$ be a cover of $\Pi_T$ by open subsets of $\Pi_T$, and call a
subset of $\Pi_T$ \emph{thin} when finitely many members of $\mathcal{U}$ cover it. A finite union
of thin sets is thin. Suppose $\Pi_T$ is not thin. The finitely many sets $C_{0,i}$,
$i \in [\sz(0)]$, have union $\Pi_T$, so one of them is not thin; let $i_0$ be the least such $i$.
Given $i_n$ with $C_{n,i_n}$ not thin, the set $C_{n,i_n}$ is the finite union of the $C_{n+1,j}$
over $j \in (\rf_n)^{-1}(i_n)$, so one of these is not thin; let $i_{n+1}$ be the least such $j$.
Then $\rf_n(i_{n+1}) = i_n$ for every $n$, so $p := (i_n)_n$ lies in $\Pi_T$, and $p \in U$ for some
$U \in \mathcal{U}$. As $U$ is open in $\Pi_T$, there are a finite $F \subseteq \Nn$ and sets
$B_k$ with $p \in \mathcal{C}(F,B) \cap \Pi_T \subseteq U$. Let $n$ be at least every element of $F$.
If $q \in C_{n,i_n}$ then $q_n = p_n$, so $q_k = p_k \in B_k$ for every $k \in F$ by
Lemma~\ref{lem:coarser}; hence $C_{n,i_n} \subseteq U$, and $C_{n,i_n}$ is thin, a contradiction. So
$\Pi_T$ is thin, which is compactness.

\emph{Hausdorff and totally disconnected.} Each $C_{n,i}=\pi_n^{-1}(i)$ is open and closed, $\pi_n$
being continuous into a discrete space. If $q \neq r$ in $\Pi_T$, the two sequences differ at
some coordinate $n$, and $C_{n,q_n}$, $C_{n,r_n}$ are disjoint open sets containing $q$ and $r$.
Moreover, if $Y \subseteq \Pi_T$ contains two points $q \neq r$, then with $n$ as before the sets
$Y \cap C_{n,q_n}$ and $Y \setminus C_{n,q_n}$ are open in $Y$, because $C_{n,q_n}$ is open and
closed; they are disjoint, cover $Y$, and contain $q$ and $r$ respectively. So $Y$ is not
connected, and the only connected subsets are points and the empty set.

So $\Pi_T$ is a profinite set. It is a subset of a countable product of finite sets, so its
cardinality is at most $2^{\aleph_0}$, which is below $\kappa$ because $\aleph_0<\kappa$ and
$\kappa$ is a strong limit; so $\Pi_T$ lies in the site and $\underline{\Pi_T}$ is defined.
\end{proof}

\begin{proposition}\label{prop:invlimit}
If $X$ is a topological space and $h_n : X \to [\sz(n)]$ are continuous maps with
$\rf_n \circ h_{n+1} = h_n$ for every $n$, there is exactly one continuous $h : X \to \Pi_T$ with
$\pi_n \circ h = h_n$ for every $n$.
\end{proposition}

\begin{proof}
$h(x) := (h_n(x))_n$ is a point of $\Pi_T$ for every $x$, by the compatibility of the $h_n$; the map
$h$ is continuous because each of its coordinates is; and $\pi_n \circ h = h_n$. Any map with this
last property has the coordinates $h_n$ and so equals $h$.
\end{proof}

The boundary is therefore the inverse limit of the finite discrete stages along the refinements,
which is how \cite{Scholze2026} writes a profinite set, as a cofiltered limit of finite sets. Passing
to condensed sets loses no map between boundaries, for every uncountable strong limit cardinal
$\kappa$. The same holds in
the universe-based formulation, sheaves on the compact Hausdorff spaces of one Grothendieck
universe with values in the sets of the next: for $\kappa$ the strongly inaccessible cardinal of the
smaller universe these are, by restriction to profinite sets, the $\kappa$-condensed sets
\cite{Scholze2026}.

\begin{theorem}\label{thm:screen-ff}
Let $T_1, T_2$ be towers.
\begin{enumerate}[label=(\roman*)]
\item For every uncountable strong limit cardinal $\kappa$, the map
$\mathrm{Hom}_{\mathrm{Prof}}(\Pi_{T_1}, \Pi_{T_2}) \to \mathrm{Hom}_{\mathrm{Cond}}(\underline{\Pi_{T_1}}, \underline{\Pi_{T_2}})$,
$h\mapsto\underline{h}$, into the $\kappa$-condensed sets of \cite{Scholze2026} is a bijection.
\item The same map is a bijection into the sheaves, for finite jointly surjective covers, on the
compact Hausdorff spaces of a Grothendieck universe with values in the sets of the next universe;
when $\kappa$ is the strongly inaccessible cardinal of the smaller universe, these are the
$\kappa$-condensed sets \cite{Scholze2026}.
\end{enumerate}
\end{theorem}

\begin{proof}
\emph{(i).} Let $\kappa$ be an uncountable strong limit cardinal. By
Theorem~\ref{thm:screen-profinite} both boundaries are profinite sets in the site, and both
condensed boundaries are defined. Condensed sets are a full subcategory of presheaves on profinite sets, so a morphism
$\underline{\Pi_{T_1}} \to \underline{\Pi_{T_2}}$ is a natural transformation of the underlying
presheaves. Since $\underline{\Pi_{T_1}}$ is the presheaf represented by $\Pi_{T_1}$, the Yoneda
lemma says that
\[
y : \mathrm{Hom}_{\mathrm{Cond}}(\underline{\Pi_{T_1}}, \underline{\Pi_{T_2}}) \longrightarrow
\underline{\Pi_{T_2}}(\Pi_{T_1}) = C(\Pi_{T_1}, \Pi_{T_2}), \qquad
\xi \longmapsto \xi_{\Pi_{T_1}}(\mathrm{id}),
\]
is a bijection. For a continuous $h : \Pi_{T_1} \to \Pi_{T_2}$,
$y(\underline{h}) = h \circ \mathrm{id} = h$, so the map
$h \mapsto \underline{h}$ of the statement, followed by $y$, is the identity of
$C(\Pi_{T_1}, \Pi_{T_2}) = \mathrm{Hom}_{\mathrm{Prof}}(\Pi_{T_1}, \Pi_{T_2})$. Hence
$h \mapsto \underline{h}$ equals $y^{-1}$ and is a bijection. This is a case of the full
faithfulness of $X\mapsto\underline{X}$ on $\kappa$-compactly generated spaces
\cite[Proposition~1.7]{Scholze2026}, read at the two boundaries: a compact Hausdorff space of
cardinality below $\kappa$ carries the quotient topology of its own identity map, and so is
$\kappa$-compactly generated.

\emph{(ii).} Let $\kappa$ be the strongly inaccessible cardinal of the smaller of two Grothendieck
universes. Up to homeomorphism, the compact Hausdorff spaces of that universe are those of cardinality
below $\kappa$, and
sheaves on $\kappa$-small compact Hausdorff spaces for finite jointly surjective covers are
equivalent, by restriction to profinite sets, to $\kappa$-condensed sets \cite{Scholze2026}, which in the
larger universe are the formulation of Barwick and Haine recorded in the same source. The
equivalence restricts to the identity on profinite sets, so it carries $\underline h$ to $\underline h$; a
strongly inaccessible cardinal is in particular an uncountable strong limit cardinal, so (ii) is (i)
at this $\kappa$.
\end{proof}

\begin{definition}\label{def:pathThrough}
For $m \leq K$ let $\delta^K_m : [\sz(K)] \to [\sz(m)]$ be given by $\delta^m_m = \mathrm{id}$ and
$\delta^{K+1}_m = \delta^{K}_m \circ \rf_K$. Choose for each $n$ and $x \in [\sz(n)]$ a site
$u_n(x)$ with $\rf_n(u_n(x)) = x$. The \emph{chain above} $i \in [\sz(n)]$ is $c_0 = i$,
$c_{k+1} = u_{n+k}(c_k)$, and $P(n,i)_m := \delta^{\,n+m}_{m}(c_m)$ for $m \in \Nn$.
\end{definition}

The site $u_n(x)$ exists because $\rf_n$ is onto. By induction on $k$, $c_k \in [\sz(n+k)]$: $c_0=i$
lies in $[\sz(n)]$, and if $c_k\in[\sz(n+k)]$ then $c_{k+1}=u_{n+k}(c_k)$ lies in $[\sz(n+k+1)]$, the
domain of $\rf_{n+k}$; so $P(n,i)_m=\delta^{n+m}_m(c_m)\in[\sz(m)]$. Since $u_{n+k}$ was chosen with
$\rf_{n+k}(u_{n+k}(x))=x$, the chain satisfies $\rf_{n+k}(c_{k+1})=\rf_{n+k}(u_{n+k}(c_k))=c_k$ for every $k$. The map $\delta^K_m$ carries a site of stage $K$ down to stage $m$ along the refinements, and
along a point of the boundary it returns that point's own coordinates.

\begin{lemma}\label{lem:drop}
For every $p \in \Pi_T$ and $m \leq K$, $\delta^K_m(p_K) = p_m$.
\end{lemma}

\begin{proof}
By induction on $K$: $\delta^m_m(p_m) = p_m$, and
$\delta^{K+1}_m(p_{K+1}) = \delta^K_m(\rf_K(p_{K+1})) = \delta^K_m(p_K)$ by the compatibility of $p$.
\end{proof}

\begin{theorem}\label{thm:hpath}
For every tower $T$, every $n$ and every $i \in [\sz(n)]$, $P(n,i)\in\Pi_T$ and $P(n,i)_n = i$. Hence
every $\pi_n$ is onto.
\end{theorem}

\begin{proof}
The recursion defining $\delta$ removes a refinement at the top. We first show that a refinement can
also be removed at the bottom:
\[
\rf_m \circ \delta^K_{m+1} = \delta^K_m \qquad \text{for all } K \geq m+1 .
\]
By induction on $K$. For $K = m+1$, $\delta^{m+1}_{m+1} = \mathrm{id}$ and
$\delta^{m+1}_m = \delta^m_m \circ \rf_m = \rf_m$, so both sides equal $\rf_m$. If the identity holds
at $K$, then
$\rf_m \circ \delta^{K+1}_{m+1} = \rf_m \circ \delta^K_{m+1} \circ \rf_K = \delta^K_m \circ \rf_K
= \delta^{K+1}_m$.

\emph{Compatibility.} For every $m$,
\[
\begin{aligned}
\rf_m\bigl(P(n,i)_{m+1}\bigr)
&= \rf_m\bigl(\delta^{n+m+1}_{m+1}(c_{m+1})\bigr)
= \delta^{n+m+1}_m(c_{m+1})\\
&= \delta^{n+m}_m\bigl(\rf_{n+m}(c_{m+1})\bigr)
= \delta^{n+m}_m(c_m)
= P(n,i)_m ,
\end{aligned}
\]
by the bottom identity (with $K = n+m+1 \geq m+1$), the defining recursion, and the chain property
$\rf_{n+m}(c_{m+1}) = c_m$. So $P(n,i) \in \Pi_T$.

\emph{Passage through $i$.} We show $\delta^{n+k}_n(c_k) = i$ for every $k$, by induction. At $k=0$
it is $\delta^n_n(i) = i$. If it holds at $k$, then
$\delta^{n+k+1}_n(c_{k+1}) = \delta^{n+k}_n(\rf_{n+k}(c_{k+1})) = \delta^{n+k}_n(c_k) = i$. At
$k = n$ this reads $\delta^{2n}_n(c_n) = i$, which is $P(n,i)_n = i$.

Finally, for every $i \in [\sz(n)]$ the point $P(n,i)$ of $\Pi_T$ satisfies $\pi_n(P(n,i)) = i$, so
$\pi_n$ is onto. Since $\sz(0) > 0$, in particular $\Pi_T$ is not empty.
\end{proof}

\begin{definition}\label{def:descends}
Write $\mathfrak{S}_X$ for the group of permutations of a finite set $X$. Let $g : [N] \to [m]$ be
onto, $\sigma \in \mathfrak{S}_{[N]}$, $\upsilon \in \mathfrak{S}_{[m]}$. Then $\sigma$ \emph{descends
to} $\upsilon$ along $g$ when $g \circ \sigma = \upsilon \circ g$. The \emph{block stabiliser}
$\mathfrak{S}_g\subseteq\mathfrak{S}_{[N]}$ is the set of $\sigma$ that descend along $g$ to some
$\upsilon$, and $d_g(\sigma)$ denotes the permutation $\sigma$ descends to.
\end{definition}

\begin{proposition}\label{prop:descend}
Let $g : [N] \to [m]$ be onto. If $\sigma$ descends along $g$ to $\upsilon$ and to $\upsilon'$ then
$\upsilon = \upsilon'$, so $d_g$ is a function on $\mathfrak{S}_g$. The permutation $\sigma$ descends to
$\upsilon$ exactly when $\sigma$ carries each block $g^{-1}(j)$ into $g^{-1}(\upsilon(j))$, and then onto
it; $\sigma \in \mathfrak{S}_g$ exactly when $\sigma$ carries each block into some block. The block
stabiliser $\mathfrak{S}_g$ is a subgroup of $\mathfrak{S}_{[N]}$ and $d_g$ a homomorphism.
\end{proposition}

\begin{proof}
\emph{Uniqueness.} For $j \in [m]$ pick $i \in g^{-1}(j)$, nonempty as $g$ is onto; then
$\upsilon(j) = g(\sigma(i)) = \upsilon'(j)$.
\emph{Blocks.} That $\sigma$ descends to $\upsilon$ says that $i \in g^{-1}(j)$ implies
$\sigma(i) \in g^{-1}(\upsilon(j))$; and a site $k$ of $g^{-1}(\upsilon(j))$ is $\sigma(\sigma^{-1}(k))$ with
$\upsilon(g(\sigma^{-1}(k))) = g(k) = \upsilon(j)$, so $\sigma^{-1}(k) \in g^{-1}(j)$, $\upsilon$ being injective, and the block is carried
onto. Conversely, if $\sigma$ carries each block into some block, then $\upsilon(j) := g(\sigma(i))$ for
any $i \in g^{-1}(j)$ is well defined and satisfies $g \circ \sigma = \upsilon \circ g$; it is onto, since
for $j' \in [m]$ and any $i' \in g^{-1}(j')$ we have
$j' = g(\sigma(\sigma^{-1}(i'))) = \upsilon(g(\sigma^{-1}(i')))$, and an onto map of a finite set to itself
is a permutation.
\emph{Group.} The identity descends to the identity; if $\sigma$ descends to $\upsilon$ and $\sigma'$
to $\upsilon'$ then $g\sigma\sigma' = \upsilon g \sigma' = \upsilon\upsilon' g$; and
composing $g\sigma = \upsilon g$ with $\sigma^{-1}$ on the right and $\upsilon^{-1}$ on the left gives
$\upsilon^{-1} g = g \sigma^{-1}$. By uniqueness these identities say $d_g(\mathrm{id})=\mathrm{id}$,
$d_g(\sigma\sigma')=d_g(\sigma)d_g(\sigma')$ and $d_g(\sigma^{-1})=d_g(\sigma)^{-1}$.
\end{proof}

\begin{proposition}\label{prop:no-descent}
Let $g_{22}$ be as in Definition~\ref{def:two-groupings} and $c := (1\ 2) \in \mathfrak{S}_{[4]}$.
The transposition $c$ descends along $g_{22}$ to no $\upsilon$; consequently no function
$\Upsilon : \mathfrak{S}_{[4]} \to \mathfrak{S}_{[2]}$ has every $\sigma$ descending along $g_{22}$ to
$\Upsilon(\sigma)$.
\end{proposition}

\begin{proof}
Recall $g_{22}(0)=g_{22}(1)=0$ and $g_{22}(2)=g_{22}(3)=1$, and $c$ exchanges the sites $1$ and $2$
and fixes $0$ and $3$. Suppose $\upsilon \in \mathfrak{S}_{[2]}$ satisfies $g_{22} \circ c = \upsilon \circ
g_{22}$, and evaluate at the two sites of the block $g_{22}^{-1}(0) = \{0,1\}$. At the site $0$,
$\upsilon(0) = \upsilon(g_{22}(0)) = g_{22}(c(0)) = g_{22}(0) = 0$. At the site $1$,
$\upsilon(0) = \upsilon(g_{22}(1)) = g_{22}(c(1)) = g_{22}(2) = 1$. The two evaluations give $0 = 1$, a
contradiction, so no such $\upsilon$ exists. If a function $\Upsilon$ as in the second clause existed,
$c$ would descend to $\Upsilon(c)$, which has just been excluded.
\end{proof}

So the grouping $g_{22}$ induces no map $\mathfrak{S}_{[4]}\to\mathfrak{S}_{[2]}$ along which every
permutation descends. The transposition $c$ splits the block $\{0,1\}$, sending $0$ into the first
block, $c(0)=0$, and $1$ into the second, $c(1)=2$, and by
Proposition~\ref{prop:descend} it is exactly such splitting that prevents a permutation from
descending. The permutations that a grouping lets through to the coarse list are those of the block
stabiliser, a subgroup by Proposition~\ref{prop:descend}.

\section{Readings on the boundary}\label{sec:readings}

A function $f : X \to \Rr$ on a topological space is \emph{locally constant} when every point of $X$ has an
open neighbourhood on which $f$ is constant.

\begin{definition}\label{def:cyl}
For $i \in [\sz(n)]$ the indicator of the cylinder $C_{n,i} = \pi_n^{-1}(i)$ is $\mathbf{1}_{C_{n,i}}$;
a function $f$ on $\Pi_T$ is \emph{constant at stage $n$} when $\pi_n(q) = \pi_n(q')$ implies
$f(q) = f(q')$.
\end{definition}

\begin{proposition}\label{prop:cyl}
Each $C_{n,i}$ is clopen; distinct sites of a stage give disjoint cylinders;
$C_{n,i} = \bigsqcup_{j \in (\rf_n)^{-1}(i)} C_{n+1,j}$; and for every open $U$ and $p \in U$ there is
$n$ with $C_{n,\pi_n(p)} \subseteq U$.
\end{proposition}

\begin{proof}
The map $\pi_n$ is continuous into the discrete space $[\sz(n)]$, where the point $\{i\}$ is open and
closed, so $C_{n,i} = \pi_n^{-1}(i)$ is open and closed. If $i \neq i'$, no $q$ has $\pi_n(q)$ equal to
both, so $C_{n,i} \cap C_{n,i'} = \emptyset$. For the decomposition, $q \in C_{n,i}$ iff $q_n = i$ iff
$\rf_n(q_{n+1}) = i$, by the compatibility of $q$, iff $q_{n+1} \in (\rf_n)^{-1}(i)$ iff
$q \in C_{n+1,j}$ for some $j \in (\rf_n)^{-1}(i)$; the union is disjoint by the second clause at
stage $n+1$. For the last clause, $U$ is open in the subspace $\Pi_T$, so there are a finite
$F \subseteq \Nn$ and sets $B_k \subseteq [\sz(k)]$ with
$p \in \mathcal{C}(F,B) \cap \Pi_T \subseteq U$. Choose $n$ at least every
element of $F$. If $q \in C_{n,\pi_n(p)}$ then $q_n = p_n$, so $q_k = p_k$ for every $k \leq n$ by
Lemma~\ref{lem:coarser}; in particular $q_k = p_k \in B_k$ for $k \in F$, and $q \in U$.
\end{proof}

\begin{definition}\label{def:profilefamily}
A \emph{compatible family of profiles} is $a = (a_n)_{n}$, $a_n : [\sz(n)] \to \Rr$, with
$(\rf_n)_{*}a_{n+1} = a_n$ for every $n$; $\mathcal{P}(T)$ is the set of these. For
$f : [\sz(n)] \to \Rr$ put $\int_n f \, da := \sum_{i} a_n(i)\, f(i)$.
\end{definition}

\begin{proposition}\label{prop:family}
For $a \in \mathcal{P}(T)$, $a_n(i) = \sum_{j \in (\rf_n)^{-1}(i)} a_{n+1}(j)$; the total mass is the
same at every stage; and $\int_{n+1} (f \circ \rf_n) \, da = \int_{n} f \, da$.
\end{proposition}

\begin{proof}
By Definition~\ref{def:push}, $((\rf_n)_*a_{n+1})(i) = \sum_{j \in (\rf_n)^{-1}(i)} a_{n+1}(j)$, and
compatibility sets this equal to $a_n(i)$; that is the first identity. Proposition~\ref{prop:push-basic}
applied to $g = \rf_n$ and $\alpha = a_{n+1}$ gives
$\sum_i a_n(i) = \sum_i ((\rf_n)_*a_{n+1})(i) = \sum_j a_{n+1}(j)$, so the total mass at stage $n$
equals that at stage $n+1$, and by induction on $n$ it equals the total mass at stage $0$. For the
third, every $j \in [\sz(n+1)]$ lies in exactly one block $(\rf_n)^{-1}(i)$, namely that of
$i = \rf_n(j)$ (Proposition~\ref{prop:push-basic}), so
\[
\begin{aligned}
\int_{n+1}(f\circ\rf_n)\,da
&= \sum_{j} a_{n+1}(j)\, f(\rf_n(j))
= \sum_{i} \sum_{j \in (\rf_n)^{-1}(i)} a_{n+1}(j)\, f(i)\\
&= \sum_{i} f(i)\, a_n(i)
= \int_n f\,da ,
\end{aligned}
\]
the second equality because $f(\rf_n(j)) = f(i)$ throughout the block of $i$, and the third by the
first identity.
\end{proof}

\begin{proposition}\label{prop:factor}
Every locally constant $f : \Pi_T \to \Rr$ is constant at some stage.
\end{proposition}

\begin{proof}
For each $p \in \Pi_T$ local constancy gives an open $V_p \ni p$ on which $f$ is constant, and
Proposition~\ref{prop:cyl} gives a stage $n_p$ with $C_{n_p,\pi_{n_p}(p)} \subseteq V_p$; so $f$ is
constant, with value $f(p)$, on the cylinder $C_{n_p,\pi_{n_p}(p)}$, which contains $p$ and is open.
These cylinders cover $\Pi_T$, which is compact by Theorem~\ref{thm:screen-profinite}, so finitely
many of them, at points $p^1, \dots, p^k$, already cover it. Let $n$ be the largest of
$n_{p^1}, \dots, n_{p^k}$. Suppose $\pi_n(q) = \pi_n(r)$. Some $l$ has
$q \in C_{n_{p^l},\pi_{n_{p^l}}(p^l)}$, that is $q_{n_{p^l}} = p^l_{n_{p^l}}$. As $n_{p^l} \leq n$ and
$q_n = r_n$, the coordinate at stage $n$ determining the coarser ones gives
$r_{n_{p^l}} = q_{n_{p^l}}$, so $r$ lies in the same cylinder. Hence $f(q) = f(p^l) = f(r)$, and $f$
is constant at stage $n$.
\end{proof}

\begin{lemma}\label{lem:conststage}
A function constant at stage $n$ is locally constant and constant at every stage $m \geq n$. So the
locally constant functions are exactly the functions constant at some stage, and any finitely many
of them are constant at a common stage.
\end{lemma}

\begin{proof}
A function constant at stage $n$ is constant on each cylinder of stage $n$, an open set containing
any given one of its points (Proposition~\ref{prop:cyl}), so it is locally constant. If $f$ is
constant at stage $n$ and $\pi_{n+1}(q) = \pi_{n+1}(r)$, then
$\pi_n(q) = \rf_n(q_{n+1}) = \rf_n(r_{n+1}) = \pi_n(r)$ and $f(q) = f(r)$; by induction $f$ is constant
at every stage $m \geq n$. With Proposition~\ref{prop:factor} this gives the description of the
locally constant functions, and finitely many of them are constant at the largest of their stages.
\end{proof}

\begin{definition}\label{def:principal}
Let $G$ be a topological group. A \emph{principal $G$-bundle} over $\Pi_T$ is a topological space
$P$ with a continuous map $\mathrm{pr}:P\to\Pi_T$ and a continuous right action
$(x,g)\mapsto x\cdot g$ of $G$ on $P$ with $\mathrm{pr}(x\cdot g)=\mathrm{pr}(x)$, such that every
point of $\Pi_T$ has an open neighbourhood $U$ and a homeomorphism
$e_U:\mathrm{pr}^{-1}(U)\to U\times G$ whose first component is $\mathrm{pr}$ and whose second
component satisfies $e_U(x\cdot g)_2=e_U(x)_2\,g$. The bundle is \emph{trivial} when there is such a
homeomorphism with $U=\Pi_T$.
\end{definition}

\begin{proposition}\label{prop:boundary-trivial}
\begin{enumerate}[label=(\roman*)]
\item For every cover of $\Pi_T$ by open sets there is a stage $n$ such that every cylinder
$C_{n,i}$ lies in one member of the cover.
\item For every topological group $G$, every principal $G$-bundle over $\Pi_T$ is trivial; it
therefore has a continuous section $s$ with $\mathrm{pr}\circ s=\mathrm{id}$, and over every open
$U\subseteq\Pi_T$ its continuous sections correspond bijectively, through the trivialisation, to the
continuous maps $U\to G$.
\item Every continuous map $[0,1]\to\Pi_T$ is constant.
\end{enumerate}
\end{proposition}

\begin{proof}
\emph{(i).} Each point $q$ lies in some member $U_q$ of the cover, and
Proposition~\ref{prop:cyl} gives a stage $n_q$ with $C_{n_q,\pi_{n_q}(q)}\subseteq U_q$. These
cylinders are open and cover $\Pi_T$, which is compact by Theorem~\ref{thm:screen-profinite}, so
finitely many of them, at points $q^1,\dots,q^k$, already cover it. Let $n$ be the largest of
$n_{q^1},\dots,n_{q^k}$ and let $i$ be a site of stage $n$. Theorem~\ref{thm:hpath} gives a point
$r$ with $\pi_n(r)=i$, and $r$ lies in the cylinder of some $q^l$, so $r_{n_{q^l}}=q^l_{n_{q^l}}$.
If $r'\in C_{n,i}$ then $r'_n=r_n$, so $r'_{n_{q^l}}=r_{n_{q^l}}=q^l_{n_{q^l}}$ by
Lemma~\ref{lem:coarser}, as $n_{q^l}\le n$; hence $C_{n,i}\subseteq U_{q^l}$.

\emph{(ii).} The neighbourhoods $U$ of Definition~\ref{def:principal} cover $\Pi_T$, so (i) gives
a stage $n$ and for each site $i$ of stage $n$ a neighbourhood $U(i)\supseteq C_{n,i}$ with its
homeomorphism $e_i:=e_{U(i)}$. Every $y\in\Pi_T$ lies in $C_{n,y_n}\subseteq U(y_n)$, so
\[
E(x):=e_{\mathrm{pr}(x)_n}(x),\qquad F(y,g):=e_{y_n}^{-1}(y,g)
\]
are defined on all of $P$ and of $\Pi_T\times G$. The sets $\mathrm{pr}^{-1}(C_{n,i})$ are open,
$\mathrm{pr}$ being continuous and $C_{n,i}$ clopen (Proposition~\ref{prop:cyl}), and so are the sets
$C_{n,i}\times G$; as $i$ runs over the sites of stage $n$ they cover $P$ and $\Pi_T\times G$, and on
them $E$ and $F$ agree with the continuous maps $e_i$ and $e_i^{-1}$, so $E$ and $F$ are continuous.
For $x\in P$ the site $i=\mathrm{pr}(x)_n$ gives $F(E(x))=e_i^{-1}(e_i(x))=x$, the first component of
$e_i(x)$ being $\mathrm{pr}(x)$. For $(y,g)$ put $z:=e_{y_n}^{-1}(y,g)$; the first component of
$e_{y_n}(z)=(y,g)$ is $\mathrm{pr}(z)$, so $\mathrm{pr}(z)=y$, the same site $y_n$ is read at $z$, and
$E(z)=e_{y_n}(z)=(y,g)$. So $E$ is a homeomorphism $P\to\Pi_T\times G$ with first component
$\mathrm{pr}$. Since $\mathrm{pr}(x\cdot g)=\mathrm{pr}(x)$, the points $x\cdot g$ and $x$ read the
same site $i$, and $E(x\cdot g)_2=e_i(x\cdot g)_2=e_i(x)_2\,g=E(x)_2\,g$. The map $s(y):=E^{-1}(y,1)$ is
continuous, and $\mathrm{pr}(s(y))=y$ because the first component of $E$ is $\mathrm{pr}$. Over an open
$U$ a continuous section $\sigma$ goes to the continuous map $y\mapsto E(\sigma(y))_2$, and a continuous
$g:U\to G$ goes back to $y\mapsto E^{-1}(y,g(y))$, which is continuous and a section for the same
reason as $s$; the two assignments are inverse, since
$E(\sigma(y))=(y,E(\sigma(y))_2)$ for a section $\sigma$ and $E$ is a bijection.

\emph{(iii).} The interval $[0,1]$ is connected, so the image of a continuous map
$[0,1]\to\Pi_T$ is connected, and the only connected subsets of $\Pi_T$ are points and the empty set
(Theorem~\ref{thm:screen-profinite}); the image is a single point.
\end{proof}

A frame over the boundary can therefore be chosen for all its points at once, cylinder by cylinder at
a single stage, so that the sheaf of continuous sections of a principal bundle over the boundary has a
global section and corresponds, open set by open set, to the sheaf of continuous $G$-valued maps; and the boundary has no non-constant
loop around which a connection could carry a frame.

\begin{definition}\label{def:pairing}
By Theorem~\ref{thm:hpath} choose for each $i\in[\sz(n)]$ some $q^n_i$ with $\pi_n(q^n_i) = i$ and set
$\widehat{f}_n(i) := f(q^n_i)$. For $f$ locally constant and $n$ a stage at which it is constant,
$\langle a, f \rangle := \int_n \widehat{f}_n \, da$.
\end{definition}

\begin{theorem}\label{thm:pairing-welldef}
Let $f$ be constant at stage $n$. Then $\widehat{f}_n$ does not depend on the choices,
$\widehat{f}_n(\pi_n(q)) = f(q)$ for every $q$, and $\widehat{f}_{n+1} = \widehat{f}_{n} \circ \rf_n$.
Consequently $\langle a, f\rangle$ depends on $a$ and $f$ alone and not on the stage.
\end{theorem}

\begin{proof}
\emph{Independence.} If $q$ and $r$ are two admissible choices at the site $i$, then
$\pi_n(q) = i = \pi_n(r)$, and constancy at stage $n$ gives $f(q) = f(r)$; so $\widehat f_n(i)$ is
the same for every choice.

\emph{Evaluation.} Given $q$, the point $q$ itself is an admissible choice at the site $\pi_n(q)$,
so by independence $\widehat f_n(\pi_n(q)) = f(q)$.

\emph{Refinement.} By Lemma~\ref{lem:conststage}, $f$ is also
constant at stage $n+1$, so $\widehat f_{n+1}$ is independent of choices as well. For
$j \in [\sz(n+1)]$ let $q$ be any point with $\pi_{n+1}(q) = j$. By compatibility
$\pi_n(q) = \rf_n(q_{n+1}) = \rf_n(j)$, and evaluation at stages $n+1$ and $n$ gives
$\widehat f_{n+1}(j) = f(q) = \widehat f_n(\pi_n(q)) = \widehat f_n(\rf_n(j))$.

\emph{Independence of the stage.} By the refinement identity and Proposition~\ref{prop:family},
$\int_{n+1}\widehat f_{n+1}\,da = \int_{n+1}(\widehat f_n\circ\rf_n)\,da = \int_n \widehat f_n\,da$.
Since $f$ is constant at every stage $m \geq n$, induction on $m$ gives
$\int_m \widehat f_m\,da = \int_n \widehat f_n\,da$ for all $m \geq n$. If $f$ is constant at stages
$n$ and $n'$, both integrals equal the integral at $\max(n,n')$, so they are equal, and
$\langle a, f\rangle$ is determined by $a$ and $f$.
\end{proof}

\begin{theorem}\label{thm:pairing-injective}
$\langle a, \mathbf{1}_{C_{n,i}} \rangle = a_n(i)$; hence if $\langle a, f\rangle = \langle b, f\rangle$ for
every locally constant $f$ then $a = b$.
\end{theorem}

\begin{proof}
The indicator $\mathbf{1}_{C_{n,i}}$ is constant at stage $n$: if $\pi_n(q) = \pi_n(r)$ then $q \in C_{n,i}$
iff $r \in C_{n,i}$, so $\mathbf{1}_{C_{n,i}}(q) = \mathbf{1}_{C_{n,i}}(r)$. It is therefore locally constant, by the
converse of Proposition~\ref{prop:factor}, and the pairing is defined at it and may be computed at
stage $n$. Its representative there is
$\widehat{(\mathbf{1}_{C_{n,i}})}_n(j) = \mathbf{1}_{C_{n,i}}(q^n_j) = [\pi_n(q^n_j) = i] = [j = i]$, so
\[
\langle a, \mathbf{1}_{C_{n,i}}\rangle = \sum_{j \in [\sz(n)]} a_n(j)\,[j=i] = a_n(i) .
\]
Now let $a, b \in \mathcal{P}(T)$ pair equally with every locally constant function. Taking the
function $\mathbf{1}_{C_{n,i}}$ gives $a_n(i) = \langle a, \mathbf{1}_{C_{n,i}}\rangle = \langle b, \mathbf{1}_{C_{n,i}}\rangle =
b_n(i)$, for every stage $n$ and every site $i$; so $a_n = b_n$ for every $n$, and $a = b$.
\end{proof}

\begin{proposition}\label{prop:dirac-formal}
For $q \in \Pi_T$, $\delta^q_n(i) := [i = \pi_n(q)]$ is a compatible family of total mass one, and
$\langle \delta^q, f \rangle = f(q)$ for every locally constant $f$.
\end{proposition}

\begin{proof}
For compatibility, fix $n$ and $i$. The sum
$\sum_{j\in(\rf_n)^{-1}(i)}[j=\pi_{n+1}(q)]$ counts the sites of the block of $i$ equal to
$\pi_{n+1}(q)$, so it is $1$ if $\pi_{n+1}(q)$ lies in that block and $0$ otherwise. Since
$\rf_n(\pi_{n+1}(q)) = \pi_n(q)$ by the compatibility of $q$, the site $\pi_{n+1}(q)$ lies in the
block of $i$ exactly when $\pi_n(q) = i$; so the sum is $[i = \pi_n(q)] = \delta^q_n(i)$, which is
$((\rf_n)_*\delta^q_{n+1})(i) = \delta^q_n(i)$. The total mass at stage $n$ is
$\sum_i [i = \pi_n(q)] = 1$, exactly one site being equal to $\pi_n(q)$. For the pairing, let $f$ be
locally constant and, by Proposition~\ref{prop:factor}, constant at stage $n$. Then
\[
\langle\delta^q,f\rangle = \sum_i [i = \pi_n(q)]\,\widehat f_n(i) = \widehat f_n(\pi_n(q)) = f(q),
\]
the last equality by Theorem~\ref{thm:pairing-welldef}.
\end{proof}

\begin{definition}\label{def:obssheaf}
For $U \subseteq \Pi_T$ open let $\mathcal{O}_T(U)$ be the locally constant functions $U \to \Rr$,
with restriction of functions. Write $\Ob$ for the locally constant functions $\Pi_T \to \Rr$, the
\emph{objects} of $T$.
\end{definition}

\begin{proposition}\label{prop:obssheaf}
$\mathcal{O}_T$ is a sheaf of sets on $\Pi_T$: restriction to a smaller open preserves local
constancy, sections over an open cover that agree on each member are equal, and sections over the
members of an open cover that agree on every overlap glue to a section over the union.
\end{proposition}

\begin{proof}
Each $U$ carries the subspace topology. \emph{Restriction}: if $f$ is locally constant on $U$ and
$U' \subseteq U$ is open, a point of $U'$ has an open $V \subseteq U$ on which $f$ is constant, and $f|_{U'}$ is constant on the open
$V \cap U'$; restriction is functorial because restricting twice is restricting once.
\emph{Locality}: if $U = \bigcup_k U_k$ and $f, f' \in \mathcal{O}_T(U)$ agree on each $U_k$, they
agree at every point of $U$, so $f = f'$. \emph{Gluing}: if $f_k \in \mathcal{O}_T(U_k)$ agree on
every overlap $U_k \cap U_l$, then $f(x) := f_k(x)$ for any $k$ with $x \in U_k$ is a well defined
function on $U = \bigcup_k U_k$ restricting to each $f_k$; a point $x \in U_k$ has an open
$V \subseteq U_k$ on which $f_k$, hence $f$, is constant, so $f$ is locally constant.
\end{proof}

\begin{theorem}\label{thm:sheaf-functional}
For $a \in \mathcal{P}(T)$, $f, f' \in \Ob$ and $c \in \Rr$,
$\langle a, f + f' \rangle = \langle a, f\rangle + \langle a, f'\rangle$ and
$\langle a, c f \rangle = c\langle a, f\rangle$; and $\langle \delta^q, f\rangle = f(q)$.
\end{theorem}

\begin{proof}
By Proposition~\ref{prop:factor} $f$ is constant at some stage $n_1$ and $f'$ at some stage $n_2$,
and by Lemma~\ref{lem:conststage} both are constant at $n := \max(n_1,n_2)$. Then $f+f'$ and $cf$ are
constant at stage $n$: if $\pi_n(q) = \pi_n(r)$ then $f(q) = f(r)$ and $f'(q) = f'(r)$, so
$(f+f')(q) = (f+f')(r)$ and $(cf)(q) = (cf)(r)$. They are therefore locally constant, the pairing is
defined at them, and all four pairings may be computed at stage $n$
(Theorem~\ref{thm:pairing-welldef}). Using the same points $q^n_i$ for all of them,
\[
\widehat{(f+f')}_n(i) = f(q^n_i) + f'(q^n_i) = \widehat f_n(i) + \widehat{f'}_n(i),
\qquad
\widehat{(cf)}_n(i) = c\,f(q^n_i) = c\,\widehat f_n(i).
\]
The map $u \mapsto \int_n u\,da = \sum_i a_n(i)\,u(i)$ is linear in $u$, so
$\langle a, f+f'\rangle = \int_n(\widehat f_n + \widehat{f'}_n)\,da = \langle a,f\rangle + \langle
a,f'\rangle$ and $\langle a, cf\rangle = c\,\langle a, f\rangle$. The last clause is
Proposition~\ref{prop:dirac-formal}.
\end{proof}

\begin{corollary}\label{cor:clopenmeasure}
For a clopen $U \subseteq \Pi_T$ the indicator $\mathbf{1}_U$ is an object, so
$\mu_a(U) := \langle a, \mathbf{1}_U\rangle$ is defined; $\mu_a(U \cup U') = \mu_a(U) + \mu_a(U')$ for
disjoint clopen $U, U'$; and $\mu_a(C_{n,i}) = a_n(i)$.
\end{corollary}

\begin{proof}
$\mathbf{1}_U$ is constant on the open set $U$ and on the open set $\Pi_T \setminus U$, so it is
locally constant. For disjoint $U, U'$, $\mathbf{1}_{U \cup U'} = \mathbf{1}_U + \mathbf{1}_{U'}$, so
the additivity is Theorem~\ref{thm:sheaf-functional}; and $C_{n,i}$ is clopen, so the
last clause is Theorem~\ref{thm:pairing-injective}.
\end{proof}

\begin{proposition}\label{prop:sheaf-nondegenerate}
Every stalk of $\mathcal{O}_T$ is a copy of $\Rr$, and the objects separate the points of $\Pi_T$.
\end{proposition}

\begin{proof}
The stalk at $q$ is the set of germs at $q$: pairs $(U,f)$ with $U \ni q$ open and
$f \in \mathcal{O}_T(U)$, two pairs being identified when the functions agree on some open
neighbourhood of $q$ inside both domains. The value $f(q)$ depends only on the germ, since
identified pairs agree near $q$ and in particular at $q$; so $(U,f) \mapsto f(q)$ is a well defined
map from the stalk to $\Rr$. It is onto: for $x \in \Rr$ the constant function $x$ on $\Pi_T$ is
locally constant and has value $x$ at $q$. It is injective: if $f(q) = f'(q)$ for sections on $U$ and
$U'$, local constancy gives open neighbourhoods $V \subseteq U$ and $V' \subseteq U'$ of $q$ on which
$f$ and $f'$ are constant, so on $V \cap V'$ both equal $f(q) = f'(q)$, and the two germs coincide. For
separation, let $q \neq r$; the sequences differ at some coordinate $n$, so $\pi_n(q) \neq \pi_n(r)$.
The indicator $\mathbf{1}_{C_{n,\pi_n(q)}}$ is locally constant (as in the proof of
Theorem~\ref{thm:pairing-injective}), so it is an object, and it takes the value $1$ at $q$ and $0$
at $r$.
\end{proof}

The objects were defined on the topology of the boundary, and the boundary was produced as a
condensed set. The two meet at the level of global sections. Write $\Rr_{\mathrm{disc}}$ for the
real numbers with the discrete topology. The condensed set $\underline{\Rr_{\mathrm{disc}}}$ assigns
to a profinite set $S$ the set of continuous maps $S \to \Rr_{\mathrm{disc}}$, which are the locally
constant maps $S\to\Rr$, and to a continuous $\psi : S' \to S$ the map $h \mapsto h \circ \psi$
\cite{Scholze2026}. It is not the condensed set $\underline{\Rr}$ of the real line with its usual
topology: on the profinite set $\Nn\cup\{\infty\}$, the one-point compactification of the discrete
$\Nn$, the map $n\mapsto1/(n+1)$, $\infty\mapsto0$, is continuous into $\Rr$ and constant on no
neighbourhood of $\infty$, so it lies in $\underline{\Rr}(\Nn\cup\{\infty\})$ and not in
$\underline{\Rr_{\mathrm{disc}}}(\Nn\cup\{\infty\})$.

\begin{theorem}\label{thm:bridge-formal}
There is a bijection
$\mathcal{O}_T(\Pi_T) \cong \mathrm{Hom}_{\mathrm{Cond}}(\underline{\Pi_T},
\underline{\Rr_{\mathrm{disc}}})$: the objects of $T$ are exactly the condensed morphisms from
the condensed boundary into the discrete line. In particular these condensed morphisms separate the
points of $\Pi_T$: for $q\ne r$ there is $\xi$ with
$\xi_{\Pi_T}(\mathrm{id})(q)\ne\xi_{\Pi_T}(\mathrm{id})(r)$.
\end{theorem}

\begin{proof}
The boundary $\Pi_T$ is a profinite set in the site by Theorem~\ref{thm:screen-profinite}, and $\underline{\Pi_T}$ is the
presheaf it represents. Condensed sets form a full subcategory of presheaves on profinite sets, so
the Yoneda lemma gives a bijection
\[
\mathrm{Hom}_{\mathrm{Cond}}(\underline{\Pi_T}, \underline{\Rr_{\mathrm{disc}}}) \longrightarrow
\underline{\Rr_{\mathrm{disc}}}(\Pi_T), \qquad \xi \longmapsto \xi_{\Pi_T}(\mathrm{id}),
\]
whose inverse sends $h$ to the morphism $\xi^h$ with components $\xi^h_S(\psi) = h \circ \psi$.
By the description of $\underline{\Rr_{\mathrm{disc}}}$, the target is the set of locally constant
maps $\Pi_T \to \Rr$, which is $\Ob=\mathcal{O}_T(\Pi_T)$. Composing, an object $f$ corresponds to
the condensed morphism $\xi^f$, and a condensed morphism $\xi$ to the object
$\xi_{\Pi_T}(\mathrm{id})$. For $q\ne r$ in $\Pi_T$, Proposition~\ref{prop:sheaf-nondegenerate}
gives an object $f$ with $f(q)\ne f(r)$, and the morphism $\xi^f$ corresponds to it.
\end{proof}

So the reading an observer performs on the boundary is a morphism of condensed sets
(Theorem~\ref{thm:bridge-formal}). The arrow between readings runs opposite to the
arrow between weights: a refinement pushes a profile forward from the fine list to the coarse one,
and pulls a reading back from the coarse list to the fine one.

\begin{definition}\label{def:stagepullback}
For $g : [N] \to [m]$ the \emph{restriction} of $f : [m] \to \Rr$ is $g^{*}f := f \circ g$. For a
stage $n$ and $u : [\sz(n)] \to \Rr$ the \emph{leg} of $u$ is $\pi_n^{*}u := u \circ \pi_n$.
\end{definition}

\begin{theorem}\label{thm:adjoint-formal}
For every $g : [N] \to [m]$, $\alpha : [N] \to \Rr$ and $f : [m] \to \Rr$,
$\sum_{i \in [N]} \alpha(i)\, (g^{*}f)(i) = \sum_{j \in [m]} (g_{*}\alpha)(j)\, f(j)$: the pushforward
of weights is the adjoint of restriction of readings.
\end{theorem}

\begin{proof}
Every $i \in [N]$ lies in exactly one block $g^{-1}(j)$, namely that of $j = g(i)$, so by
Proposition~\ref{prop:push-basic} the left sum may be computed block by block:
\[
\sum_{i} \alpha(i)\, f(g(i))
= \sum_{j} \sum_{i \in g^{-1}(j)} \alpha(i)\, f(g(i))
= \sum_{j} f(j) \sum_{i \in g^{-1}(j)} \alpha(i)
= \sum_{j} (g_{*}\alpha)(j)\, f(j),
\]
the second equality because $g(i) = j$ throughout the inner sum, and the third by
Definition~\ref{def:push}.
\end{proof}

\begin{proposition}\label{prop:stagepullback}
Restriction is functorial, $\mathrm{id}^{*} = \mathrm{id}$ and $(g_2 g_1)^{*} = g_1^{*} g_2^{*}$. For $g$
onto it is injective, and $f \in \mathrm{im}\, g^{*}$ iff $f$ is constant on the blocks of $g$. If
$g(j_1) = g(j_2)$ for some $j_1 \neq j_2$ then $g^{*}$ is not surjective.
\end{proposition}

\begin{proof}
\emph{Functoriality.} $\mathrm{id}^*f = f \circ \mathrm{id} = f$, and for composable
$g_1 : [N] \to [m]$, $g_2 : [m] \to [k]$ and $f : [k] \to \Rr$,
$(g_2 g_1)^* f = f \circ (g_2 \circ g_1) = (f \circ g_2) \circ g_1 = g_1^*(g_2^* f)$ by associativity
of composition.
\emph{Injectivity.} Let $g$ be onto and $f\circ g = f'\circ g$. Each $i \in [m]$ equals $g(j)$ for
some $j$, so $f(i) = f(g(j)) = f'(g(j)) = f'(i)$, and $f = f'$.
\emph{Image.} A restriction $g^*u$ is constant on blocks, since $g(j_1) = g(j_2)$ gives
$u(g(j_1)) = u(g(j_2))$. Conversely, let $f$ be constant on the blocks of $g$ and define
$u(i) := f(j)$ for any $j \in g^{-1}(i)$; the block is nonempty because $g$ is onto, and the value
does not depend on $j$ because $f$ is constant on the block. Then $(g^*u)(j) = u(g(j)) = f(j)$, as
$j$ lies in the block of $g(j)$; so $f = g^*u$.
\emph{Failure of surjectivity.} If $j_1 \neq j_2$ and $g(j_1) = g(j_2)$, the function
$f(j) := [j = j_1]$ takes the value $1$ at $j_1$ and $0$ at $j_2$, so it is not constant on the
block containing both, and by the description of the image it is not a restriction.
\end{proof}

\begin{theorem}\label{thm:legs}
$\pi_{n+1}^{*}(\rf_n)^{*}u = \pi_n^{*}u$; each $\pi_n^{*}u$ is locally constant; and every
object arises from a single finite stage and from exactly one reading there:
\[
\forall f \in \Ob,\quad \exists\, n,\ \exists!\, u : [\sz(n)] \to \Rr, \quad \pi_n^{*}u = f .
\]
\end{theorem}

\begin{proof}
\emph{Compatibility.} For $q \in \Pi_T$,
$(\pi_{n+1}^{*}(\rf_n)^*u)(q) = u(\rf_n(q_{n+1})) = u(q_n) = (\pi_n^{*}u)(q)$, by the compatibility
of $q$.
\emph{Local constancy.} $\pi_n^{*}u$ is constant at stage $n$, since $\pi_n(q) = \pi_n(r)$ gives
$u(\pi_n(q)) = u(\pi_n(r))$; so it is constant on each cylinder of stage $n$, an open set by
Proposition~\ref{prop:cyl}, and is locally constant.
\emph{Existence.} Let $f \in \Ob$. Proposition~\ref{prop:factor} gives a stage $n$ at which $f$ is
constant, and $u := \widehat f_n$ satisfies $(\pi_n^{*}u)(q) = \widehat f_n(\pi_n(q)) = f(q)$ for
every $q$, by Theorem~\ref{thm:pairing-welldef}.
\emph{Uniqueness.} If $u\circ\pi_n=v\circ\pi_n$, then for each $i \in [\sz(n)]$
Theorem~\ref{thm:hpath} supplies $q$ with $\pi_n(q)=i$, and $u(i) = u(\pi_n(q)) = v(\pi_n(q)) =
v(i)$; so $u = v$. The argument holds at every stage $n$, so $\pi_n^{*}$ is injective for every $n$.
\end{proof}

\begin{proposition}\label{prop:obsalg}
If $f$ and $f'$ are locally constant so is $x \mapsto \mathrm{op}(f(x), f'(x))$ for any binary operation;
hence $\Ob$ is an $\Rr$-subalgebra of the functions on $\Pi_T$, each map $\pi_n^{*}$ is a map of
$\Rr$-algebras, and $\Ob$ is not the zero ring.
\end{proposition}

\begin{proof}
\emph{Stability.} Given $x$, take open neighbourhoods $V$ and $V'$ of $x$ on which $f$ and $f'$ are
constant. On the open neighbourhood $V \cap V'$ of $x$ both are constant, so the function
$y \mapsto \mathrm{op}(f(y),f'(y))$ is constant there; hence it is locally constant.
\emph{Subalgebra.} The constant functions are locally constant, and applying stability to the
operations $(s,t) \mapsto s + t$ and $(s,t) \mapsto st$ shows that $\Ob$ is closed under pointwise
sums and products, hence under scalar multiples, a scalar multiple being a product with a constant.
So $\Ob$ contains $0$ and $1$ and is an $\Rr$-subalgebra of all functions $\Pi_T \to \Rr$.
\emph{Legs.} $\pi_n^{*}$ takes values in $\Ob$ by Theorem~\ref{thm:legs}, and for readings $u, v$ and
$c \in \Rr$, $(\pi_n^{*}(u+v))(q) = u(q_n) + v(q_n)$, $(\pi_n^{*}(uv))(q) = u(q_n)v(q_n)$,
$(\pi_n^{*}(cu))(q) = c\,u(q_n)$ and $(\pi_n^{*}1)(q) = 1$; so $\pi_n^{*}$ respects every operation.
\emph{Nonzero.} By Theorem~\ref{thm:hpath} $\Pi_T$ has a point $q$, at which the constant functions
$0$ and $1$ take different values; so $0 \neq 1$ in $\Ob$.
\end{proof}

The algebra of the boundary is the direct limit of the stage algebras along restriction, which is
the algebra Stone duality attaches to a profinite set \cite{Gregoric2024}.

\begin{proposition}\label{prop:dirlim}
For $n \leq K$ and $u : [\sz(n)] \to \Rr$, $\pi_K^{*}(\delta^K_n)^*u = \pi_n^{*}u$. For every
$\Rr$-algebra $B$ and $\Rr$-algebra maps $\varphi_n : \Rr^{[\sz(n)]} \to B$ with
$\varphi_{n+1} \circ (\rf_n)^* = \varphi_n$ for every $n$, there is exactly one $\Rr$-algebra map
$\Phi : \Ob \to B$ with $\Phi \circ \pi_n^{*} = \varphi_n$ for every $n$: the objects, with the legs,
are the direct limit of the stage algebras.
\end{proposition}

\begin{proof}
\emph{Compatibility.} For $q \in \Pi_T$,
$(\pi_K^{*}(\delta^K_n)^*u)(q) = u(\delta^K_n(q_K)) = u(q_n)$ by Lemma~\ref{lem:drop}.
\emph{Existence.} Since $(\delta^{K+1}_n)^* = (\rf_K)^* \circ (\delta^K_n)^*$, induction on $K$ gives
$\varphi_K \circ (\delta^K_n)^* = \varphi_n$ for $K \geq n$. Every object is $\pi_n^{*}u$ for some
$n$ and $u$ (Theorem~\ref{thm:legs}); put $\Phi(\pi_n^{*}u) := \varphi_n(u)$. This does not depend
on the choice: if $\pi_n^{*}u = \pi_{n'}^{*}v$, then at $K := \max(n,n')$ the readings
$(\delta^K_n)^*u$ and $(\delta^K_{n'})^*v$ have the same leg by the compatibility, so they are equal by
the injectivity of $\pi_K^{*}$ (Theorem~\ref{thm:legs}), and
$\varphi_n(u) = \varphi_K((\delta^K_n)^*u) = \varphi_K((\delta^K_{n'})^*v) = \varphi_{n'}(v)$. Any two
objects are legs at one common stage $K$, and $\pi_K^{*}$ and $\varphi_K$ are $\Rr$-algebra maps
(Proposition~\ref{prop:obsalg}), so $\Phi$ respects sums, products, scalars and $1$.
\emph{Uniqueness.} Every object is a leg, and $\Phi \circ \pi_n^{*} = \varphi_n$ fixes $\Phi$ on it.
\end{proof}

These constructions start from the finite stages and groupings of Definition~\ref{def:tower} alone, and
the boundary they produce is a profinite set on which every path is constant
(Theorem~\ref{thm:screen-profinite}, Proposition~\ref{prop:boundary-trivial}). The entropy of the
screen is the number $N$ of its qubits, a Planck area entering only when
the screen is given a geometry \cite{FieldsGlazebrookMarciano2022}. We take the qubits of a physical
screen as its finest stage, each implementing one measurable spin \cite{FieldsGlazebrookMarciano2022},
and read the boundary of a tower as the idealisation of a refinement that never stops. No metric is among
the data of a tower (Definition~\ref{def:tower}), and the dimension that a probe reads off a stage is
fixed by the profile alone (Corollary~\ref{cor:split}). This
agrees with \cite{Oggad2026}, where the two that a diffusion probe returns on a dust is a reading of
the probe itself.

\section{Objects and the distance between them}\label{sec:objects}

The objects an observer identifies on the boundary are sections of a sheaf over the boundary itself;
no spacetime is assumed as a base. They are also outcome assignments, in the sense in which
contextuality speaks of them. In the sheaf-theoretic account of contextuality a finite set of
measurements carries the event sheaf, whose sections over a set of measurements are the joint
assignments of outcomes to them \cite{AbramskyBrandenburger2011}. At a stage of the tower the
measurements are the sites, an assignment over the whole stage is a reading at that stage, and the
question is how such assignments relate to the objects of $T$. The answer is that they are
the objects, each read at the first stage fine enough to resolve it, and that refining the stage
leaves them in place.

\begin{definition}\label{def:persist}
For a finite set $X$ of measurements with outcomes in a set $O$, the event sheaf sends
$U \subseteq X$ to $\Ee(U) := O^U$, the set of functions $U\to O$, with restriction of functions.
For $X = [\sz(n)]$ and $O=\Rr$ a global section
$s\in\Ee(X)$ defines the object $\pi_n^{*}s = s \circ \pi_n$, its leg
(Definition~\ref{def:stagepullback}). An object $f$ \emph{persists from stage $n$} when
$f=\pi_n^{*}s$ for some global section $s$ over stage $n$.
\end{definition}

\begin{theorem}\label{thm:objects-formal}
For every $f : \Pi_T \to \Rr$, $f \in \Ob$ iff $f$ persists from some stage, that is, iff there are
$n$ and a global section $s$ over stage $n$ with $\pi_n^{*}s = f$; and each map
$s\mapsto\pi_n^{*}s$ on global sections over stage $n$ is injective.
\end{theorem}

\begin{proof}
A global section over stage $n$ is a function $s : [\sz(n)] \to \Rr$.

\emph{Only if.} Let $f \in \Ob$. Theorem~\ref{thm:legs} gives a stage $n$ and a reading
$u : [\sz(n)] \to \Rr$ with $\pi_n^{*}u = f$; then $s := u$ is a global section over stage $n$ with
$\pi_n^{*}s = f$.

\emph{If.} If $f = \pi_n^{*}s$, then $f$ is locally constant by the second
clause of Theorem~\ref{thm:legs}, so $f \in \Ob$.

\emph{Injectivity.} Fix $n$ and let $\pi_n^{*}s = \pi_n^{*}s'$. For each site
$i \in [\sz(n)]$, Theorem~\ref{thm:hpath} supplies $q \in \Pi_T$ with $\pi_n(q) = i$, and
$s(i) = s(\pi_n(q)) = (\pi_n^{*}s)(q) = (\pi_n^{*}s')(q) = s'(i)$. Hence $s = s'$.
\end{proof}

\begin{theorem}\label{thm:persist-refine}
An object persisting from stage $n$ persists from stage $n+1$:
$\pi_{n+1}^{*}(\rf_n)^{*}s = \pi_n^{*}s$ for every $n$ and every global section $s$ over stage
$n$.
\end{theorem}

\begin{proof}
For $q \in \Pi_T$,
\[
(\pi_{n+1}^{*}(\rf_n)^{*}s)(q) = s\bigl(\rf_n(q_{n+1})\bigr) = s(q_n) = (\pi_n^{*}s)(q),
\]
by the compatibility $\rf_n(q_{n+1}) = q_n$ of the point $q$.
\end{proof}

By induction an object persisting from stage $n$ persists from every later stage, and
\emph{persistence}, in the sense used throughout this paper, is this invariance of an object under
refinement of the stage at which it is read, and nothing more. At stage $n$ the $2^{\sz(n)}$ joint eigenvectors of the sign observables are
indexed by the sign vectors $s:[\sz(n)]\to\{\pm1\}$ (Proposition~\ref{prop:cube}), each giving an
object $\pi_n^{*}s$ that persists from stage $n$, and $s\mapsto(\rf_n)^{*}s$ carries the sign
vectors of stage $n$ injectively into those of stage $n+1$, and not onto when $\sz(n)<\sz(n+1)$
(Proposition~\ref{prop:refstage} with $O=\{\pm1\}$): the sign vectors run from coarse to fine while
the refinements run onto from fine to coarse. An assignment at stage $n$ and its
restriction along $\delta^K_n$ to any finer stage $K$ give the same object
(Proposition~\ref{prop:dirlim}), and distinct points of the boundary are told apart by objects
(Proposition~\ref{prop:sheaf-nondegenerate}). The distance between two objects is the mass, under a compatible family of profiles, of
the amount by which they disagree. Its arguments are objects, which are functions on
the boundary (Definitions~\ref{def:obssheaf} and~\ref{def:obsdist}), so it measures how two readings differ and assigns no distance to points.
It adds up, over the sites of any stage at which both objects are constant, the weight of each site
times the size of their disagreement there (Theorem~\ref{thm:pairing-welldef}).

\begin{definition}\label{def:obsdist}
For $a \in \mathcal{P}(T)$ and $f, f' \in \Ob$ put $d_a(f,f') := \langle a, |f - f'| \rangle$; call $f$
and $f'$ \emph{indistinguishable} when $d_a(f,f') = 0$.
\end{definition}

The function $|f-f'|$ is locally constant by Proposition~\ref{prop:obsalg}, applied to the operation
$(s,t) \mapsto |s-t|$, so the pairing is defined at it. Write $a \geq 0$ when $a_n(i) \geq 0$ for
every stage $n$ and site $i$, and $a > 0$ when every $a_n(i) > 0$.

\begin{lemma}\label{lem:pairing-mono}
$\langle a, 0\rangle = 0$; and if $a \geq 0$, $u, v \in \Ob$ and $u \leq v$ pointwise, then
$\langle a, u\rangle \leq \langle a, v\rangle$.
\end{lemma}

\begin{proof}
The zero function is constant at stage $0$ with representative $0$. For monotonicity, choose by
Lemma~\ref{lem:conststage} a stage $n$ at which both $u$ and $v$ are constant; with the same points
$q^n_i$, $\widehat u_n(i) = u(q^n_i) \leq v(q^n_i) = \widehat v_n(i)$, and multiplying by
$a_n(i) \geq 0$ and summing over $i$ gives $\int_n \widehat u_n\,da \leq \int_n \widehat v_n\,da$.
\end{proof}

\begin{theorem}\label{thm:metric-formal}
If $a \geq 0$ then $d_a$ is a pseudometric on $\Ob$.
\end{theorem}

\begin{proof}
A pseudometric is a function $d : \Ob \times \Ob \to \Rr$ with $d(f,f) = 0$, $d(f,f') = d(f',f)$,
$d(f,f') \geq 0$ and $d(f,f'') \leq d(f,f') + d(f',f'')$. Every $|f-f'|$ is in $\Ob$, as
noted after Definition~\ref{def:obsdist}, so $d_a$ is defined on all pairs.
\emph{Zero on the diagonal.} $|f-f|$ is the zero function, and $\langle a, 0\rangle = 0$.
\emph{Symmetry.} $|f-f'|$ and $|f'-f|$ are the same function, so they have the same pairing.
\emph{Nonnegativity.} $0 \leq |f-f'|$ pointwise, so monotonicity gives
$0 = \langle a, 0\rangle \leq \langle a, |f-f'|\rangle = d_a(f,f')$.
\emph{Triangle inequality.} For every point $x$, $|f(x)-f''(x)| \leq |f(x)-f'(x)| + |f'(x)-f''(x)|$, the
triangle inequality of the real line. Both sides are objects, the right-hand side by
Proposition~\ref{prop:obsalg}, so monotonicity and the additivity of
Theorem~\ref{thm:sheaf-functional} give
$d_a(f,f'') \leq \langle a, |f-f'| + |f'-f''|\rangle = d_a(f,f') + d_a(f',f'')$.
\end{proof}

\begin{theorem}\label{thm:metric-degenerate}
$d_{\delta^q}(f,f') = |f(q) - f'(q)|$, so distinct objects agreeing at $q$ stand at distance zero.
\end{theorem}

\begin{proof}
By Proposition~\ref{prop:dirac-formal}, $\delta^q$ is a compatible family, so $\delta^q \in
\mathcal{P}(T)$ and $d_{\delta^q}$ is defined; its values are $0$ or $1$, so $\delta^q \geq 0$ and
Theorem~\ref{thm:metric-formal} applies to it. The function $|f-f'|$ is locally constant
(Proposition~\ref{prop:obsalg}), so the evaluation clause of Proposition~\ref{prop:dirac-formal}
applies to it and gives
$d_{\delta^q}(f,f') = \langle \delta^q, |f-f'|\rangle = |f-f'|(q) = |f(q)-f'(q)|$. If $f(q) = f'(q)$ this
is $0$, whether or not $f$ and $f'$ agree elsewhere.
\end{proof}

\begin{theorem}\label{thm:metric-separates}
If $a > 0$ pointwise then $d_a(f,f') = 0$ iff $f = f'$. For any $a \geq 0$ indistinguishability is an
equivalence relation, and the relations between objects present the groupoid of that relation.
\end{theorem}

\begin{proof}
\emph{Separation.} Let $a > 0$. If $f = f'$ then $d_a(f,f') = 0$ by Theorem~\ref{thm:metric-formal},
since $a > 0$ implies $a \geq 0$. Conversely let $d_a(f,f')=0$, and write $w := |f-f'| \in \Ob$. By
Proposition~\ref{prop:factor} $w$ is constant at some stage $n$, and
\[
0 = d_a(f,f') = \langle a, w\rangle = \sum_i a_n(i)\, \widehat w_n(i) .
\]
Each term is a product of $a_n(i) > 0$ and $\widehat w_n(i) = w(q^n_i) \geq 0$, so each term is
nonnegative; a sum of nonnegative terms vanishes only when every term does, so
$a_n(i)\,\widehat w_n(i) = 0$, and since $a_n(i) > 0$, $\widehat w_n(i) = 0$ for every $i$. For any
$q \in \Pi_T$, Theorem~\ref{thm:pairing-welldef} gives $|f(q)-f'(q)| = w(q) = \widehat w_n(\pi_n(q)) =
0$, so $f(q) = f'(q)$. Hence $f = f'$.

\emph{Equivalence.} Let $a \geq 0$ and write $f \sim f'$ for $d_a(f,f') = 0$. Reflexivity is
$d_a(f,f) = 0$ and symmetry is $d_a(f,f') = d_a(f',f)$, both from Theorem~\ref{thm:metric-formal}. For
transitivity, if $f \sim f'$ and $f' \sim f''$, the triangle inequality and nonnegativity give
$0 \leq d_a(f,f'') \leq d_a(f,f') + d_a(f',f'') = 0$, so $f \sim f''$.

\emph{Groupoid.} The groupoid of $\sim$ has the objects of $T$ as its objects and exactly one
arrow $f \to f'$ when $f \sim f'$, none otherwise. Reflexivity supplies the identities, transitivity
the composite of $f \to f'$ and $f' \to f''$, and symmetry an arrow $f' \to f$ inverse to $f \to f'$. The
associativity, unit and inverse laws hold because between two objects there is at most one arrow,
so any two parallel composites coincide.
\end{proof}

\begin{lemma}\label{lem:positive-family}
Every tower carries a strictly positive compatible family of profiles of total mass one, namely
$s_0(i):=1/\sz(0)$ and $s_{n+1}(j):=s_n(\rf_n(j))/|\rf_n^{-1}(\rf_n(j))|$. Consequently two objects
$f,f'$ differ if and only if $d_a(f,f')\neq0$ for every compatible family $a>0$.
\end{lemma}

\begin{proof}
Every block $\rf_n^{-1}(i)$ is nonempty, $\rf_n$ being onto, so $s_{n+1}$ is defined, and by
induction on $n$ it is strictly positive, $s_0$ being so as $\sz(0)>0$. For a site $i$ of stage $n$,
every $j\in\rf_n^{-1}(i)$ has $s_{n+1}(j)=s_n(i)/|\rf_n^{-1}(i)|$, so
$((\rf_n)_*s_{n+1})(i)=s_n(i)$ and $s$ is compatible; its total at stage $0$ is
$\sz(0)\cdot1/\sz(0)=1$, and the total is the same at every stage (Proposition~\ref{prop:family}).
If $f\neq f'$, Theorem~\ref{thm:metric-separates} gives $d_a(f,f')\neq0$ for every compatible
$a>0$. If $f=f'$, then $d_s(f,f')=0$ by Theorem~\ref{thm:metric-formal} while $s>0$, so not every
compatible $a>0$ puts them at nonzero distance.
\end{proof}

Both cases occur on one tower. The \emph{binary tower} has $\sz(n)=2^n$ and $\rf_n(j)=j\bmod2^n$;
these data form a tower, since $2^n>0$ and $j\mapsto j\bmod2^n$ maps $[2^{n+1}]$ onto $[2^n]$, each
$j<2^n$ being its own image.

\begin{proposition}\label{prop:metric-witness}
At the binary tower, $b_n(i) := 2^{-n}$ is a compatible family with $b > 0$, and there are a point
$q$ and two distinct objects $f, f'$ with $d_{\delta^q}(f,f') = 0$ and $d_b(f,f') \neq 0$.
\end{proposition}

\begin{proof}
Every block of $j \mapsto j \bmod 2^n$ is $\{i, i + 2^n\}$, so
$((\rf_n)_*b_{n+1})(i) = 2 \cdot 2^{-(n+1)} = b_n(i)$, and $b > 0$. Take $q := P(1,0)$, the point
through the site $0$ of stage one given by Theorem~\ref{thm:hpath}, and the objects $f := 0$ and
$f' := \mathbf{1}_{C_{1,1}}$. The point $P(1,1)$ lies in $C_{1,1}$, so $f'$ takes the value $1$ there and
$f' \neq f$; and $q \notin C_{1,1}$, so $f'(q) = 0 = f(q)$ and $d_{\delta^q}(f,f') = 0$ by
Theorem~\ref{thm:metric-degenerate}. Since $b > 0$ and $f \neq f'$, $d_b(f,f') \neq 0$ by
Theorem~\ref{thm:metric-separates}.
\end{proof}

A profile concentrated at a point cannot tell apart two objects agreeing there, and against a
strictly positive one the objects form a metric space. By Theorems~\ref{thm:metric-formal}
and~\ref{thm:metric-separates}, $d_a$ for $a > 0$ satisfies the axioms of a pseudometric together
with $d_a(f,f') = 0 \Rightarrow f = f'$, which are the axioms of a metric; the family $b$ of
Proposition~\ref{prop:metric-witness} realises this at the binary tower, while the family
$\delta^{P(1,0)}$ on the same tower identifies the two distinct objects $0$ and $\mathbf{1}_{C_{1,1}}$:
they differ, since $P(1,1)\in C_{1,1}$ by Theorem~\ref{thm:hpath}, and $|0-\mathbf{1}_{C_{1,1}}|$ vanishes at
$P(1,0)$, whose stage-one site is $0$ (Theorem~\ref{thm:hpath}), so the pairing of
Proposition~\ref{prop:dirac-formal} gives $d_{\delta^{P(1,0)}}(0,\mathbf{1}_{C_{1,1}})=0$. The two
families are compatible and both have total mass one, $2^n\cdot2^{-n}=1$ at every stage for $b$; and
$d_a$ is defined from $a$ and the objects alone (Definition~\ref{def:obsdist}).

\section{Contextuality}\label{sec:strata}

The objects of Theorem~\ref{thm:objects-formal} are outcome assignments, and they form a sheaf:
assignments that agree on overlaps always glue. In the sheaf-theoretic account contextuality is a
property not of assignments but of families of distributions over them: a family of distributions, one per
context and compatible on overlaps, is contextual when no single distribution on global assignments
reproduces it, and Abramsky and Brandenburger distinguish three strengths of this failure,
probabilistic, possibilistic and strong \cite{AbramskyBrandenburger2011}; the strongest is the
absence of any global assignment compatible with the supports. A finite set $X$ of measurements
with outcomes in a finite set $O$ is covered by contexts $\Mm$, an antichain of subsets, and carries
the event sheaf $\Ee(U)=O^U$ of Definition~\ref{def:persist}. An empirical model $\mathsf e$
assigns to each $C\in\Mm$ a distribution $\mathsf e_C$ on $\Ee(C)$, compatible on overlaps. It is
\emph{probabilistically extendable} when a distribution $d$ on $\Ee(X)$ has $d|_C=\mathsf e_C$ for every
$C$, and \emph{probabilistically contextual} otherwise; \emph{possibilistically extendable} when a nonempty $S\subseteq\Ee(X)$ has
$S|_C=\operatorname{supp}(\mathsf e_C)$ for every $C$, and \emph{possibilistically contextual} otherwise; and \emph{strongly contextual} when
$\Se:=\{s\in\Ee(X):s|_C\in\operatorname{supp}(\mathsf e_C)\ \forall C\}$ is empty. Probabilistic extendability is the existence of a nonnegative solution of a linear system, with rows indexed by pairs $(C,s)$, $s\in\Ee(C)$, columns by global
assignments $t$, incidence matrix $\mathbf{M}_{(C,s),t}=[t|_C=s]$ and right side $v_{(C,s)}=\mathsf e_C(s)$.

\begin{lemma}\label{lem:linsys}
For $x$ indexed by global assignments, $(\mathbf{M}x)_{(C,s)}$ is the value at $s$ of the marginal
$x|_C$, so a distribution $d$ extends $\mathsf e$ iff $x=d$ solves $\mathbf{M}x=v$ with $x\ge0$; and if
$\Mm\neq\varnothing$, every nonnegative solution of $\mathbf{M}x=v$ is a probability distribution.
If a row vector $y$ has $y^{\mathsf T}\mathbf{M}\ge0$ componentwise and $y^{\mathsf T}v<0$, no
$x\ge0$ solves $\mathbf{M}x=v$; and if a subsystem $\mathbf{M}_Rx=v_R$, formed by a set $R$ of rows,
has no nonnegative solution, neither has $\mathbf{M}x=v$.
\end{lemma}

\begin{proof}
$(\mathbf{M}x)_{(C,s)}=\sum_{t\,:\,t|_C=s}x_t$, which is the value at $s$ of the marginal $x|_C$; so
$\mathbf{M}x=v$ says exactly that every marginal of $x$ is the corresponding $\mathsf e_C$. Summing the rows
of one context $C$ gives $\sum_{s\in\Ee(C)}(\mathbf{M}x)_{(C,s)}=\sum_tx_t$, since every global
assignment restricts to exactly one section over $C$, while $\sum_sv_{(C,s)}=\sum_s\mathsf e_C(s)=1$; so a
nonnegative solution sums to one. If $y^{\mathsf T}\mathbf{M}\ge0$ and $y^{\mathsf T}v<0$, an
$x\ge0$ with $\mathbf{M}x=v$ would give
$0\le(y^{\mathsf T}\mathbf{M})x=y^{\mathsf T}(\mathbf{M}x)=y^{\mathsf T}v<0$. A subsystem is implied by the full system,
so a solution of the whole would solve it.
\end{proof}

The $(2,2,2)$ Bell model of \cite{AbramskyBrandenburger2011} has $X=\{a,a',b,b'\}$, $O=\{0,1\}$,
contexts $\{a,b\},\{a,b'\},\{a',b\},\{a',b'\}$, and distributions, on the outcomes $00,01,10,11$,
$(\tfrac12,0,0,\tfrac12)$, $(\tfrac38,\tfrac18,\tfrac18,\tfrac38)$,
$(\tfrac38,\tfrac18,\tfrac18,\tfrac38)$, $(\tfrac18,\tfrac38,\tfrac38,\tfrac18)$ respectively.
Number the sixteen rows in that order and the global assignments by
$x_{1+8s(a)+4s(b)+2s(a')+s(b')}$.

In each context the first digit of an outcome is the value of the measurement named first. Row
$4(c-1)+o$ is the $o$-th outcome of the $c$-th context, and the global assignment with values
$s(a),s(b),s(a'),s(b')$ is column $1+8s(a)+4s(b)+2s(a')+s(b')$. This is a bijection of
$\{0,1\}^4$ onto $\{1,\dots,16\}$, every integer $0\le c\le15$ having exactly one binary expansion
$c=8x_1+4x_2+2x_3+x_4$ with $x_i\in\{0,1\}$, since the sixteen choices of $x$ give values in
$\{0,\dots,15\}$ that are distinct, at the first index $k$ where two choices differ the larger
contributing $2^{4-k}$, more than the at most $2^{4-k}-1$ that the later digits can make up.

\begin{lemma}\label{lem:bellmodel}
The four distributions are probability distributions, compatible on overlaps: two distinct contexts
share at most one measurement, and each distribution gives each of its two measurements the
marginal $(\tfrac12,\tfrac12)$.
\end{lemma}

\begin{proof}
Each table has nonnegative entries summing to one. The contexts are four distinct pairs, each with
one measurement from $\{a,a'\}$ and one from $\{b,b'\}$, so two of them share at most one. The value
$0$ of the first measurement receives $\tfrac12+0$, $\tfrac38+\tfrac18$, $\tfrac38+\tfrac18$ or
$\tfrac18+\tfrac38$, and the value $0$ of the second the same sums; each is $\tfrac12$, and the value
$1$ receives the remaining $\tfrac12$. On a shared measurement both marginals are therefore
$(\tfrac12,\tfrac12)$.
\end{proof}

\begin{theorem}\label{thm:bell}
The four rows $R=\{1,6,11,13\}$ of the Bell system,
\begin{align}
x_1 + x_2 + x_3 + x_4        &= \tfrac12,   \label{eq:r1}\\
x_2 + x_4 + x_6 + x_8        &= \tfrac18,   \label{eq:r6}\\
x_3 + x_4 + x_{11} + x_{12}  &= \tfrac18,   \label{eq:r11}\\
x_1 + x_5 + x_9 + x_{13}     &= \tfrac18,   \label{eq:r13}
\end{align}
are already inconsistent over the nonnegative reals.
\end{theorem}

\begin{proof}
\emph{The rows.} Row $1$ is the outcome $00$ of $\{a,b\}$, with right side $\tfrac12$; its global
assignments have $s(a)=s(b)=0$, so their columns are $1+2s(a')+s(b')\in\{1,2,3,4\}$. Row $6$ is the
outcome $01$ of $\{a,b'\}$, with right side $\tfrac18$; its assignments have $s(a)=0$, $s(b')=1$, so
their columns are $2+4s(b)+2s(a')\in\{2,4,6,8\}$. Row $11$ is the outcome $10$ of $\{a',b\}$, with
right side $\tfrac18$; its assignments have $s(a')=1$, $s(b)=0$, so their columns are
$3+8s(a)+s(b')\in\{3,4,11,12\}$. Row $13$ is the outcome $00$ of $\{a',b'\}$, with right side
$\tfrac18$; its assignments have $s(a')=s(b')=0$, so their columns are
$1+8s(a)+4s(b)\in\{1,5,9,13\}$. These are \eqref{eq:r1}--\eqref{eq:r13}.

\emph{The inconsistency.} The four columns of \eqref{eq:r1}, the assignments with
$s(a)=s(b)=0$, are distinguished by $(s(a'),s(b'))$. Column $1$, with $(0,0)$, occurs in
\eqref{eq:r13}; column $2$, with $(0,1)$, in \eqref{eq:r6}; column $3$, with $(1,0)$, in
\eqref{eq:r11}; and column $4$, with $(1,1)$, in both \eqref{eq:r6} and \eqref{eq:r11}. For $x\ge0$
every term on the left of \eqref{eq:r1} therefore reappears on the left of one of the other three
equations, distinct terms in distinct places, and the remaining terms there are nonnegative, so
\[
\tfrac12=x_1+x_2+x_3+x_4\le(x_2+x_4+x_6+x_8)+(x_3+x_4+x_{11}+x_{12})+(x_1+x_5+x_9+x_{13})
=\tfrac38 ,
\]
which is false.

\emph{The certificate.} The same argument is the dual certificate $y=(-1,1,1,1)$, read against the
rows in the order \eqref{eq:r1}, \eqref{eq:r6}, \eqref{eq:r11}, \eqref{eq:r13}. The coefficient of
$x_j$ in $y^{\mathsf T}\mathbf{M}_R$ is
$-[j\in\{1,2,3,4\}]+[j\in\{2,4,6,8\}]+[j\in\{3,4,11,12\}]+[j\in\{1,5,9,13\}]$. Columns $1,2,3$
receive $-1$ and exactly one $+1$; column $4$ receives $-1$ and two $+1$; columns
$5,6,8,9,11,12,13$ receive one $+1$ and no $-1$; columns $7,10,14,15,16$ receive nothing. Hence
\[
y^{\mathsf T}\mathbf{M}_R=(0,0,0,1,1,1,0,1,1,0,1,1,1,0,0,0)\ge0,
\qquad
y^{\mathsf T}v_R=-\tfrac12+\tfrac18+\tfrac18+\tfrac18=-\tfrac18<0,
\]
and the dual inequality above excludes every $x\ge0$. All arithmetic is exact in $\Qq$.
\end{proof}

\begin{corollary}\label{cor:bellfull}
The full sixteen-row system $\mathbf{M}x=v$ of the Bell model has no nonnegative solution, and the Bell model
is not probabilistically extendable.
\end{corollary}

\begin{proof}
The four rows of Theorem~\ref{thm:bell} are rows of the full system, so by Lemma~\ref{lem:linsys} the
full system has no nonnegative solution; the certificate of that theorem, extended by zeros on the
other twelve rows, certifies it directly. By Lemma~\ref{lem:linsys} a probabilistic extension would be
such a solution.
\end{proof}

\begin{theorem}\label{thm:bellsupport}
For the Bell model, $\Se = \{ s : s(a) = s(b) \}$, so $|\Se| = 8$; the model is not strongly
contextual.
\end{theorem}

\begin{proof}
The distribution at $\{a,b\}$ is $(\tfrac12,0,0,\tfrac12)$, whose support is $\{00,11\}$. The other
three distributions have every entry in $\{\tfrac18,\tfrac38\}$, so their supports are the whole of
$O^C$, and the condition $s|_C\in\operatorname{supp}(\mathsf e_C)$ holds for every $s$ at those three
contexts. A global assignment therefore lies in $\Se$ exactly when $(s(a),s(b))\in\{00,11\}$, that is
when $s(a)=s(b)$. Such an $s$ is fixed by the common value of $s(a)$ and $s(b)$ and by the values
$s(a'),s(b')$, which are unconstrained, so $|\Se|=2\cdot2\cdot2=8$. In particular $\Se\ne\varnothing$,
and strong contextuality, which is $\Se=\varnothing$, fails.
\end{proof}

The set $\Se$ also restricts onto every support, so the Bell model is possibilistically extendable.
The distribution at $\{a,b\}$ is $(\tfrac12,0,0,\tfrac12)$ \cite{AbramskyBrandenburger2011} with support $\{00,11\}$, which $\Se$ gives.
The other three distributions have entries $\tfrac18$ and $\tfrac38$ only
\cite{AbramskyBrandenburger2011}, so their supports hold all four sections. Each of those contexts contains at most one of $a$ and $b$, and $\Se$ leaves the common
value of $s(a)$ and $s(b)$ free together with $s(a')$ and $s(b')$, so it gives all four sections there
as well. Being probabilistically contextual (Corollary~\ref{cor:bellfull}), the Bell model therefore
belongs to the weakest of the three levels of \cite{AbramskyBrandenburger2011}.

\begin{theorem}\label{thm:cabello}
Let $\Mm$ cover a set $V$ with $|V| = 18$, $|\Mm| = 9$, $|C| = 4$ for $C \in \Mm$, every $v\in V$ lying
in exactly two contexts, and let each support be the exactly-one-true predicate. Then $\Se = \varnothing$:
no $s : V \to \{0,1\}$ has exactly one $1$ in every context.
\end{theorem}

\begin{proof}
Suppose $s:V\to\{0,1\}$ lies in $\Se$, so that its restriction to each context has exactly one $1$.
Count the pairs $(v,C)$ with $v\in C\in\Mm$, each weighted by $s(v)$, in two orders. Summing first
over $v\in C$ and then over $C$ gives
\[
\sum_{C\in\Mm}\sum_{v\in C}s(v)=\sum_{C\in\Mm}1=|\Mm|=9 .
\]
Summing first over the contexts containing a fixed $v$ and then over $v$ gives
\[
\sum_{v\in V}s(v)\,\bigl|\{C\in\Mm:v\in C\}\bigr|=2\sum_{v\in V}s(v),
\]
an even integer. The two sums count the same weighted pairs, so $9$ would be even, which it is not;
no such $s$ exists.
\end{proof}

\begin{corollary}\label{cor:parity}
Let $\Mm$ be a finite family of contexts over a finite set $V$, every element lying in exactly $k$
contexts, with $k\nmid|\Mm|$. Then no $s:V\to\{0,1\}$ has exactly one $1$ in every context.
\end{corollary}

\begin{proof}
The double count of the proof of Theorem~\ref{thm:cabello} gives
$|\Mm|=\sum_{C\in\Mm}\sum_{v\in C}s(v)=k\sum_{v\in V}s(v)$, a multiple of $k$.
\end{proof}

Theorem~\ref{thm:cabello} is the case $k=2$, $|\Mm|=9$. Abramsky and Brandenburger prove the general
form, that a global section of the exactly-one-true predicate forces every common divisor of the
numbers $|\{C\in\Mm:v\in C\}|$ to divide $|\Mm|$ \cite{AbramskyBrandenburger2011}.

These are the parameters of the eighteen-vector configuration of Cabello, Estebaranz and
Garc\'{\i}a-Alcaine, nine orthogonal tetrads of eighteen vectors in $\Rr^4$ in which every vector
occurs twice \cite{CabelloEstebaranzGarciaAlcaine1996}, which \cite{AbramskyBrandenburger2011}
present as a measurement cover. The hypothesis is also met without vectors.

\begin{proposition}\label{prop:circulant}
Let $G$ be the graph on $\mathbb{Z}/9\mathbb{Z}$ whose edges are $\{i,i+1\}$ and $\{i,i+2\}$, let $V$ be
its edges and let $\Mm$ consist of the nine edge-stars. Then $|V|=18$, $|\Mm|=9$, every star has four
elements, every edge lies in exactly two stars, and $\Mm$ is an antichain cover of $V$; so
Theorem~\ref{thm:cabello} applies to it.
\end{proposition}

\begin{proof}
The eighteen edges are distinct, since $\pm1\not\equiv\pm2$ modulo $9$; each vertex $i$ meets exactly
the four edges $\{i,i\pm1\},\{i,i\pm2\}$, so every star has four elements; each edge lies in the stars
of its two ends and no other, so every element lies in exactly two contexts. Two stars share at most
the edge joining their centres, so the nine stars are distinct sets of equal size, and of two sets of
equal size neither contains the other unless they are equal.
\end{proof}

\begin{proposition}\label{prop:circulant-model}
On the cover of Proposition~\ref{prop:circulant}, with $O=\{0,1\}$, let $\mathsf e_C$ give weight
$\tfrac14$ to each of the four sections $s\in\Ee(C)$ with exactly one $1$, and weight $0$ to the
others. Then $\mathsf e$ is an empirical model, compatible on overlaps, whose support at every context is
the exactly-one-true predicate; it is strongly contextual.
\end{proposition}

\begin{proof}
\emph{Distributions and supports.} A section of $\Ee(C)$ with exactly one $1$ is fixed by the
element of $C$ carrying the $1$, so there are $|C|=4$ of them. Each receives $\tfrac14$ and every
other section $0$, so $\mathsf e_C$ is nonnegative with total $4\cdot\tfrac14=1$, and its support is
exactly the set of sections with exactly one $1$.

\emph{Marginals at one element.} For $v\in C$ exactly one of the four sections with exactly one $1$
has the value $1$ at $v$, namely the one whose $1$ is at $v$. So the marginal of $\mathsf e_C$ at $v$ gives
$1$ the weight $\tfrac14$ and $0$ the weight $\tfrac34$, whichever context $C\ni v$ is used.

\emph{Compatibility.} By the proof of Proposition~\ref{prop:circulant} two distinct stars share at
most one edge. If $C\cap C'=\varnothing$ both marginals on it are the unit mass on the empty section.
If $C\cap C'=\{v\}$, a section over $\{v\}$ is its value at $v$, so the marginals of $\mathsf e_C$ and
$\mathsf e_{C'}$ on $C\cap C'$ are the marginals at $v$, which agree by the previous step. So $\mathsf e$ is
compatible on overlaps, and it is an empirical model. Its supports are those of
Theorem~\ref{thm:cabello}, which gives $\Se=\varnothing$.
\end{proof}

Write $\{0,1\}$ for the boolean semiring $(\{0,1\},\vee,0,\wedge,1)$, and let $\operatorname{sg}:\Rr_{\ge0}\to\{0,1\}$ send $0$ to $0$ and every
positive real to $1$, so that the support of a nonnegative entry is its image under
$\operatorname{sg}$; applied to a matrix or a vector it acts entrywise.

\begin{proposition}[generalising the corresponding result of {\cite{AbramskyBrandenburger2011}} for the incidence matrix of a cover]\label{prop:vvb}
If a system $\mathbf{M}x = v$ with nonnegative entries, augmented by the row $\sum_tx_t=1$, has a
solution over the nonnegative reals, then the system with every entry replaced by its support,
$\operatorname{sg}(\mathbf{M})x = v_b$ with $v_b:=\operatorname{sg}(v)$, augmented in the same way, has
a solution over the booleans. Hence strong contextuality implies possibilistic non-extendability, which
implies probabilistic non-extendability.
\end{proposition}

\begin{proof}
\emph{Transport.} For $a,b\ge0$ one has $a+b=0$ iff $a=b=0$, so
$\operatorname{sg}(a+b)=\operatorname{sg}(a)\vee\operatorname{sg}(b)$; and $ab=0$ iff $a=0$ or $b=0$,
the reals having no zero divisors, so $\operatorname{sg}(ab)=\operatorname{sg}(a)\wedge\operatorname{sg}(b)$.
Induction on the number of terms extends the first identity to finite sums. Let $x\ge0$ solve the
augmented system. Row $i$ of $\mathbf{M}x=v$ gives
\[
\operatorname{sg}(v_i)=\operatorname{sg}\Bigl(\sum_t\mathbf{M}_{it}x_t\Bigr)
=\bigvee_t\operatorname{sg}(\mathbf{M}_{it})\wedge\operatorname{sg}(x_t),
\]
which is row $i$ of the support system evaluated at $\operatorname{sg}(x)$; and the augmenting row
gives $\bigvee_t\operatorname{sg}(x_t)=\operatorname{sg}(1)=1$. So $\operatorname{sg}(x)$ is a boolean
solution of the augmented support system, which is the first sentence.

\emph{Boolean solutions are possibilistic extensions.} For the incidence system of an empirical model
the matrix is already $0$--$1$ and $v_b$ has entries $[s\in\operatorname{supp}(\mathsf e_C)]$. A vector
$z\in\{0,1\}^{\Ee(X)}$ is the indicator of $S=\{t:z_t=1\}$. The augmenting row reads
$\bigvee_tz_t=1$, that is $S\ne\varnothing$. The row $(C,s)$ reads
$\bigvee_{t\,:\,t|_C=s}z_t=[s\in\operatorname{supp}(\mathsf e_C)]$, that is: some $t\in S$ restricts to $s$
exactly when $s$ lies in the support. Over all $s\in\Ee(C)$ this says $S|_C=\operatorname{supp}(\mathsf e_C)$.
So the boolean solutions of the augmented support system are exactly the possibilistic extensions.

\emph{Possibilistic extensions meet $\Se$.} If $S$ is a possibilistic extension and $t\in S$, which
exists as $S\ne\varnothing$, then for every $C$, $t|_C\in S|_C=\operatorname{supp}(\mathsf e_C)$, so $t\in\Se$.

\emph{The chain.} A probabilistic extension $d$ is a nonnegative solution of the augmented incidence
system, the augmenting row being $\sum_td(t)=1$; by the transport $\operatorname{sg}(d)$ is a boolean
solution,
hence a possibilistic extension, namely $\operatorname{supp}(d)$. A possibilistic extension forces
$\Se\ne\varnothing$. Read contrapositively, $\Se=\varnothing$ excludes every possibilistic extension,
and the absence of a possibilistic extension excludes every probabilistic one.
\end{proof}

Applied context by context, $\operatorname{sg}$ sends $\mathsf e_C$ to the indicator of its
support, by its definition: $0\mapsto0$ and every positive value $\mapsto1$. It is a homomorphism of semirings, but Proposition~\ref{prop:noeffect} below shows that it
is not a morphism of effect algebras, and that no map from $[0,1]$ to $\{0,1\}$ is. An effect algebra is a partial
commutative monoid $(E,\oplus,0,1)$ with orthosupplements in which $x\oplus1$ is defined only for
$x=0$; a morphism preserves $1$ and every defined sum.

Here a partial commutative monoid has a partial operation $\oplus$, commutative and associative
wherever defined, with unit $0$, and an orthosupplement of $x$ is the unique $x^\perp$ with
$x\oplus x^\perp=1$. The unit interval is an effect algebra with $a\oplus b=a+b$ defined exactly when
$a+b\le1$ and $a^\perp=1-a$. The two-element effect algebra $\{0,1\}$ has $0\oplus0=0$,
$0\oplus1=1\oplus0=1$, and $1\oplus1$ undefined, as the last axiom requires. A morphism $f$ is
required to satisfy $f(1)=1$ and, whenever $a\oplus b$ is defined, that $f(a)\oplus f(b)$ be defined
and equal to $f(a\oplus b)$. These definitions and the example $[0,1]$ are those of
\cite{Abramsky2026}, and $\{0,1\}$ is the two-element case of his example of a Boolean algebra, in
which disjunction is defined only on disjoint pairs.

\begin{proposition}\label{prop:noeffect}
$\operatorname{sg} : \Rr_{\geq 0} \to \{0,1\}$ is a surjective homomorphism of semirings, the join of
$\{0,1\}$ being total, and its restriction to $[0,1]$ is surjective and carries every defined sum
$a + b \leq 1$ to $\operatorname{sg}(a) \vee \operatorname{sg}(b)$; but that restriction is not a
morphism of effect algebras $[0,1] \to \{0,1\}$, and no map $[0,1] \to \{0,1\}$ is one.
\end{proposition}

\begin{proof}
The semiring clauses are $\operatorname{sg}(a+b)=\operatorname{sg}(a)\vee\operatorname{sg}(b)$ and
$\operatorname{sg}(ab)=\operatorname{sg}(a)\operatorname{sg}(b)$ on nonnegative reals, both proved in
the transport step of Proposition~\ref{prop:vvb}, together with $\operatorname{sg}(0)=0$ and
$\operatorname{sg}(1)=1$; the join $\vee$ is defined on every pair, so the first clause needs no
side condition. Both values are attained, at $0$ and at $1$, which lie in $[0,1]$, so
$\operatorname{sg}$ and its restriction are surjective. A defined sum $a+b\le1$ in $[0,1]$ is in
particular a sum of nonnegative reals, so it goes to $\operatorname{sg}(a)\vee\operatorname{sg}(b)$.

Let $f:[0,1]\to\{0,1\}$ be any map preserving $1$ and defined sums. In $[0,1]$ the sum
$\tfrac12\oplus\tfrac12=1$ is defined, so $f(\tfrac12)\oplus f(\tfrac12)$ must be defined and equal to
$f(1)=1$. If $f(\tfrac12)=1$ the sum $1\oplus1$ is undefined; if $f(\tfrac12)=0$ the sum is
$0\oplus0=0\ne1$. Neither case is possible, so no such $f$ exists; in particular the restriction of
$\operatorname{sg}$, which has $\operatorname{sg}(\tfrac12)=1$, is not a morphism.
\end{proof}

Over the effect algebra $\{0,1\}$ the models are the deterministic assignments and none is
contextual (Proposition~\ref{prop:boolmodels} below), and \cite{Abramsky2026} accommodates the
possibilistic models by dropping the last effect-algebra axiom, which gives what it calls a soft
effect algebra.

\begin{proposition}[\cite{Abramsky2026}]\label{prop:boolmodels}
Let $\Mm$ cover $X$, and let each $C\in\Mm$ carry a $\{0,1\}$-valued distribution $\mathsf e_C$ on $\Ee(C)$,
the marginals of these, $\oplus$-sums over fibres, agreeing on overlaps. Then each $\mathsf e_C$ is the point
mass at one section $s_C$, and there is a global assignment $t$ with $t|_C=s_C$ for every $C$, so the
point mass at $t$ extends the model and it is not contextual.
\end{proposition}

\begin{proof}
A $\{0,1\}$-valued distribution on $\Ee(C)$ is a family whose $\oplus$-sum is defined and equal to
$1$; two entries equal to $1$ would require $1\oplus1$, and entries all $0$ sum to $0$, so exactly one
section $s_C$ receives $1$. Compatibility on overlaps then says that $s_C$ and $s_{C'}$ agree on
$C\cap C'$, since the marginal of a point mass at $s_C$ is the point mass at its restriction.
Sections of $O^{(-)}$ agreeing on overlaps glue, the contexts covering $X$, to a global $t$ with
$t|_C=s_C$: put $t(x):=s_C(x)$ for any $C\ni x$. The marginal of the point mass at $t$ on $C$ is the
point mass at $t|_C=s_C$, which is $\mathsf e_C$.
\end{proof}

With the boolean join made total, the models over $\{0,1\}$ are exactly the possibilistic models
\cite{Abramsky2026}, and the support of every probabilistic model is one of them: $\operatorname{sg}$
preserves finite sums (transport step of Proposition~\ref{prop:vvb}), so the support of a marginal
is the marginal of the support, the supports are therefore compatible on overlaps, and each is
nonempty, its join being $\operatorname{sg}(1)=1$ \cite{AbramskyBrandenburger2011}.

\begin{corollary}\label{cor:placement}
Probabilistic contextuality and strong contextuality both occur and are ranked rather than
exclusive. Writing $B$ for the Bell system and $\Mm$ for a cover as in Theorem~\ref{thm:cabello}:
\begin{enumerate}[label=(\roman*)]
\item no $x \geq 0$ solves $\mathbf{M}_R x = v_R$, so $B$ fails probabilistic extendability;
\item $\Se \neq \varnothing$ for $B$, which is not strongly contextual;
\item no $s : V \to \{0,1\}$ has exactly one $1$ on every $C \in \Mm$, so every empirical model on
the cover with the exactly-one-true supports, such as that of Proposition~\ref{prop:circulant-model},
is strongly contextual;
\item for every nonnegative-real system $(A, v)$, if $\operatorname{sg}(A)x = \operatorname{sg}(v)$ has
no boolean solution then $Ax = v$ has no nonnegative solution.
\end{enumerate}
So $B$ occupies the probabilistic stratum, the model on the cover the strong one, and every strongly contextual
model is also probabilistically contextual. The ordering is one way: probabilistic
non-extendability and $\Se = \varnothing$ are not equivalent.
\end{corollary}

\begin{proof}
Clause (i) is Theorem~\ref{thm:bell}, and a probabilistic extension of $B$ would be a nonnegative
solution of the full system, hence of its rows $R$. Clause (ii) is Theorem~\ref{thm:bellsupport}.
Clause (iii) is Theorem~\ref{thm:cabello}, and it is not vacuous: the cover of edge-stars satisfies
its hypothesis (Proposition~\ref{prop:circulant}) and carries the empirical model of
Proposition~\ref{prop:circulant-model}. For (iv), let $x\ge0$ solve $Ax=v$. By the identities of
Proposition~\ref{prop:noeffect}, row $i$ gives
$\operatorname{sg}(v_i)=\bigvee_t\operatorname{sg}(A_{it})\wedge\operatorname{sg}(x_t)$, so
$\operatorname{sg}(x)$ is a boolean solution of $\operatorname{sg}(A)x=\operatorname{sg}(v)$;
contrapositively, no boolean solution leaves no nonnegative one.

The placement follows. $B$ fails probabilistic extendability by (i) while $\Se\ne\varnothing$ by (ii),
which is the probabilistic stratum; any model whose supports are those of Theorem~\ref{thm:cabello} has
$\Se=\varnothing$ by (iii), which is the strong one. If a model has $\Se=\varnothing$, the possibilistic
extensions are excluded by the third step of the proof of Proposition~\ref{prop:vvb}, so the augmented
support system has no boolean solution, and by (iv) the augmented incidence system has no nonnegative
solution: the model also fails probabilistic extendability.
Finally, if probabilistic non-extendability were equivalent to $\Se=\varnothing$, then $B$, which is
probabilistically non-extendable by (i), would have $\Se=\varnothing$, against (ii).
\end{proof}

A second reading of the Kochen--Specker theorem is Bohrification. For a unital $C^*$-algebra $A$,
Heunen, Landsman and Spitters take the poset $\mathcal C(A)$ of unital
commutative $C^*$-subalgebras ordered by inclusion and the topos of functors $\mathcal C(A)\to
\mathbf{Set}$; the Bohrification of $A$ is the tautological functor sending each $C$ to itself, a
commutative $C^*$-algebra inside that topos, and their form of the Kochen--Specker theorem is that its
Gelfand spectrum has no points when $A$ is the algebra of bounded operators on a Hilbert space of
dimension greater than two \cite{HeunenLandsmanSpitters2009}. Being defined for every unital
$C^*$-algebra, Bohrification is available at every finite stage $M_n(\Cc)$, the $C^*$-algebra of $n\times n$ complex matrices with unit the identity matrix, of a tower of matrix
algebras, such as that of the fuzzy sphere in Section~\ref{sec:towers}, and its index is already
large at a single stage, of at least the cardinality of the continuum from $n=2$ on
(Proposition~\ref{prop:continuum}). We take $A=M_n(\Cc)$, which is the algebra of bounded operators on
$\Cc^n$.

\begin{lemma}\label{lem:closedsub}
Every $*$-subalgebra of $M_n(\Cc)$ is closed, so the unital commutative $C^*$-subalgebras of
$M_n(\Cc)$ are its unital commutative $*$-subalgebras.
\end{lemma}

\begin{proof}
A $*$-subalgebra is a linear subspace of the finite-dimensional space $M_n(\Cc)$, and a
finite-dimensional subspace of a Hausdorff topological vector space over $\Cc$ is closed.
\end{proof}

\begin{proposition}\label{prop:continuum}
For $n \geq 2$ the unital commutative $*$-subalgebras of $M_n(\Cc)$ have at least the cardinality of
the continuum.
\end{proposition}

\begin{proof}
For $t\in(0,1)$ let $v(t)=(1,t,0,\dots,0)\in\Cc^n$, with coordinates indexed from $0$, and let
$P_t=(1+t^2)^{-1}v(t)v(t)^{\mathsf T}$, the orthogonal projection onto $\Cc v(t)$: it is real
symmetric, so self-adjoint, and $P_t^2=(1+t^2)^{-2}v(t)\bigl(v(t)^{\mathsf T}v(t)\bigr)v(t)^{\mathsf T}
=P_t$ because $v(t)^{\mathsf T}v(t)=1+t^2$. Its entries are $(P_t)_{jk}=v(t)_jv(t)_k/(1+t^2)$, so
$(P_t)_{00}=1/(1+t^2)$ and $(P_t)_{01}=t/(1+t^2)$. Put $A_t=\{a1+bP_t:a,b\in\Cc\}$. It is a linear
subspace containing $1$; it is closed under multiplication, since
$(a1+bP_t)(c1+dP_t)=ac\,1+(ad+bc+bd)P_t$ by $P_t^2=P_t$; it is closed under the adjoint, since
$(a1+bP_t)^*=\bar a1+\bar bP_t$; and it is commutative, the product formula being symmetric in the two
factors. So $A_t$ is a unital commutative $*$-subalgebra.

The elements $1$ and $P_t$ are linearly independent: $P_t$ has rank one and $n\ge2$, while a scalar
matrix has rank $0$ or $n$. Suppose $A_t=A_u$ for $t,u\in(0,1)$. Then $P_u\in A_t$, so $P_u=a1+bP_t$
for some $a,b$, and squaring,
\[
P_u=P_u^2=a^2\,1+(2ab+b^2)P_t .
\]
Comparing coefficients in the basis $1,P_t$ gives $a^2=a$ and $2ab+b^2=b$. If $a=1$ then $b^2+b=0$,
so $b\in\{0,-1\}$. The case $b=0$ makes $P_u=1$, which has rank $n\ge2$, not one. The case $b=-1$
makes $P_u=1-P_t$, whence $P_uP_t=P_t-P_t^2=0$ and $P_uv(t)=0$; but
$P_uv(t)=(1+u^2)^{-1}\bigl(v(u)^{\mathsf T}v(t)\bigr)v(u)$ with $v(u)^{\mathsf T}v(t)=1+ut>0$, a
contradiction. So $a=0$, then $b^2=b$, and $b=0$ would make $P_u=0$; hence $b=1$ and $P_u=P_t$.
Reading the ratio of the entries $(0,1)$ and $(0,0)$ on both sides gives $u=t$. The map
$t\mapsto A_t$ is therefore injective on $(0,1)$, which has the cardinality of the continuum.
\end{proof}

The poset that indexes the topos is therefore uncountable from $n=2$ on, and the spectrum it
indexes is pointless from $n=3$ on, $M_n(\Cc)$ being the bounded operators on a Hilbert space of
dimension $n>2$ \cite{HeunenLandsmanSpitters2009}, while the cover of a finite scenario has finitely
many contexts, its contexts being subsets of a finite set.

We return to the finite stages of a tower and their outcome assignments. The comparison
of two stages of a tower asks whether a coarse stage and a refinement of it, read through their
outcome assignments, are the same system or one only embedded in the other. Let $O$ be a finite outcome set and
$g:[N]\to[m]$ onto.

\begin{definition}\label{def:lts}
Fix a set $\mathcal{L}$ of contexts $V\subseteq[m]$. On the coarse sections $O^{[m]}$ and the fine
sections $O^{[N]}$ define labelled graphs $\Gamma_{\mathrm c}$ and $\Gamma_{\mathrm f}$: for $s\neq t$,
$s\sim_V t$ in $\Gamma_{\mathrm c}$ iff $s|_V=t|_V$, and $s'\sim_V t'$ in $\Gamma_{\mathrm f}$ iff
$s'|_{g^{-1}(V)}=t'|_{g^{-1}(V)}$, so a label names the same context on both sides. The induced
map is the restriction $g^{*}s := s \circ g$, as in Definition~\ref{def:stagepullback} with $O$ in
place of $\Rr$. For a map $f:O^{[m]}\to O^{[N]}$ from coarse to fine sections, the
\emph{multiplicity} $\mathrm{mult}_f(e)$ of a fine labelled edge $e$ is the number of coarse labelled
edges with its label mapped onto it; with $\deg$ the number of labelled neighbours, the \emph{mass
ratio} is $\mathrm{mr}_f(x)=\sum_{f(u)=x}\deg_{\mathrm c}(u)/\deg_{\mathrm f}(x)$; and
$c_f := \min_{e} \mathrm{mult}_f(e)/\max_{x} \mathrm{mr}_f(x)$.
\end{definition}

A labelled edge is a pair $(V,\{s,t\})$ with $V\in\mathcal L$ and $s\sim_Vt$; the coarse labelled
edge $(V,\{s,t\})$ is mapped onto $(V,\{f(s),f(t)\})$, and $\deg(u)=\sum_{V\in\mathcal L}
|\{t:t\sim_Vu\}|$ counts neighbours with their labels. The ratio $\mathrm{mr}_f$ is taken where
$\deg_{\mathrm f}$ does not vanish, and $c_f$ where the fine system has an edge.

\begin{proposition}\label{prop:homologous}
Let $f$ be a bijection of states that preserves and reflects every labelled edge, every fine state
having a labelled neighbour. Then every multiplicity and every mass ratio of $f$ equals one, and
$c_f=1$.
\end{proposition}

\begin{proof}
A fine edge $(V,\{x,y\})$ receives the coarse edge $(V,\{f^{-1}(x),f^{-1}(y)\})$, which is an edge by
reflection, and no other, since by injectivity a coarse pair mapped onto $\{x,y\}$ is that pair; so
every multiplicity is $1$. The fibre over $x$ is the single state $f^{-1}(x)$, whose neighbours at $V$
are carried bijectively onto those of $x$, so $\deg_{\mathrm c}(f^{-1}(x))=\deg_{\mathrm f}(x)\neq0$ and
$\mathrm{mr}_f(x)=1$. Hence $c_f=1/1=1$.
\end{proof}

Call a bijection of states preserving and reflecting every labelled edge an \emph{identification}
of the two stages. Call the stages \emph{identified} when an identification exists, and
\emph{embedded} when none exists and the coarse system embeds as a full labelled subsystem through
$g^{*}$.

\begin{proposition}\label{prop:refstage}
Let $g : [N] \to [m]$ be onto with $m < N$ and $|O| \geq 2$. Then $g^{*}$ is injective and not
surjective, and $|O|^{m} \neq |O|^{N}$, so the two section sets admit no bijection. At $|O|=1$ the
map $g^{*}$ is a bijection.
\end{proposition}

\begin{proof}
Let $s\circ g=t\circ g$. For a coarse site $c\in[m]$ pick $j\in[N]$ with $g(j)=c$, which exists because $g$
is onto; then $s(c)=s(g(j))=t(g(j))=t(c)$. So $s=t$ and $g^{*}$ is injective. A function from an
$m$-element set to $O$ is a choice of one of $|O|$ values at each of $m$ points, so
$|O^{[m]}|=|O|^m$ and likewise $|O^{[N]}|=|O|^N$. With $|O|\ge2$ and $m<N$,
$|O|^N=|O|^m|O|^{N-m}\ge2|O|^m>|O|^m$. A map from a finite set into a strictly larger one is not onto,
and finite sets of different cardinalities admit no bijection. At $|O|=1$ both $O^{[m]}$ and $O^{[N]}$
consist of the single constant function, and the unique map between them is a bijection.
\end{proof}

\begin{theorem}\label{thm:lts-embedding}
Let $g$ be onto. Then $g^{*}$ is injective and preserves and reflects labelled edges label by
label, $g^{*}s \sim_{V} g^{*}t$ iff $s \sim_{V} t$; if moreover $|O| > 1$ and $m < N$ then
$g^{*}$ is not onto.
\end{theorem}

\begin{proof}
Injectivity is the first step of the proof of Proposition~\ref{prop:refstage}, where only the
surjectivity of $g$ enters. Fix $V\in\mathcal L$ and coarse sections $s,t$. By injectivity, $s\ne t$
iff $g^{*}s\ne g^{*}t$, so the two edge relations have the same distinctness clause. For the
agreement clause, $g^{*}s$ and $g^{*}t$ agree on $g^{-1}(V)$ iff $s(g(j))=t(g(j))$ for every
$j\in g^{-1}(V)$. As $j$ ranges over $g^{-1}(V)$, $g(j)$ ranges over all of $V$, every coarse site of $V$
having a preimage; so the condition is $s(c)=t(c)$ for every $c\in V$, which is $s|_V=t|_V$. The two
clauses together give $g^{*}s\sim_Vg^{*}t$ iff $s\sim_Vt$. Non-surjectivity under
$|O|>1$ and $m<N$ is Proposition~\ref{prop:refstage}.
\end{proof}

\begin{theorem}\label{thm:verdict}
Let $g$ be onto, $|O| > 1$, $m < N$, and $V_0 \in\mathcal{L}$ a proper context, $V_0 \neq [m]$. Then
the fine system carries a labelled edge at $V_0$ with an endpoint outside the image of $g^{*}$,
every such edge has multiplicity zero, and $c_{g^{*}} = 0$. No identification of the two stages
exists, and the two stages are embedded and not identified.
\end{theorem}

\begin{proof}
\emph{An edge at a missed section.} By Theorem~\ref{thm:lts-embedding} there is a fine section $u$
outside the image of $g^{*}$. As $V_0\ne[m]$ there is a coarse site $c\notin V_0$, and as $g$ is onto a site
$i$ with $g(i)=c$, so $i\notin g^{-1}(V_0)$. As $|O|\ge2$, let $u'$ agree with $u$ except at $i$, where
it takes a different value. Then $u'\ne u$ and $u'|_{g^{-1}(V_0)}=u|_{g^{-1}(V_0)}$, so
$(V_0,\{u,u'\})$ is a labelled edge of the fine system with the endpoint $u$ outside the image.

\emph{Multiplicity zero.} Let $(V_0,\{x,y\})$ be any fine edge with, say, $x$ outside the image. A
coarse edge $(V_0,\{s,t\})$ is mapped onto it only if $\{g^{*}s,g^{*}t\}=\{x,y\}$, which
requires $x\in\{g^{*}s,g^{*}t\}$ and is impossible. So its multiplicity is $0$, and the least
multiplicity over all fine edges, a minimum of nonnegative integers attaining $0$, is $0$.

\emph{The denominator.} The alteration above applies to every fine section $x$, giving a neighbour
of $x$ at $V_0$; so $\deg_{\mathrm f}(x)\ge1$ for every $x$, and every ratio $\mathrm{mr}_{g^{*}}(x)$ is
defined. Likewise each coarse section $s$ has a neighbour at $V_0$, obtained by changing its value at
the coarse site $c\notin V_0$, so $\deg_{\mathrm c}(s)\ge1$. For $x=g^{*}s$ in the image the fibre is
$\{s\}$, by injectivity, and $\mathrm{mr}_{g^{*}}(x)=\deg_{\mathrm c}(s)/\deg_{\mathrm f}(x)>0$. The image is
nonempty, so the maximum of $\mathrm{mr}_{g^{*}}$ over the finitely many fine sections is positive.

\emph{The verdict.} Hence $c_{g^{*}}=0/\max_x\mathrm{mr}_{g^{*}}(x)=0$. An identification would be a
bijection between $O^{[m]}$ and $O^{[N]}$, and by Proposition~\ref{prop:refstage} none exists, so the
stages are not identified; by Theorem~\ref{thm:lts-embedding} the coarse system embeds as a full
labelled subsystem through $g^{*}$, so they are embedded.
\end{proof}

\begin{theorem}\label{thm:verdict-limit}
Let $|O|\ge2$, let $n$ be a stage with $\sz(n) < \sz(n+1)$, and $V \subsetneq [\sz(n)]$. Write
$\Gamma_O(T)$ for the labelled graph whose vertices are the functions $\Pi_T \to O$ read off a single
stage, two distinct ones joined at $V$ when they agree at every point $p$ with $p_n\in V$. Then
$u \mapsto u \circ \pi_n$ is injective and preserves and reflects labelled edges at every context, and
it misses a labelled edge at $V$: some labelled edge of $\Gamma_O(T)$ has an endpoint no stage-$n$
reading reaches.
\end{theorem}

\begin{proof}
A function read off stage $k$ is one of the form $w\circ\pi_k$ with $w:[\sz(k)]\to O$; two distinct
such functions are joined at a context $U\subseteq[\sz(n)]$ when they agree on every $p$ with $p_n\in
U$. Stage $n$ carries the labelled graph of Definition~\ref{def:lts} with $m=\sz(n)$: two distinct
readings $u,u'$ are joined at $U$ when $u|_U=u'|_U$.

\emph{Injectivity.} If $u\circ\pi_n=u'\circ\pi_n$, then for each site $i$ of stage $n$
Theorem~\ref{thm:hpath} gives a point $p$ with $p_n=i$, and $u(i)=u(p_n)=u'(p_n)=u'(i)$.

\emph{Labelled edges.} Fix $U\subseteq[\sz(n)]$. If $u|_U=u'|_U$, then every point $p$ with $p_n\in U$
has $u(p_n)=u'(p_n)$. Conversely, if $u\circ\pi_n$ and $u'\circ\pi_n$ agree at every point $p$ with
$p_n\in U$, then for $i\in U$ the point of Theorem~\ref{thm:hpath} with $p_n=i$ gives $u(i)=u'(i)$. With
injectivity carrying the clause $u\ne u'$ in both directions, $u\circ\pi_n$ and $u'\circ\pi_n$ are
joined at $U$ iff $u$ and $u'$ are.

\emph{A reading of stage $n+1$ that stage $n$ misses.} The refinement $\rf_n:[\sz(n+1)]\to[\sz(n)]$ is
onto with $\sz(n)<\sz(n+1)$, so Proposition~\ref{prop:refstage} gives $w:[\sz(n+1)]\to O$ not of the form
$u\circ\rf_n$. Suppose $w\circ\pi_{n+1}=u\circ\pi_n$ for some $u$. For a site $i$ of stage $n+1$ take
the point $p$ of Theorem~\ref{thm:hpath} with $p_{n+1}=i$; since $p\in\Pi_T$, $p_n=\rf_n(p_{n+1})=\rf_n(i)$,
and evaluating at $p$ gives $w(i)=u(\rf_n(i))$. So $w=u\circ\rf_n$, against the choice of $w$; hence
$f:=w\circ\pi_{n+1}$ is a vertex of $\Gamma_O(T)$ outside the image of $u\mapsto u\circ\pi_n$.

\emph{The edge.} Since $V\ne[\sz(n)]$, pick a site $j\notin V$ of stage $n$, and, $\rf_n$ being onto, a site $i$ of
stage $n+1$ with $\rf_n(i)=j$. Let $w'$ agree with $w$ except at $i$, where it takes another value, and
put $f':=w'\circ\pi_{n+1}$. The point of Theorem~\ref{thm:hpath} with $p_{n+1}=i$ gives
$f(p)=w(i)\ne w'(i)=f'(p)$, so $f\ne f'$. A point $p$ with $p_n\in V$ has $\rf_n(p_{n+1})=p_n\in V$, so
$p_{n+1}\ne i$, as $\rf_n(i)=j\notin V$; there $f(p)=w(p_{n+1})=w'(p_{n+1})=f'(p)$. So $f$ and $f'$
are joined at $V$ in $\Gamma_O(T)$, and the endpoint $f$ is reached by no stage-$n$ reading.
\end{proof}

\begin{corollary}\label{cor:limitcoeff}
In the setting of Theorem~\ref{thm:verdict-limit}, a labelled edge of $\Gamma_O(T)$ with an endpoint
outside the image of $u\mapsto u\circ\pi_n$ receives no labelled edge of stage $n$, and the least
multiplicity over the labelled edges of $\Gamma_O(T)$ at $V$ is zero.
\end{corollary}

\begin{proof}
A labelled edge of stage $n$ is carried onto a pair of image points, which cannot contain an endpoint
outside the image. Theorem~\ref{thm:verdict-limit} supplies such an edge at $V$, whose
multiplicity is therefore zero.
\end{proof}

\section{From the pair to the web}\label{sec:monogamy}

The operational $\mathrm{ER}=\mathrm{EPR}$ theorem concerns one pair of agents sharing entanglement
at the Tsirelson bound through a decoherence-free channel \cite{FieldsGlazebrookMarcianoZappala2025}.
Whether it lifts from the pair to a web of many parties is a question about how entanglement may be
distributed, and we take the additive form of monogamy as the defining constraint. For $n$ qubits
$A_1,\dots,A_n$ in a state with reduced density operators $\mathsf{d}_{A_1A_j}$ and
$\mathsf{d}_{A_1(A_2\cdots A_n)}$, the tangle $\mathfrak{t}$, the squared concurrence, satisfies
$\sum_{j\ge2}\mathfrak{t}(\mathsf{d}_{A_1A_j})\le\mathfrak{t}(\mathsf{d}_{A_1(A_2\cdots A_n)})$,
conjectured for pure states by Coffman, Kundu and Wootters \cite{CoffmanKunduWootters2000} and
proved for all states of $n$ qubits by Osborne and Verstraete \cite{OsborneVerstraete2006}.

A web is our model of such a distribution: one nonnegative number per pair of parties, read as the
entanglement the pair shares, with what a single party spreads over all of its partners at most
one. For a pure state of three qubits the right side of the inequality is
$4\det\mathsf{d}_{A_1}$ \cite{CoffmanKunduWootters2000}, which is at most one: the one-qubit density
operator $\mathsf{d}_{A_1}$ is positive semidefinite with trace one, so its eigenvalues are some
$p\in[0,1]$ and $1-p$ and
$\det\mathsf{d}_{A_1}=p(1-p)$, while $1-4p(1-p)=(2p-1)^2\ge0$. A web is thus a symmetric table of real numbers with bounded row sums. A party is an index $i\in[m]$,
a thread at $i$ is an entry $w(i,j)$ with $j\ne i$, and a pair $\{i,j\}$ always has $i\ne j$.

\begin{definition}\label{def:web}
A \emph{monogamous web} on $m$ parties is $w : [m] \times [m] \to \Rr$ with $w(i,j) = w(j,i)$,
$w(i,i) = 0$, $w(i,j) \geq 0$ and $\sum_{j} w(i,j) \leq 1$ for every $i$.
\end{definition}

\begin{proposition}\label{prop:monogamy-load}
For $m \geq 3$ the table $w(i,j) := [ i \neq j ]$ satisfies every clause of Definition~\ref{def:web}
but the last, violates the last, and has no thread of weight at most $\tfrac1{m-1}$.
\end{proposition}

\begin{proof}
The predicate $i\ne j$ is symmetric in $i$ and $j$, so $w(i,j)=w(j,i)$; it fails at $i=j$, so
$w(i,i)=0$; and every value lies in $\{0,1\}$, so $w\ge0$. At a party $i$ the row sum counts the
$j\in[m]$ with $j\ne i$, of which there are $m-1$; for $m\ge3$ this is at least $2>1$, so the last
clause fails at every party. Every thread has $w(i,j)=1$, while $m\ge3$ gives
$\tfrac1{m-1}\le\tfrac12<1$; so no thread has weight at most $\tfrac1{m-1}$.
\end{proof}

\begin{theorem}\label{thm:thin}
In a web on $m > 1$ parties every party has a thin thread,
$\forall i\ \exists j \neq i : w(i,j) \leq \tfrac{1}{m-1}$, and the bound is attained: the
\emph{equidistant} web $w_{\mathrm{u}}(i,j) := \tfrac{1}{m-1}$ for $i \neq j$ is a web meeting it
with equality. For every $\varepsilon > 0$ there is $m_0$ such that every party of every web on
$m \geq m_0$ parties has a thread thinner than $\varepsilon$.
\end{theorem}

\begin{proof}
\emph{The pigeonhole.} Fix $i$. The diagonal entry vanishes, so
$\sum_{j\ne i}w(i,j)=\sum_jw(i,j)\le1$, a sum of $m-1\ge1$ terms. If every term exceeded
$\tfrac1{m-1}$, adding the $m-1$ strict inequalities would give
$\sum_{j\ne i}w(i,j)>(m-1)\cdot\tfrac1{m-1}=1$, which is false. Hence some $j\ne i$ has
$w(i,j)\le\tfrac1{m-1}$.

\emph{Attainment.} Set $w_{\mathrm u}(i,i)=0$. The table is symmetric and nonnegative, and its row sum
at every party is $(m-1)\cdot\tfrac1{m-1}=1\le1$, so it is a web. Every one of its threads equals
$\tfrac1{m-1}$, so the thin thread the pigeonhole provides has weight exactly $\tfrac1{m-1}$, and no
constant smaller than $\tfrac1{m-1}$ can replace the bound: $w_{\mathrm u}$ has no thread below it.

\emph{The limit.} Given $\varepsilon>0$, choose an integer $m_0\ge2$ with $m_0-1>1/\varepsilon$. For
$m\ge m_0$ the first part applies, $m$ being greater than one, and the thin thread at $i$ has
$w(i,j)\le\tfrac1{m-1}\le\tfrac1{m_0-1}<\varepsilon$.
\end{proof}

\begin{definition}\label{def:bridge}
For $t\ge\tfrac12$, $\{i,j\}$ is a \emph{bridge at $t$} in $w$ when $w(i,j) > t$.
\end{definition}

We model the monogamous pair of the two-agent $\mathrm{ER}=\mathrm{EPR}$ statement
\cite{FieldsGlazebrookMarcianoZappala2025} by a thread of weight one, $w(i,j)=1$.

\begin{proposition}\label{prop:below-half}
On three parties the table with $w(0,1)=w(0,2)=\tfrac12$, its symmetric entries, and all other
entries $0$ is a web, and party $0$ holds two threads above every $t<\tfrac12$.
\end{proposition}

\begin{proof}
The table is symmetric, nonnegative and zero on the diagonal, and its row sums are
$1,\tfrac12,\tfrac12$, each at most one. Both threads at party $0$ have weight $\tfrac12>t$.
\end{proof}

\begin{theorem}\label{thm:bridge-matching}
In any web a party has at most one bridge: $w(i,j) > \tfrac12$ and $w(i,k) > \tfrac12$ imply $j = k$;
the bridges form a partial matching. For $m \geq 3$ no web has every pair as a bridge.
\end{theorem}

\begin{proof}
\emph{One bridge per party.} Let $w(i,j)>\tfrac12$ and $w(i,k)>\tfrac12$, and suppose $j\ne k$. Then
$j$ and $k$ are two distinct indices of the row of $i$, every other entry of that row is nonnegative,
and
\[
\sum_lw(i,l)\;\ge\;w(i,j)+w(i,k)\;>\;\tfrac12+\tfrac12\;=\;1,
\]
against the last clause of Definition~\ref{def:web}. So $j=k$.

\emph{A partial matching.} Fix $t\ge\tfrac12$. A bridge at $t$ is a pair $\{i,j\}$ with $i\ne j$,
since $w(i,i)=0\le t$; and $w(i,j)>t\ge\tfrac12$. If a party $i$ lay in two distinct bridges
$\{i,j\}$ and $\{i,k\}$, then $j\ne k$ and both weights would exceed $\tfrac12$, which the first part
excludes. So the bridges at $t$ are pairwise disjoint two-element sets, which is a partial matching.

\emph{No complete graph.} Let $m\ge3$ and take the parties $0,1,2$. If every pair were a bridge at
some $t\ge\tfrac12$, then $\{0,1\}$ and $\{0,2\}$ would both be bridges, so $w(0,1)>\tfrac12$ and
$w(0,2)>\tfrac12$, and the first part would give $1=2$.
\end{proof}

\begin{theorem}\label{thm:matching}
For $m$ even let $w_{\mathrm{m}}(i,j) := 1$ if $i \neq j$ and $\lfloor i/2 \rfloor = \lfloor j/2
\rfloor$, and $0$ otherwise. Then $w_{\mathrm{m}}$ is a web, and for every $t < 1$ every party holds a
bridge: $\forall i\ \exists j \neq i : w_{\mathrm{m}}(i,j) > t$. A web of any even size carries $m/2$
simultaneous bridges.
\end{theorem}

\begin{proof}
\emph{A web.} The defining condition is symmetric in $i$ and $j$, so $w_{\mathrm m}$ is symmetric; it
requires $i\ne j$, so the diagonal vanishes; and the values are $0$ and $1$, so the table is
nonnegative. At a party $i$ put $q=\lfloor i/2\rfloor$. The integers $j\ge0$ with $\lfloor j/2\rfloor=q$
are exactly $2q$ and $2q+1$. Both lie in $[m]$: $2q\le i\le m-1$, and since $2q$ is even and $m-1$ is
odd, $2q\le m-2$, so $2q+1\le m-1$. One of the two is $i$, so the row sum at $i$ counts exactly one
$j$ and equals $1\le1$.

\emph{A bridge at every party.} Let $p(i)=i+1$ for $i$ even and $p(i)=i-1$ for $i$ odd; this is the
element of $\{2q,2q+1\}$ other than $i$, so $p(i)\ne i$, $p(i)\in[m]$, and
$w_{\mathrm m}(i,p(i))=1>t$ for every $t<1$.

\emph{Simultaneous bridges.} The pairs $\{2q,2q+1\}$ for $q=0,\dots,m/2-1$ are $m/2$ pairwise disjoint
pairs, each of weight $1$, so for every $t\in[\tfrac12,1)$ they are $m/2$ bridges of the one web
$w_{\mathrm m}$, every party lying in exactly one of them.
\end{proof}

\begin{proposition}\label{prop:matching-bridges}
For every partial matching $\mathcal{W}$ on $[m]$, the table $w_{\mathcal{W}}(i,j):=[\{i,j\}\in\mathcal{W}]$
is a web whose bridges at every $t\in[\tfrac12,1)$ are exactly the pairs of $\mathcal{W}$. The bridge
sets of webs at such $t$ are therefore exactly the partial matchings.
\end{proposition}

\begin{proof}
The table is symmetric, vanishes on the diagonal and is nonnegative, and its row sum at $i$ is the
number of pairs of $\mathcal{W}$ containing $i$, at most one; so it is a web. Its entries are $1>t$ on
the pairs of $\mathcal{W}$ and $0\le t$ elsewhere, so its bridges at $t$ are the pairs of
$\mathcal{W}$. Conversely the
bridges of any web at $t\ge\tfrac12$ form a partial matching by Theorem~\ref{thm:bridge-matching}.
\end{proof}

\begin{proposition}\label{prop:uniform-transition}
At two parties the equidistant web is entirely bridge, $w_{\mathrm{u}}(0,1) = 1 > \tfrac12$; at three
or more it has no bridge, $w_{\mathrm{u}}(i,j) = \tfrac{1}{m-1} \leq \tfrac12$.
\end{proposition}

\begin{proof}
At $m=2$, $w_{\mathrm u}(0,1)=\tfrac1{2-1}=1>\tfrac12$, so its only pair is a bridge at every
$t\in[\tfrac12,1)$. At $m\ge3$, $m-1\ge2$ gives $w_{\mathrm u}(i,j)=\tfrac1{m-1}\le\tfrac12$ for every
pair, while a bridge at $t$ needs a weight above $t\ge\tfrac12$; so no pair is a bridge at any
threshold.
\end{proof}

So, for a fixed factorisation into parties, the lift goes through in both halves: the thin thread to every party of every web, and a thread
of weight one, the monogamous pair that the two-agent $\mathrm{ER}=\mathrm{EPR}$ statement takes as
its hypothesis \cite[Corollary~2]{FieldsGlazebrookMarcianoZappala2025}, at every pair of a partial matching. For every bridge threshold in $[\tfrac12,1)$ monogamy forbids a party
two bridges and nothing more, the bridge sets of webs being exactly the partial matchings
(Proposition~\ref{prop:matching-bridges}), so the complete graph of bridges on three or more parties
is excluded and every partial matching is allowed. The equidistant web passes from all bridge to no bridge
between two and three parties (Proposition~\ref{prop:uniform-transition}).

\section{Two towers}\label{sec:towers}

This section sets two towers of algebras side by side: the matrix algebras of the fuzzy sphere and
the finite-dimensional stages of an AF algebra. In this section $N$ and $M$ denote the stage and the
matrix size of the fuzzy sphere, not the size of a screen or its sign observables.
Rieffel's convergence
of matrix algebras to the sphere is stated without a rate, an explicit stage being only remarked to
follow from careful bookkeeping \cite{Rieffel2004}; the quantized distance gives an
$\varepsilon$--$N$ convergence with no rate \cite{Wu2006}; and spectral truncations of tori converge
with an approximation constant shown only to tend to zero as the truncation is lifted
\cite{LeimbachVanSuijlekom2024}.

At stage $N$ of the fuzzy sphere of \cite{DAndreaLizziVarilly2013} the algebra is
$M_{N+1}=\mathrm{End}(V_{N/2})$, on which $SU(2)$ acts by conjugation in the spin-$N/2$
representation $V_{N/2}$, with the Dirac operator $D_N$ built from it and the spin-$\tfrac12$
representation, and the Connes spectral distance between the two polar coherent states is in closed
form \cite{DAndreaLizziVarilly2013}
\begin{equation}\label{eq:pole}
\rho_N(\pi) \;=\; \sum_{k=1}^{N} \frac{1}{\sqrt{k(N-k+1)}} .
\end{equation}
Written in the number $M=N+1$ of subintervals rather than the number $N$ of interior nodes,
\eqref{eq:pole} is the Riemann sum of step $1/M$ at the interior nodes for the density
$(x(1-x))^{-1/2}$, since $k(N-k+1)=M^2\,\tfrac kM\bigl(1-\tfrac kM\bigr)$, and the integral of that
density over $(0,1)$ is $\pi$, the length of a half great circle of the unit sphere, by the
substitution $x=\sin^2y$ (Step~1 of the proof below).

\begin{theorem}\label{thm:zeta}
With $M = N+1$,
\[
\pi - \rho_N(\pi) \;=\; \frac{-2\zeta(1/2)}{\sqrt{M}} \;+\; O\!\left(M^{-3/2}\right),
\qquad -2\zeta(1/2) \;\in\; \bigl[\, 2,\; 4\sqrt{2}-2 \,\bigr] .
\]
\end{theorem}

\begin{proof}
Throughout $N\ge1$, so $M\ge2$; $h=1/M\le\tfrac12$; and $S_n=\sum_{k=1}^nk^{-1/2}$.

\emph{Step 1, the variable.} Since $k(N-k+1)=k(M-k)=M^2\,kh\,(1-kh)$, with
$g(x)=(x(1-x))^{-1/2}$ each summand of \eqref{eq:pole} is $M^{-1}g(kh)=h\,g(kh)$, and
\[
\rho_N(\pi)=\sum_{k=1}^{M-1}h\,g(kh),
\]
an interior-node Riemann sum. The substitution $x=\sin^2y$, $y\in(0,\tfrac\pi2)$, has
$\mathrm dx=2\sin y\cos y\,\mathrm dy$ and $g(x)=(\sin y\cos y)^{-1}$,
so $\int_\varepsilon^{1-\varepsilon}g=\int2\,\mathrm dy$ over the corresponding interval, and letting
$\varepsilon\to0$ gives $\int_0^1g=\pi$ as an improper integral.

\emph{Step 2, splitting the endpoints.} On $(0,1)$ write
\[
g(x)=x^{-1/2}+(1-x)^{-1/2}+r(x).
\]
Since $\int_0^1x^{-1/2}\,\mathrm dx=\int_0^1(1-x)^{-1/2}\,\mathrm dx=2$, and $r$ will be shown
continuous on $[0,1]$, linearity of the improper integrals gives $\int_0^1r=\pi-4$. The first singular
term contributes $\sum_{k=1}^{M-1}h(kh)^{-1/2}=M^{-1/2}S_{M-1}$, and the second, under the bijection
$k\mapsto M-k$ of $\{1,\dots,M-1\}$, contributes the same. Hence
\begin{equation}\label{eq:rho-split}
\rho_N(\pi)=2M^{-1/2}S_{M-1}+h\sum_{k=1}^{M-1}r(kh).
\end{equation}

\emph{Step 3, the partial sums of $k^{-1/2}$.} Let $B(x)=\{x\}-\tfrac12$. For $f$ continuously
differentiable on $[k,k+1]$, integration by parts gives
$\int_k^{k+1}(x-k-\tfrac12)f'(x)\,\mathrm dx=\tfrac12f(k+1)+\tfrac12f(k)-\int_k^{k+1}f$, that is,
$f(k+1)=\int_k^{k+1}f+\tfrac12(f(k+1)-f(k))+\int_k^{k+1}Bf'$. Summing over $k=n,\dots,K-1$,
\[
\sum_{k=n+1}^{K}f(k)=\int_n^Kf+\tfrac12\bigl(f(K)-f(n)\bigr)+\int_n^KB(x)f'(x)\,\mathrm dx ,
\]
the first-order Euler--Maclaurin formula. Take $f(x)=x^{-s}$ with $s$ real, $s>1$, and let
$K\to\infty$: $f(K)\to0$, $\int_n^\infty x^{-s}\,\mathrm dx=n^{1-s}/(s-1)$ and $f'(x)=-sx^{-s-1}$, so
\begin{equation}\label{eq:em}
\zeta(s)=\sum_{k=1}^nk^{-s}+\frac{n^{1-s}}{s-1}-\frac12n^{-s}-s\int_n^\infty B(x)x^{-s-1}\,\mathrm dx .
\end{equation}
For fixed integer $n\ge1$ the right side is holomorphic on $D=\{\operatorname{Re}s>0\}\setminus\{1\}$.
The finite sum and the two powers are, away from $s=1$. For the integral, each truncation
$\int_n^KB(x)x^{-s-1}\,\mathrm dx$ is entire in $s$, being an integral over a compact interval of a
bounded integrand holomorphic in $s$; and since $|B|\le\tfrac12$, the tail beyond $K$ is at most
$\tfrac12\int_K^\infty x^{-\sigma_0-1}\,\mathrm dx=K^{-\sigma_0}/(2\sigma_0)$ on
$\operatorname{Re}s\ge\sigma_0>0$, so the truncations converge locally uniformly on
$\operatorname{Re}s>0$ and the limit is holomorphic by Weierstrass' theorem. The Riemann zeta function
is holomorphic on $\Cc\setminus\{1\}$, and the two sides of \eqref{eq:em} agree on the real
half-line $s>1$, which has accumulation points in the connected open set $D$; by the identity theorem
they agree on all of $D$. At $s=\tfrac12$, where $n^{1/2}/(-\tfrac12)=-2\sqrt n$, this reads
\begin{equation}\label{eq:sn}
S_n=2\sqrt n+\zeta(\tfrac12)+\tfrac12n^{-1/2}+\tfrac12E_n,\qquad
E_n=\int_n^\infty B(x)x^{-3/2}\,\mathrm dx .
\end{equation}
To bound $E_n$, let $P(x)=(\{x\}^2-\{x\})/2$. On each $[k,k+1]$ it is the polynomial
$((x-k)^2-(x-k))/2$, with derivative $x-k-\tfrac12=B(x)$ and with $P(k)=P(k+1)=0$; its minimum, at
the midpoint, is $-\tfrac18$, so $P\in[-\tfrac18,0]$. Integrating by parts on each $[k,k+1]$, the
boundary terms vanish and
$\int_k^{k+1}Bx^{-3/2}\,\mathrm dx=\tfrac32\int_k^{k+1}Px^{-5/2}\,\mathrm dx$. Summing over $k\ge n$,
\[
|E_n|=\tfrac32\Bigl|\int_n^\infty P(x)x^{-5/2}\,\mathrm dx\Bigr|\le\tfrac32\cdot\tfrac18\cdot\tfrac23n^{-3/2}
=\tfrac18n^{-3/2}.
\]

\emph{Step 4, the endpoint sums.} Apply \eqref{eq:sn} at $n=M-1\ge1$ and multiply by $2M^{-1/2}$. With
$u=1/M\le\tfrac12$,
\[
2M^{-1/2}S_{M-1}=4\sqrt{1-u}+2\zeta(\tfrac12)M^{-1/2}+u(1-u)^{-1/2}+M^{-1/2}E_{M-1}.
\]
Three elementary bounds expand the terms. First, $\phi(u)=1-\tfrac u2-\sqrt{1-u}$ has $\phi(0)=\phi'(0)=0$
and $0\le\phi''(u)=\tfrac14(1-u)^{-3/2}\le\tfrac14\cdot2^{3/2}<1$ on $[0,\tfrac12]$, so by Taylor's
theorem $0\le\phi(u)\le\tfrac12u^2$ and $4\sqrt{1-u}=4-2u-4\phi(u)$ with $0\le4\phi(u)\le2u^2$.
Second, by the mean value theorem $0\le(1-u)^{-1/2}-1\le\tfrac12(1-u)^{-3/2}u\le\sqrt2\,u$, so
$u(1-u)^{-1/2}=u+\theta_1$ with $0\le\theta_1\le\sqrt2u^2$. Third, $M-1\ge M/2$, so
$M^{-1/2}|E_{M-1}|\le\tfrac18M^{-1/2}(M/2)^{-3/2}=\tfrac{\sqrt2}{4}M^{-2}$. Collecting, with $u=M^{-1}$,
\begin{equation}\label{eq:endpoint}
2M^{-1/2}S_{M-1}=4+2\zeta(\tfrac12)M^{-1/2}-M^{-1}+R_1,\qquad|R_1|\le\bigl(2+\sqrt2+\tfrac{\sqrt2}4\bigr)M^{-2}\le4M^{-2}.
\end{equation}

\emph{Step 5, the remainder function.} The binomial series
$(1-x)^{-1/2}=\sum_{j\ge0}c_jx^j$, with $c_j=\binom{2j}{j}4^{-j}>0$ and $c_0=1$, converges for $|x|<1$.
Hence $\Psi(x)=((1-x)^{-1/2}-1)/x=\sum_{j\ge1}c_jx^{j-1}$ is a power series of radius one; on $[0,\tfrac12]$
it and its first two derivatives are nonnegative and increasing, so bounded by their values at
$\tfrac12$, and $\Psi(\tfrac12)=2\sqrt2-2<1$. For $x\in(0,1)$,
$g(x)-x^{-1/2}=x^{-1/2}((1-x)^{-1/2}-1)=x^{1/2}\Psi(x)$, so
\[
r(x)=x^{1/2}\Psi(x)-(1-x)^{-1/2}\qquad(0<x<1).
\]
As $x\to0^+$ this tends to $-1$. Since $g(1-x)=g(x)$ and the two singular terms are exchanged by
$x\mapsto1-x$, $r(1-x)=r(x)$, and $r$ tends to $-1$ at $1$ as well. Setting $r(0)=r(1)=-1$ makes $r$
continuous on $[0,1]$ and twice continuously differentiable on $(0,1)$. Two estimates hold.
\begin{enumerate}[label=(\roman*)]
\item For $0<x\le\tfrac12$, $r(x)-r(0)=x^{1/2}\Psi(x)-\bigl((1-x)^{-1/2}-1\bigr)=x^{1/2}\Psi(x)-x\Psi(x)
=x^{1/2}(1-x^{1/2})\Psi(x)$, so $0\le r(x)-r(0)\le x^{1/2}$.
\item On $(0,\tfrac12]$, $(x^{1/2}\Psi)''=-\tfrac14x^{-3/2}\Psi+x^{-1/2}\Psi'+x^{1/2}\Psi''$; since
$x^{-1/2}=x\cdot x^{-3/2}\le\tfrac12x^{-3/2}$ and $x^{1/2}=x^2x^{-3/2}\le\tfrac14x^{-3/2}$ there,
$|(x^{1/2}\Psi)''|\le(\tfrac14\Psi(\tfrac12)+\tfrac12\Psi'(\tfrac12)+\tfrac14\Psi''(\tfrac12))x^{-3/2}$.
Also $|((1-x)^{-1/2})''|=\tfrac34(1-x)^{-5/2}\le\tfrac34\cdot2^{5/2}=3\sqrt2\le\tfrac32x^{-3/2}$, as
$x^{-3/2}\ge2^{3/2}$. So $|r''(x)|\le Cx^{-3/2}$ on $(0,\tfrac12]$ with
$C=\tfrac14\Psi(\tfrac12)+\tfrac12\Psi'(\tfrac12)+\tfrac14\Psi''(\tfrac12)+\tfrac32$, and by the symmetry
$|r''(x)|\le C\min(x,1-x)^{-3/2}$ on $(0,1)$.
\end{enumerate}

\emph{Step 6, the trapezoid estimate.} For $f$ twice continuously differentiable on $[a,b]$, two
integrations by parts against $\mathcal{K}(x)=\tfrac12(x-a)(b-x)$, which vanishes at both ends and
has $\mathcal{K}''=-1$, give
\[
\int_a^bf-\frac{b-a}2\bigl(f(a)+f(b)\bigr)=-\int_a^b\mathcal{K}(x)f''(x)\,\mathrm dx,
\]
and since $\mathcal{K}\ge0$ with $\int_a^b\mathcal{K}=(b-a)^3/12$, the left side is at most $\tfrac{(b-a)^3}{12}\sup|f''|$ in
absolute value. Let
$T_h=h\bigl[\tfrac12r(0)+\sum_{k=1}^{M-1}r(kh)+\tfrac12r(1)\bigr]$, the trapezoid sum of $r$ on the
nodes $kh$. Its error $T_h-\int_0^1r$ is the sum of the errors on the $M$ intervals.
On $[kh,(k+1)h]$ with $1\le k\le M-2$, every $x$ has $x\ge kh$ and $1-x\ge(M-k-1)h$, so by (ii)
$\sup|r''|\le C\bigl((kh)^{-3/2}+((M-k-1)h)^{-3/2}\bigr)$, and the error there is at most
$\tfrac C{12}h^{3/2}\bigl(k^{-3/2}+(M-k-1)^{-3/2}\bigr)$. As $k$ runs from $1$ to $M-2$ both $k$ and
$M-k-1$ run over $\{1,\dots,M-2\}$, so these errors total at most
$\tfrac C6\zeta(\tfrac32)h^{3/2}$, the series $\sum k^{-3/2}$ converging. On $[0,h]$, where $r''$ is
unbounded, write the error as $\int_0^h(r-r(0))-\tfrac h2(r(h)-r(0))$; by (i), with $h\le\tfrac12$, it
is at most $\int_0^hx^{1/2}\,\mathrm dx+\tfrac h2h^{1/2}=\tfrac76h^{3/2}$. By the symmetry
$r(1-x)=r(x)$ the same bound holds on $[1-h,1]$. Hence
\[
\Bigl|T_h-(\pi-4)\Bigr|\le C_2M^{-3/2},\qquad C_2=\tfrac73+\tfrac C6\zeta(\tfrac32).
\]
Since $r(0)+r(1)=-2$,
\begin{equation}\label{eq:remainder}
h\sum_{k=1}^{M-1}r(kh)=T_h-\tfrac h2\bigl(r(0)+r(1)\bigr)=\pi-4+M^{-1}+R_2,\qquad|R_2|\le C_2M^{-3/2}.
\end{equation}

\emph{Step 7, adding.} Substituting \eqref{eq:endpoint} and \eqref{eq:remainder} into
\eqref{eq:rho-split}, the constants $4$ and $\pi-4$ add to $\pi$ and the terms $-M^{-1}$ and $+M^{-1}$
cancel, so
\[
\rho_N(\pi)=\pi+2\zeta(\tfrac12)M^{-1/2}+R_1+R_2,\qquad
\Bigl|\pi-\rho_N(\pi)-\frac{-2\zeta(1/2)}{\sqrt M}\Bigr|\le(4+C_2)M^{-3/2}
\]
for every $N\ge1$, using $M^{-2}\le M^{-3/2}$. This is the expansion with its error term.

\emph{Step 8, the bracket.} Let $a_n=S_n-2\sqrt n$ and $b_n=S_n-2\sqrt{n+1}$. By \eqref{eq:sn} and
$|E_n|\le\tfrac18n^{-3/2}$, $a_n\to\zeta(\tfrac12)$, and $b_n-a_n=-2(\sqrt{n+1}-\sqrt n)\to0$, so
$b_n\to\zeta(\tfrac12)$ as well. Rationalising, $2(\sqrt{n+1}-\sqrt n)=2/(\sqrt{n+1}+\sqrt n)$, and
since $\sqrt n\le\sqrt{n+1}\le\sqrt{n+2}$,
\begin{gather*}
a_{n+1}-a_n=(n+1)^{-1/2}-\frac2{\sqrt{n+1}+\sqrt n}\le0,\\
b_{n+1}-b_n=(n+1)^{-1/2}-\frac2{\sqrt{n+2}+\sqrt{n+1}}\ge0 .
\end{gather*}
So $a_n$ decreases to $\zeta(\tfrac12)$ and $b_n$ increases to it, and
$1-2\sqrt2=b_1\le\zeta(\tfrac12)\le a_1=-1$. Multiplying by $-2$ gives
$2\le-2\zeta(\tfrac12)\le4\sqrt2-2$.
\end{proof}

On the other side the comparison is with an AF algebra, a $C^*$-inductive limit of a
sequence of finite-dimensional $C^*$-algebras, on which Aguilar and Latr\'emoli\`ere build a
metric from a faithful tracial state and a sequence $\beta$ \cite{AguilarLatremoliere2015}. A $(C,D)$-quasi-Leibniz quantum compact
metric space is a unital $C^*$-algebra with a lower semicontinuous seminorm $L$ defined on a dense
Jordan--Lie subalgebra of the self-adjoint part, vanishing exactly on the scalars, whose dual metric
on states metrises the weak$^*$ topology, and which bounds the Jordan and Lie products by
$C(\|a\|L(b)+\|b\|L(a))+DL(a)L(b)$, with $C\ge1$ and $D\ge0$ \cite{AguilarLatremoliere2015}. For
fixed $C$ and $D$ the quantum propinquity $\Lambda$ is a metric, up to isometric isomorphism, on
the class of such spaces \cite{AguilarLatremoliere2015}.

Let $\Aa$ be an AF algebra with faithful tracial state $\operatorname{tr}$, the $C^*$-inductive limit
of finite-dimensional $(\Aa_n,\iota_n)$ with $\Aa_0 = \Cc$ and every $\iota_n$ unital and injective,
$\mathbb{E}_n$ the $\operatorname{tr}$-preserving conditional expectation onto the image of $\Aa_n$, and
$\iota^{(n)}$ the induced map of $\Aa_n$ into that image. Call $\beta : \Nn \to (0,\infty)$
\emph{admissible} when it has limit $0$, monotonicity not being required, and put
$L^{\beta}(a) = \sup_n \|a - \mathbb{E}_n(a)\|/\beta(n)$ on the self-adjoint part. For every
admissible $\beta$, Aguilar and Latr\'emoli\`ere prove that $(\Aa,L^{\beta})$ is a
$(2,0)$-quasi-Leibniz quantum compact metric space and that the quantum propinquity obeys
$\Lambda((\Aa_n, L^{\beta}\circ\iota^{(n)}),(\Aa,L^{\beta})) \le \beta(n)$ at every stage
\cite{AguilarLatremoliere2015}.

\begin{proposition}\label{prop:af}
There is no $w : \Nn \to \Rr$ with $\beta(n) = w(n)$ for every admissible $\beta$ and every $n$.
\end{proposition}

\begin{proof}
The sequences
$\beta_1(n)=\tfrac12 2^{-n}$, $\beta_2(n)=\tfrac12 3^{-n}$ and $\beta_3(n)=1/(n+1)$ take positive
values and tend to $0$, since $2^{-n}$, $3^{-n}$ and $1/(n+1)$ do; so all three are admissible. At $n=3$ they
take the values $\tfrac12\cdot\tfrac18=\tfrac1{16}$, $\tfrac12\cdot\tfrac1{27}=\tfrac1{54}$ and
$\tfrac14$, which are pairwise distinct. If some $w:\Nn\to\Rr$ had $\beta(n)=w(n)$ for every admissible
$\beta$ and every $n$, then $w(3)=\beta_1(3)=\tfrac1{16}$ and $w(3)=\beta_2(3)=\tfrac1{54}$, which is
impossible. So no such $w$ exists.
\end{proof}

The Aguilar--Latr\'emoli\`ere bound at stage $n$ is the value $\beta(n)$ of the admissible sequence
chosen \cite{AguilarLatremoliere2015}, and by Proposition~\ref{prop:af} no single sequence of
numbers is that value for every admissible choice.

\begin{corollary}\label{cor:kind}
With $M=N+1$, the noncommutative side is determined by its own Dirac operator,
\[
\sqrt{M}\bigl(\pi - \rho_N(\pi)\bigr) \;\longrightarrow\; -2\zeta(1/2),
\]
while for every $d : \Nn \to \Rr$ and every stage $n$ the AF side is not: not every
admissible $\beta$ has $\beta(n)=d(n)$.
\end{corollary}

\begin{proof}
\emph{The noncommutative side.} Step 7 of the proof of Theorem~\ref{thm:zeta} gives
$|\pi-\rho_N(\pi)-(-2\zeta(\tfrac12))M^{-1/2}|\le(4+C_2)M^{-3/2}$ for every $N\ge1$. Multiplying by
$\sqrt M$,
\[
\bigl|\sqrt M\bigl(\pi-\rho_N(\pi)\bigr)-\bigl(-2\zeta(\tfrac12)\bigr)\bigr|\le(4+C_2)M^{-1}\longrightarrow0 .
\]

\emph{The AF side.} The admissible sequences $\beta_1(n)=\tfrac12 2^{-n}=2^{-(n+1)}$ and
$\beta_3(n)=1/(n+1)$ of the proof of Proposition~\ref{prop:af} differ at every $n$, because
$2^{n+1}>n+1$ for every $n\ge0$: at $n=0$ this is $2>1$, and $2^{n+1}>n+1$ gives
$2^{n+2}=2\cdot2^{n+1}>2(n+1)\ge n+2$. Fix $d$ and $n$. If every admissible $\beta$ had
$\beta(n)=d(n)$, then $\beta_1(n)=d(n)=\beta_3(n)$, which is false; so some admissible $\beta$ has
$\beta(n)\ne d(n)$.
\end{proof}

One side supplies a distance between two states of one algebra, fixed with its rate by the Dirac
operator; the other a bound from a stage to its limit, equal to the value $\beta(n)$ of the admissible
sequence chosen \cite{AguilarLatremoliere2015}.

\section{Discussion}\label{sec:discussion}

With nonnegative weights, a unit prefactor and joint outcomes that each pick out a single state, the spectrum of the quantum-scale sky depends on the brightness of its stars alone; two stargazers differ only through the constellations by which each of them reads it, since the levels a readout reads are fixed by the weights and its own sets (Definition~\ref{def:readout}). An observer reads faithfully for every choice of brightnesses exactly when the different nonempty lists the stars carry are linearly independent over the field with two elements; once its choices are recoded, such an observer sees the sky in non-overlapping pieces. Resolved without end, the sky becomes dust where the observer keeps the settled values; any two different ones are told apart by every positive weighing that stays consistent from one stage to the next, and reality for the observer, as we took it in the introduction, is the separation and concordance of its readings as its view is refined. The limits of the observer's constellations are, to adapt Wittgenstein's phrase \cite{Wittgenstein1922}, the
limits of its world.

Under the hypothesis of Proposition~\ref{prop:cube} the spectrum is a function of the profile alone
(Corollary~\ref{cor:split}), whatever basis, realisation of the observables or reference angle per
qubit is chosen (Theorem~\ref{thm:signfamily}, Proposition~\ref{prop:phi}); no relabelling moves it
(Proposition~\ref{prop:relabel}) while a regrouping can (Proposition~\ref{prop:groupings-separate}).
Carrying a profile at every weighting is the linear independence of the distinct nonzero traces over the
field with two elements (Theorem~\ref{thm:dichotomy}), which yields a partition after an invertible
recoding (Corollary~\ref{cor:partition}); a readout whose distinct nonzero traces are dependent carries
a profile only on finitely many proper linear subspaces of weightings (Corollary~\ref{cor:thin}). The
dust is the profinite boundary of a tower of groupings (Theorem~\ref{thm:screen-profinite}), and every
strictly positive compatible profile separates the objects on that boundary
(Theorem~\ref{thm:metric-separates}).

Descriptions developed apart meet on this one screen. The qubit screen of Fields, Glazebrook and
Marcian\`o \cite{FieldsGlazebrookMarciano2022} is read by linear algebra over the field with two
elements (Theorem~\ref{thm:dichotomy}), and refined without end it has a boundary that is a condensed
set in the sense of Clausen and Scholze \cite{Scholze2026} (Theorem~\ref{thm:screen-profinite}). The
objects that a nonnegative profile cannot tell apart form the groupoid of an equivalence relation
(Theorem~\ref{thm:metric-separates}). The levels of contextuality of Abramsky and Brandenburger
\cite{AbramskyBrandenburger2011} order the models read here (Corollary~\ref{cor:placement}), and the
Bohrification of Heunen, Landsman and Spitters \cite{HeunenLandsmanSpitters2009} is available at
every finite stage of a tower of matrix algebras. The monogamy inequality
\cite{CoffmanKunduWootters2000,OsborneVerstraete2006} confines the weight-one threads that model the
pairs of the operational $\mathrm{ER}=\mathrm{EPR}$ \cite{FieldsGlazebrookMarcianoZappala2025} to
partial matchings (Theorem~\ref{thm:bridge-matching}), and the fuzzy sphere of D'Andrea, Lizzi and
V\'arilly \cite{DAndreaLizziVarilly2013} supplies a tower whose pole distance is computed from its own
Dirac operator (Theorem~\ref{thm:zeta}). The Markov blanket of the free energy principle, which in
many applications is partitioned into sensory and active states, is implemented by the screen
$\mathscr{B}$ itself \cite{FieldsGlazebrookMarciano2022}. In the classical limit, where the
entanglement entropies of the two systems vanish, the states of $\mathscr{B}$ become the blanket states
of a Markov blanket that acts as a classical information channel \cite{FieldsEtAl2023}. The
partition of a blanket's states into sensory and active ones is drawn in the causal-network
representation, where Pearl first defined the blanket \cite{FieldsGlazebrookMarciano2022}. On the
screen of Figure~\ref{fig:screen} the same operators measure and prepare, each qubit being prepared by
one system and then measured by the other \cite{FieldsGlazebrookMarciano2022}. We therefore read
the sensory and active roles of $\Omega$ as its measurements and its preparations through one and the
same quantum reference frame, told apart in time as $\Omega$ prepares and $\mho$ measures or the reverse.
In this reading the state of each qubit is sensory for $\Omega$ while $\Omega$ measures it and active
while $\Omega$ prepares it.

Oggad \cite{Oggad2026} shows that on a structure with a single relaxation scale a diffusion probe
returns two without any dynamics, at that scale and in the limit of many samples, and gives a spectral
test based on the density of eigenvalues near zero that tells such a reading from a dimension the
structure carries. Here the plateau that Convention~\ref{conv:dial} reads off a geometric profile can
be set to any positive value by the ratio of the weights (Theorem~\ref{thm:plateau}). At equal weights summing to one the running dimension stays under Oggad's line $D_S=2t$ and falls short of two at unit time by a margin whose
bound shrinks as qubits are added (Corollary~\ref{cor:uniform-line}). For every number of qubits the halving weights normalised to total one are collected by grouping some equal-weight screen (Lemma~\ref{lem:rational}); with
those weights the running dimension stays below two at every time, and once the screen has enough
qubits for a prescribed margin it comes within that margin of two on a stretch of time that grows with
the number of qubits (Theorem~\ref{thm:dyadic}).

\bibliographystyle{plain}
\bibliography{references}

\end{document}